\documentclass[11pt]{article}
\usepackage{amsmath,amsfonts,amssymb,eucal}
\usepackage[all]{xy}
\advance \textheight by \topskip 
\usepackage[colorlinks=true,urlcolor=blue]{hyperref}

\hypersetup{				
backref = true,				
pagebackref = true,			
hyperindex = true, 			
colorlinks = true, 			
breaklinks = true, 			
urlcolor = blue, 			
linkcolor = blue, 			
bookmarks = true,			
bookmarksopen = true, 		        
}
\makeatletter
\@addtoreset{equation}{section}
 
\makeatother

\newtheorem{theorem}{Theorem}[section]

\newtheorem{definition}[theorem]{Definition}

\def\NN{\mathbb N}
\def\ZZ{\mathbb Z}

\def\RR{\mathbb R}

\def\g{\mathfrak{g}}
\def\D{{\cal D}}
\def\tvi{\vrule height 12pt depth 6pt width 0pt}
\def\tv{\tvi\vrule}
\def\cc#1{\kern .7em\hfill #1 \hfill\kern .7em}
\newcommand{\beq}{\begin{equation}}
\newcommand{\eeq}{\end{equation}}
\newcommand{\beqa}{\begin{eqnarray}}
\newcommand{\eeqa}{\end{eqnarray}}
\newcommand{\e}{\varepsilon}
\newcommand{\ttA}{{\tilde {\tilde A}}}
\newcommand{\tA}{\tilde  A}

\newcommand{\tB}{\tilde  B}
\newcommand{\hA}{\hat  A}
\newcommand{\hhA}{{\hat {\hat A}}}
\newcommand{\hB}{\hat  B}
\newcommand{\hhB}{{\hat {\hat B}}}

\newcommand{\noi}{\noindent}

\def\>{\rangle}
\def\<{\langle}

\date{\today}

\begin{document}
\begin{titlepage}

\title{
{\bf Some Results on Cubic and Higher Order Extensions of the Poincar\'e 
Algebra
}}

\author{
{\sf M. Rausch de Traubenberg}\thanks{e-mail:
rausch@lpt1.u-strasbg.fr}\,\,
\\
{\small {\it
Laboratoire de Physique Th\'eorique, CNRS UMR  7085,
Universit\'e Louis Pasteur}}\\
{\small {\it  3 rue de l'Universit\'e, 67084 Strasbourg Cedex, France}}\\ 
}
\date{}
\maketitle
\vskip-1.5cm

\vspace{2truecm}

\begin{abstract}
\noindent
In these lectures we study  some possible  higher order
(of degree greater than two)  extensions of
the Poincar\'e algebra.
We first give some general properties
of  Lie superalgebras with some emphasis on the supersymmetric extension
of the Poincar\'e algebra or Supersymmetry.
Some general features on the so-called
Wess-Zumino model (the simplest field theory invariant under Supersymmetry)
are then given.
We further introduce an additional
algebraic structure called Lie algebras of order F, which naturally comprise
 the concepts
of ordinary Lie algebras and superalgebras.
This structure enables
us to define various non-trivial  extensions of the Poincar\'e algebra.
These extensions are
studied more precisely in two different contexts.
The first algebra  we are considering
is shown to be
 an (infinite dimensional) higher order extension
of the Poincar\'e algebra in $(1+2)-$dimensions  and turns out to induce
a symmetry which connects relativistic anyons.
The second extension we are studying is related
to a specific finite dimensional Lie algebra of order three, which
is a cubic extension of the Poincar\'e algebra in $D-$space-time
dimensions. Invariant Lagrangians are constructed.

\end{abstract}

\bigskip \bigskip \bigskip

\noi
{\it Mini course given at the Workshop
higher symmetries in physics,
 Madrid, Spain,  November 6-8, 2008.}
\end{titlepage}
\newpage
\begin{titlepage}
\tableofcontents
\end{titlepage}
\newpage

\vspace{2cm}

\newpage
\section{Introduction}

Describing   the laws of physics in terms of  underlying symmetries has 
always been a powerful tool. 
For instance the Casimir operators of the Poincar\'e algebra 
are related to the mass and the spin of elementary particles. 
Moreover, it has been understood that all the fundamental
interactions (electromagnetic, weak and strong) are related to
the Lie algebra $\mathfrak{u}(1)_Y \times \mathfrak{su}(2)_L 
\times \mathfrak{su}(3)_c$,
in the so-called Standard Model.  The Standard Model is then 
described by the Lie algebra
$ I\mathfrak{so}(1,3) \times \mathfrak{u}(1)_Y \times 
\mathfrak{su}(2)_L \times \mathfrak{su}(3)_c$, where $I\mathfrak{so}(1,3)$ is related
to space-time symmetries and  $\mathfrak{u}(1)_Y \times 
\mathfrak{su}(2)_L \times \mathfrak{su}(3)_c$ to internal symmetries.
The elementary particles, namely, the quarks, the leptons and the Higgs boson
are in appropriate
representations  of $\mathfrak{u}(1)_Y \times 
\mathfrak{su}(2)_L \times \mathfrak{su}(3)_c$.
Even if the Standard  Model is the physical theory were the confrontation
between experimental results and theoretical predictions is in an extremely good
agreement, there are
strong arguments (which cannot be summarised here) that it is not the final theory.
 
Thus, to understand the properties of elementary particles,  it is then interesting to study
the kind of symmetries which are allowed in space-time. Within the
framework of Quantum Field Theory (unitarity of the $S$ matrix {\it etc}.),
S. Coleman and  J. Mandula have  shown  \cite{cm} that if the symmetries are
described in terms of Lie algebras,  only trivial extensions
of the Poincar\'e algebra can be obtained. Namely, the fundamental symmetries are
based on $ I\mathfrak{so}(1,3) \times \mathfrak{g}$
with $  \mathfrak{g}  \supseteq \mathfrak{u}(1)_Y \times \mathfrak{su}(2)_L \times \mathfrak{su}(3)_c$
a compact  Lie algebra describing the fundamental interactions and
$[ I\mathfrak{so}(1,3),\mathfrak{g}]=0$. Several algebras, in relation to
phenomenology, have been investigated (see {\it e.g.} \cite{ gut}) such as $\mathfrak{su}(5), \mathfrak{so}(10), 
\mathfrak{e}_6$ 
{\it etc.} Such theories are usually refer to  ``Grand-Unified-Theories'' or  GUT, {\it i.e.}
theories  which unify all the fundamental interactions. The fact that elements of
$\mathfrak{g}$ and $I\mathfrak{so}(1,3)$ commute means that we have a trivial
extension of the Poincar\'e algebra.

Then, R.~Haag, J.~T.~Lopuszanski and M.~F.~Sohnius \cite{hls} understood that is was
possible to extend in a non-trivial way,
{\it i.e.} to introduce new generators
which are not Lorentz scalars but spinors,  the symmetries of space-time within
the framework of Lie superalgebras (see Definition \ref{superlie}).
This extension is  unique
 (up to the number of supercharges) and  called supersymmetry.
For a discussion of the Coleman \& Mandula theorem see also
the Appendix B of \cite{wei}.
 
In the context of the two above mentioned theorems, two mathematical
structures emerge naturally in the description of the symmetries
in physics: the Lie algebras and the Lie superalgebras. 
Since, the
properties of elementary particles are dictated by its symmetries, it
should be interesting to know,  whether or not other types of symmetries
might be possible.
The main feature of the above descriptions of physical symmetries
 are based upon algebras, {\it i.e.} vector spaces equipped
with a binary multiplication, the commutator or the anticommutator. 
This observation  leads immediately to
the simple question: would it
be possible to obtain higher order algebras in Quantum Field Theory?
 Of course these new types of symmetries should not
be in conflict with the basic principle of Quantum Field Theory.
Furthermore, since we do not want to leave aside the basic principles of 
Relativity this new structure, if acceptable, should in some sense reproduce
the Poincar\'e symmetries.
Ternary algebras  and higher order algebras
have been considered in physics only occasionally
(see for instance \cite{nam,bg1,bg2,k1,k2,k3,az,r} and references therein).
For some mathematical references one can see 
\cite{fi,g,mv,gr}. Recently they was some revival of interest in 
ternary algebras when it has been realised that a
 ternary algebra defined by a fully
antisymmetric product appears in the description
of  multiple M2-branes \cite{bala} (see also \cite{ff}).

In a series of papers \cite{flie1,flie2,flie3} a specific   
$F-$ary algebra, called Lie algebra of order $F$
was introduced. This algebra presents the interesting feature to have
a Lie algebra as a subalgebra. It is thus a kind of hybrid mathematical
structure with two products: one binary and the second of order $F$. 
Furthermore, Lie algebras of order $F$
can be seen as a possible
generalisation of Lie (super)algebras.
A Lie algebra of order $F$  admits a $\mathbb Z_F-$grading, 
the zero-graded part
being a Lie algebra. An $F-$fold symmetric product (playing the role of
the anticommutator in the case $F=2$)  expresses the zero graded  part
in terms of the non-zero graded part. 
In the same manner than the Lie (super)algebras lead to description
of (extended) space-time symmetries, Lie algebras of order $F$ allow
to construct higher order extensions of the Poincar\'e algebras.
See also \cite{extpo} for different extensions
 of the Poincar\'e algebra.

These lectures are devoted to give some examples of higher order extensions of
the Poincar\'e algebra. These extensions will be studied with
many details, and  a collection of
useful technical points will be given in some appendices.
We also give some emphasise on the way we by-pass the no-go theorems.
Since these types of structures can be seen as a possible ``generalisations''
of supersymmetry, and in order to stress on the analogy on both
structures, we give in   Section 2 of  this lecture some well-known
results on supersymmetry. Some results upon  the Wess-Zumino model are given.
Section three is the main subject of these lectures.
We firstly give the precise definition of Lie algebra of order $F$ that we
apply in two different contexts to extend the 
Poincar\'e symmetries. 
It is known that small dimensional space-times present exceptional 
properties. In particular, in three space-time dimensions there exists states
which are neither bosons nor fermions but anyons. We construct
a higher order extension of the Poincar\'e algebra in three space-time
dimensions \cite{fsusy3d}.  Studying its representations explicitly shows
 that this extension
induces a symmetry between relativistic anyons in a straight analogy
with supersymmetry, which is a Fermi-Bose symmetry. In the next subsection,
we construct a cubic extension of the Poincar\'e algebra in
any space-time dimensions. In this case, the cubic extension considered
turns out to induce a symmetry on generalised gauge fields or $p-$forms.
The transformation laws have a  geometrical interpretation in terms of
the natural operations on  $p-$forms. Then invariant Lagrangians are
constructed \cite{p-form}. In a series of appendices useful technical details
are given, in order to illustrate some points, or to give all the
needed definitions and identities
to prove the results. In Appendix A we give our 
conventions for spinors in four space-times dimensions. In Appendix
B, some emphasis on  relativistic wave equations for anyons 
in three space-times dimensions is given. 
The main point of our algebras is that
they are partially $F-$ary. This means that studying their representations
goes along the same lines of studying the representations of supersymmetric
algebras, but in the context of  Clifford algebras of polynomial's
instead of Clifford algebras. In Appendix C, we give some results on Clifford
algebras of polynomial's. In Appendix D,  infinite
dimensional representations of Lie algebras are studied with some emphasis
on Verma modules or
indecomposable representations. 
Appendix E is a collection of useful identities on 
spinors and $p-$forms in $4n-$dimensional space-time.

\section{Lie superalgebras and supersymmetry}

The purpose of this section is to give some general features of Lie
superalgebras
and to show its implementation in Quantum Field Theory. The basic idea is very
simple. It is a consequence of   the Noether  and spin-statistics theorems.
The Noether theorem 
establishes a deep relation between symmetries and conservation laws.
The spin-statistics theorem is related to the quantisation of fields.
It is known that, when the space-time dimension is higher than three, 
there are two types of fields,
the bosons of integer spin and the fermions of half-integer spin. The
former are quantised using commutation relations whilst the latter are 
quantised with anticommutation relations. Thus, if we assume that we have
some conserved charges of integer spin $X_1, \cdots, X_m$ and of half-integer
spin $Y_1, \cdots, Y_n$ that form a {\it closed algebra}, the only allowed
possibility, is given by an algebraic structure involving commutation
relations and anticommutation relations. Furthermore, since both sides of
the equality have to behave in the same manner with respect to the
Poincar\'e algebra, the only possibility is given by

\beqa
\label{spin-noether}
\left[X_i,X_j\right]=f_{ij}{}^k X_k, \ \
\left[X_i,Y_a\right]=R_{ia}{}^b Y_b, \ \ 
\left\{Y_a, Y_b \right\}=C_{ab}{}^i X_i.
\eeqa

\noi
The mathematical structure hidden behind relations \eqref{spin-noether} is 
called a Lie superalgebra. In fact it is the discovery of supersymmetry
in Quantum Field Theory \cite{susy,wzmod} which gave rise to the concept of Lie
superalgebra and its subsequent classification  \cite{super}.

\begin{definition}
\label{superlie}
A Lie (complex or real) superalgebra is a $\mathbb{Z}_2-$graded vector space 
$\mathfrak{g}=\mathfrak{g}_0 \oplus \mathfrak{g}_1$ endowed with the following structure
\begin{enumerate}
\item $\mathfrak{g}_0$ is a Lie algebra, we denote by $[\ ,\ ]$ the bracket on  $\mathfrak{g}_0$
($\left[\mathfrak{g}_0, \mathfrak{g}_0 \right] \subseteq \mathfrak{g}_0$);
\item $\mathfrak{g}_1$  is a representation of $\mathfrak{g}_0$
($\left[\mathfrak{g}_0, \mathfrak{g}_1 \right] \subseteq \mathfrak{g}_1$);
\item there exits a $\mathfrak{g}_0-$equivariant mapping 
$\left\{ \ ,\  \right\}: S^2\left(\mathfrak{g}_1\right) \to \mathfrak{g}_0$ where 
$S^2\left(\mathfrak{g}_1\right)$ denotes the two-fold symmetric product of $\mathfrak{g}_1$
($\left\{\mathfrak{g}_1, \mathfrak{g}_1 \right\} \subseteq \mathfrak{g}_0$);
\item The following Jacobi identities hold ($\forall \ b_1,b_2,b_3 \in \mathfrak{g}_0, \forall \  f_1,f_2,f_3 \in
 \mathfrak{g}_1$)
\beqa
\label{jacobi}
&&\left[\left[b_1,b_2\right],b_3\right] +
\left[\left[b_2,b_3\right],b_1\right] +
\left[\left[b_3,b_1\right],b_2\right] =0 \nonumber \\
&&\left[\left[b_1,b_2\right],f_3\right] +
\left[\left[b_2,f_3\right],b_1\right] +
\left[\left[f_3,b_1\right],b_2\right]  =0  \\
&&\left[b_1,\left\{f_2,f_3\right\}\right] -
\left\{\left[b_1,f_2 \right],f_3\right\}  -
\left\{f_2,\left[b_1,f_3\right] \right\} =0 \nonumber \\
&& \left[ f_1,\left\{f_2,f_{3}\right\} \right] + \left[ f_2,\left\{f_3,f_{1}\right\} \right] 
 + \left[ f_3,\left\{f_1,f_{2}\right\} \right]  =0.
 \nonumber
\eeqa
\end{enumerate}
\end{definition}

In the definition above,
the generators of zero  (resp. one) gradation are called the bosonic 
(resp. fermionic)  generators or
$\mathfrak{g}_0$ (resp. $\mathfrak{g}_1$),  is called the bosonic 
(resp.  fermionic) part of the Lie
superalgebra. The first Jacobi identity is the usual Jacobi identity for 
Lie algebras, the second says 
that $\mathfrak{g}_1$ is a representation of  $\mathfrak{g}_0$, 
the third identity is the equivariance
of $\{ \  ,\  \}$. These identities are just consequences of  
$1., 2.$ and $3.$, respectively. 
However, the fourth Jacobi identity,
which is an extra constraint, is just the $\mathbb Z_2-$graded Leibniz rule.
This means that in Definition \ref{superlie} we could have avoided
the three first Jacobi identities. Let us mention however that,
for some authors the Jacobi identities are given in the definition of Lie
superalgebras, and as a consequence they deduce that $\g_0$ is a Lie algebra
and $\g_1$ is a representation of $\g_0$. Finally all the Jacobi
identities unify. Indeed,
if for homogeneous elements $X, Y, Z$ of 
$\g$ we denote by $|X|$ {\it etc} their $\ZZ_2-$grade
and define the graded-commutator by $[X,Y]_\pm=-(-1)^{|X| |Y|} [Y,X]_\pm$, 
we have

$$
(-1)^{|Z||X|}[X,[Y,Z]_\pm]_\pm+
(-1)^{|X||Y|}[Y,[Z,X]_\pm]_\pm+
(-1)^{|Y||Z|}[Z,[X,Y]_\pm]_\pm=0.
$$

\subsection{The super-Poincar\'e algebra}
The supersymmetric extension of the Poincar\'e algebra is constructed,
in the framework of Lie superalgebras, by adjoining to the 
Poincar\'e generators anticommuting elements, called supercharges
(we denote $Q$), which belong to the spinor
representation of the Poincar\'e algebra. Thus the supersymmetric algebra is
a Lie superalgebra  $\mathfrak{g}= I\mathfrak{so}(1,d-1) \oplus {\cal S}$
with  brackets

\beqa
[L,L] =L, \ [L,P]=P,\  [L,Q]=Q,\ [P,Q]=0,\ \{Q,Q\}=P,
\eeqa

\noi
with $(L,P)$ the generators of the Poincar\'e algebra that belong to the 
bosonic part of the algebra
and $Q$  the fermionic part of the algebra which is called the superchages.
This extension is non-trivial, because the supercharges $Q$ are spinors, 
and thus 
do not commute with the generators of the Lorentz algebra.
In these lectures we are considering the simplest supersymmetric extension
of the Poincar\'e algebra, {\it i.e.} where there is only one 
spinor supercharge.
For $N-$extended  supersymmetry ($N\le 8$),
{\it i.e.} where $N$ spinor-supercharges are introduced,
the reader is referred  to the literature.
There are many good text books on the subject. For historical references
one can see \cite{west,wb,so,f}. For a more modern presentation see
{\it e.g.} \cite{terning}, and for the construction of supersymmetric
models in particle physics see for instance \cite{nilles,martin,aitch}.

We now set up our conventions.
We recall that the left- (right-)handed spinors are respectively in the
${\bf 2}$ (resp. $\overline {\bf 2}$) representation of $SL(2,\mathbb C)$.
We denote ${\bf 2}= \left<\psi_\alpha, \alpha=1,2\right>$ and  
$\overline{\bf 2}= \left<\bar \chi^{\dot \alpha}, \dot \alpha=1,2\right>$
the two-dimensional representation of $SL(2,\mathbb C)$ and its complex
conjugate representation. 
A Majorana spinor is given by 

$$\Psi_M{}= \begin{pmatrix}
\psi_\alpha \\ \bar \psi^{\dot \alpha}
\end{pmatrix},
$$

\noi 
with  $\left(\psi_\alpha\right)^*= \bar \psi_{\dot \alpha}$
(where $^*$ denotes the complex conjugation).
See  Appendix \ref{conventions} for our conventions and useful identities.
With the notations above the supersymmetric algebra is given by
$
\g= I\mathfrak{so}(1,3) \oplus {\bf 2} \oplus \overline {\bf 2}
$, with $I\mathfrak{so}(1,3)= \left< L_{\mu \nu}=-L_{\nu \mu}, P_\mu, \
\mu,\nu=0,\cdots,3 \right>$
 the Poincar\'e algebra and ${\bf 2} = \left<Q_\alpha,
\alpha=1,2\right>$,  $\overline {\bf 2}= \left<\bar Q^{\dot \alpha},
\dot \alpha=1,2\right>$
the supercharges that we take Majorana  
$(Q_\alpha)^\dag = \bar Q_{\dot \alpha}$. We now determine the various
brackets of the supersymmetric algebra:

\begin{enumerate}
\item The even-even part of the algebra is just the Lie algebra structure of
the Poincar\'e algebra
\beqa
\label{susy00}
\left[L_{\alpha \beta}, L_{\mu \nu}\right]&=&
\eta_{\beta \nu} L_{\mu \alpha} - \eta_{\alpha \nu} L_{\mu \beta} +
\eta_{\beta \mu} L_{\alpha \nu} - \eta_{\alpha \mu} L_{\beta \nu},
\nonumber  \\ 
\left[L_{\alpha \beta},P_\mu\right]&=& 
\eta_{\beta \mu} P_\alpha -\eta_{\alpha \mu} P_\beta,
\\
 \left[P_{\alpha },P_\beta\right]&=& 0. \nonumber
\eeqa

\noi with $\eta=\text{diag}(1,-1,-1,-1)$ the Minkowski metric.
We  have to emphasise that with our conventions, since our structure constants
are real, for a unitary representation, the operators $L_{\mu \nu}$ and
$P_\mu$ are antihermitian. The usual quadri-momentum
and angular momentum of  physical applications are given by 
$-iP_\mu, -iL_{\mu \nu}$, and are thus hermitian for unitary representations.

\item The odd-even part of the algebra is given by the action of the Poincar\'e
algebra on the spinors $Q_\alpha$ and $\bar Q^{\dot \alpha}$. 
The action of the Lorentz generator is known.
For the action
of $P$ on $Q$ we have {\it a priori}

$$
\left[P_\mu,Q_\alpha\right]= p \sigma_\mu{}_{\alpha \dot \alpha} 
\bar Q^{\dot \alpha},  \ \ 
\left[P_\mu,\bar Q^{\dot \alpha}\right]= p' 
\bar\sigma_\mu{}^{\dot \alpha  \alpha} 
 Q_{ \alpha}.
$$

\noi To write the R.H.S. of the equation above we have  identified
 the only tensors
with the appropriate index structure. 
But the Jacobi identity with $(P_\mu, P_\nu,
Q_\alpha)$ gives $pp'=0$, and since $Q$ is the complex conjugate of 
$\bar Q$, $p=0$ automatically implies that $p'=0$ and conversely.
Thus we have the non-vanishing brackets

\beqa
\label{susy01}
\left[L_{\mu \nu}, Q_{\alpha}\right]&=& \sigma_{\mu \nu}{}_\alpha{}^\beta
Q_\beta, \\
\left[L_{\mu \nu}, \bar Q^{\dot \alpha}\right]&=& 
\bar \sigma_{\mu \nu}{}^{\dot \alpha}{}_{\dot \beta}
\bar Q^{\dot \beta}. \nonumber
\eeqa

We have to pose for a while with this part of the algebra. Indeed, the
relations \eqref{susy01} seem to be incompatible with the fact that
we are dealing
with a real Lie superalgebra since the structure constants are complex.
\noi
However, it is known that when the space-time dimension is four one
can find a representation where the Dirac $\Gamma-$matrices are purely
imaginary ($\Gamma^M_0=\sigma_2 \otimes \sigma_1, \Gamma^M_1=i\sigma_3 \otimes
\sigma_0, \Gamma^M_2=-i \sigma_2 \otimes \sigma_2, 
\Gamma^M_3=-i\sigma_1 \otimes \sigma_0$), the Majorana representation.
This means that in this representation
the matrices $\Gamma^M_{\mu \nu}$ are real.
Thus  we clearly see, in this representation, that the structure constants
become real.

\item The odd-odd part of the algebra
 has to close on the even part of the algebra.
Since $ {\bf 2} \otimes {\bf 2} = {\bf 1} \oplus {\bf 3}_+$,
$ \overline {\bf 2} \otimes \overline {\bf 2} = {\bf 1} \oplus {\bf 3}_-$
and ${\bf 2} \otimes \overline{\bf 2}= {\bf 4}$
with ${\bf 1}$ the scalar representation, ${\bf 3}_\pm$ the 
(anti-)self-dual two-forms  and ${\bf 4}$ the vector representation 
of $SL(2,\mathbb C)$, we have
{\it a priori}

\beqa
\label{susy11-}
\left\{Q_\alpha, Q_\beta\right\}&=& a \sigma^{\mu \nu}{}_{\alpha \beta} 
L_{\mu \nu}, \nonumber \\
\left\{\bar Q_{\dot \alpha}, \bar Q_{\dot \beta}\right\}&=& 
b \bar \sigma^{\mu \nu}{}_{\dot \alpha \dot \beta} 
L_{\mu \nu},  \\  
\left\{Q_\alpha, Q_\beta\right\}&=&-ic \sigma^\mu{}_{\alpha \dot \alpha} P_\mu.
\nonumber
\eeqa

\noi The right handed part of \eqref{susy11-} is obtained by means of the
natural tensors acting on the spinor space. Now using the Jacobi identity
with $(P,Q,Q)$ gives $a=0$, since $P$ commute with $Q$ and do not commute
with $L$. Similarly we have $b=0$. We identify now the constant $c$.
The relation above gives $\left\{Q_1,\bar Q_{1}\right\} +
\left\{Q_2,\bar Q_{2}\right\}= -2i c P_0$. 
Now if we assume that the supercharges
act on a Hilbert space, since $\bar Q_{\dot \alpha} = Q^\dag_\alpha$, for
any element $| \psi>$ we have 
$\sum_{\alpha}<\psi|\left\{ Q^\dag_{\alpha}, Q_\alpha \right\}
|\psi> = -2i c <\psi|P_0| \psi> $. Thus
$ -2i c <\psi|P_0| \psi>  $ is positive. Assuming $-iP_0$ is a
positive operator (the energy with our convention)
 gives $ c>0$. The conventional 
normalisation is to take $ c=2$. The odd-odd part of the algebra is then

\beqa
\label{susy11}
\left\{Q_\alpha, \bar Q_{\dot \alpha}\right\}
&=&-2i \sigma^\mu_{\alpha \dot \alpha} P_\mu.
\eeqa

\noi
The relation above seems to be in conflict with the reality of the Lie
superalgebra.
To solve this discrepancy, write the relations \eqref{susy11} in the
four dimensional representation

\beqa
\label{4-comp}
\left\{ \begin{pmatrix} Q_\alpha \\ \bar Q^{\dot \alpha}\end{pmatrix},
\begin{pmatrix} Q_\beta & \bar Q^{\dot \beta} \end{pmatrix}\right\}=-2i
\begin{pmatrix}
0&\sigma_\mu{}_{\alpha \dot \gamma}\\
\bar \sigma_\mu{}^{\dot \alpha \gamma}&0
\end{pmatrix}
\begin{pmatrix}
-\epsilon_{\gamma \beta}&0 \\
0&-\bar \epsilon^{\dot \gamma \dot \beta} 
\end{pmatrix} P^\mu,
\eeqa

\noi introduce 

$$C=-i \Gamma_0 \Gamma_2=\begin{pmatrix} -\e&0 \\ 0& -\bar \e 
\end{pmatrix}$$

\noi
the charge conjugation matrix. Denoting $Q_a$ the components of the 
four dimensional Majorana spinor, the relations \eqref{4-comp} become

$$
\left\{Q_a, Q_b \right\}= -2i (\Gamma^\mu C)_{ab} P_\mu. 
$$

Now,  we write the equation above in the Majorana
representation for which the Dirac $\Gamma-$matrices 
and $C$ are purely imaginary. If we denote $Q^M_a$ the supercharge in
 the Majorana 
representation and substitute $Q^M_a \to S_a = e^{i \frac{\pi}{4}}Q^M_a$, then 
we obtain

$$
\left\{S_a,S_b\right\}=-2 (\Gamma^M_\mu C^M)_{ab} P^\mu,
$$

\noi
and the structure constant becomes real.  We however prefer to write the
odd-odd part of the algebra in the form \eqref{susy11}.

\end{enumerate}

The algebra given by equation \eqref{susy00}, \eqref{susy01} and \eqref{susy11}
defines the simplest supersymmetric extension of the Poincar\'e algebra.
Extended supersymmetric algebras 
with a number of spinor charges $N \le 8$ \cite{hls,wzmod,ext,ss} 
or supersymmetric algebras in any space-times dimensions $D \le 11$
\cite{susyd} can however be defined.
As we are concerned, we just focus on the simplest extension of the
Poincar\'e symmetry \eqref{susy00}, \eqref{susy01} and \eqref{susy11}. 
We will however not follow the standard approach where irreducible
representations 
of the Poincar\'e superalgebra were systematically  investigated \cite{ss}
leading to the implementation of supersymmetry in Quantum Field Theory
and then to  construction of supersymmetric extensions of the Standard
Model of Particle Physics \cite{nilles, martin, aitch}. 
We simply quote  some results that can be found in the text book given
in references: (i) we clearly see that $P_\mu P^\mu$ is a Casimir operator,
this means that in any representation all states have the same mass; 
(ii) for any representation of the supersymmetric algebra
there is an equal number of bosonic and fermionic degrees of freedom.
Following the method of Wigner induced representations we identify to types
of relevant representations the massive representations $P^\mu P_\mu < 0$ 
(recall that with our conventions the mass operator is given by
$M^2 = - P_\mu P^\mu$) and the massless representations $P^\mu P_\mu =0$.
When supersymmetry is implemented in Quantum Field Theory, it turns out that
it is a new symmetry that mixes bosons and fermions. Beyond its
purely mathematical interest this has also some phenomenological interesting
consequences. This was probably why supersymmetry becomes so popular in 
the description of physical laws in particle physics (see for instance
\cite{nilles,martin}). Finally, observing that the symmetric product of 
two supersymmetric transformations gives rise to a space-time translation,
the construction of local supersymmetric theory is necessarily a theory 
of gravity called supergravity. Several authors introduced independently
supergravity \cite{sugra}.

\subsection{Finite dimensional matrix representation of the supersymmetric
algebra}
We now give a finite dimensional matrix representation of
the algebra above.  At first, we introduce a five dimensional representation
of the Poincar\'e algebra

\beqa
L_{\mu \nu} \longrightarrow
L_5{}_{\mu \nu}= \begin{pmatrix}
\begin{tabular}{c|c}
$J_{\mu \nu}$&$0$ \\
\hline
$0$&$0$
\end{tabular}
\end{pmatrix} \ \ 
P_\mu \longrightarrow
P_5{}_\mu = \begin{pmatrix}
\begin{tabular}{c|c}
$0$&$\delta_\mu$ \\
\hline
$0$&$0$
\end{tabular}
\end{pmatrix}, \ \
\begin{pmatrix} 4 \times 4&4 \times 1 \\ 1 \times 4 & 1\times 1 \end{pmatrix}
\eeqa

\noi
where $\delta_\mu$ is a column vector with components $\delta_\mu{}^\nu$.
In this representation, the group element's read

$$
(\Lambda,T)=e^{\frac12 \omega^{\alpha \beta} J_{5\alpha \beta}} 
e^{t^\alpha P_{5 \alpha}}= 
\begin{pmatrix}
\begin{tabular}{c|c}
$\Lambda$&$t$ \\
\hline
$0$&$1$
\end{tabular}
\end{pmatrix},
$$

\noi
and the quadrivector $x$ is embedded in a five dimensional rep. 
$X_5=\begin{pmatrix} x \\ 1 \end{pmatrix}$ such that in the Poincar\'e
transformation above we have

$$
X' =\begin{pmatrix} x' \\ 1\end{pmatrix}=\begin{pmatrix}
\begin{tabular}{c|c}
$\Lambda$&$t$ \\
\hline
$0$&$1$
\end{tabular}
\end{pmatrix}  \begin{pmatrix} x \\ 1\end{pmatrix}=
 \begin{pmatrix} \Lambda x+ t  \\ 1\end{pmatrix}.
$$

In a similar manner, we can define a 
nine-dimensional matrix representation of the supersymmetric algebra.
We define a nine-dimensional vector with indices 
$I=(\mu,4,\alpha, \dot \alpha)$ given by

$$
X_9{}^I= \begin{pmatrix} x^\mu \\ 1 \\ \theta^\alpha \\ 
\bar \theta^{\dot \alpha} \end{pmatrix}.
$$

\noi Since we consider $\theta^\alpha$ instead of $\theta_\alpha$ we have
for its transformation law under a Lorentz transformation

$$
\delta_\omega \psi_\alpha = \frac 12 \omega^{\mu \nu}
\sigma_{\mu \nu}{}_\alpha{}^\beta \psi_\beta \ \ \Rightarrow\ \ 
\delta_\omega \psi^\alpha = -\frac 12 \omega^{\mu \nu}
\sigma_{\mu \nu}{}^\alpha{}_\beta \psi^\beta 
$$

\noi
but we have $\sigma_{\mu \nu}{}^\alpha{}_\beta=
\sigma_{\mu \nu}{}_\beta{}^\alpha$. 

We define now the $9 \times 9 $ matrices (to simplify notations we keep the same
symbols)

\beqa
L_{\mu \nu} = 
\begin{pmatrix}
   \begin{tabular}{c|c|c|c}
      $J_{\mu \nu} $&$0$&$0$&$0$\\
      \hline
     $0$&$0$&$0$ &  $0$ \\
      \hline
 $0$&$0$ &$-\sigma_{\mu \nu}^t$&$0$ \\
\hline
$0$&$0$&$0$ &$\bar \sigma_{\mu \nu}$ \\
    \end{tabular}
\end{pmatrix}, \ \ 
P_\mu= 
\begin{pmatrix}
   \begin{tabular}{c|c|c|c}
      $0$&$\delta_\mu$&$0$&$0$\\
      \hline
     $0$&$0$&$0$ &  $0$ \\
      \hline
 $0$&$0$ &$0$&$0$ \\
\hline
$0$&$0$&$0$ &$0$ \\
    \end{tabular}
\end{pmatrix}, \ \ 
Q=\begin{pmatrix}
0&Q_{01} \\
Q_{10}&0
\end{pmatrix} =
\begin{pmatrix}
5 \times 5 & 5\times 4 \\
4 \times 5& 4 \times 4
\end{pmatrix}
\nonumber
\eeqa

\noi
with $J_{\mu \nu}$ the Lorentz generators in the defining vector representation
with matrix elements
$(J_{\mu \nu})^\alpha{}_\beta= \delta_\mu{}^\alpha \eta_{\nu \beta}-
\delta_\nu{}^\alpha \eta_{\mu \alpha}$.
We introduce  $Q_{01}{}^M{}_A, Q_{10}{}^A{}_M$ with $M=0,1,2,3,4$ and
$A= 1,2,\dot 1 ,\dot 2$ the components of the supercharges. 
If we note $I=(\mu \beta \dot \beta)$ and
$J=(\nu \gamma \dot \gamma)$ the indices of lines and columns of the
following matrices, we get

\beqa
\label{mat}
\begin{array}{lll}
(Q_{01}{}_\alpha)^M{}_A=
\begin{pmatrix}
0&i \sigma^\mu{}_{\alpha \dot \gamma} \\
0&0
\end{pmatrix},  &
(\bar Q_{01}{}_{\dot \alpha})^M{}_A =\begin{pmatrix}
-i \sigma^\mu{}_{\gamma \dot \alpha}&0\\
0&0
\end{pmatrix}, &
\begin{pmatrix}
 4 \times 2 & 4 \times 2 \\
1 \times 2& 1 \times 2
\end{pmatrix},
\\
(Q_{10}{}_\alpha)^A{}_M=
\begin{pmatrix}0&\delta^\beta{}_\alpha \\
0&0
\end{pmatrix},&
(\bar Q_{10}{}_{\dot \alpha})^A{}_M = \begin{pmatrix}
0&0 \\
0&-\delta^{\dot \beta}{}_{\dot \alpha}
\end{pmatrix},
&
\begin{pmatrix}
2 \times 4 & 2 \times 1 \\
2 \times 4 & 2 \times 1
\end{pmatrix}.
\end{array}
\eeqa

\noi
A direct calculation shows 
that the matrices \eqref{mat} satisfy the algebra \eqref{susy00},
\eqref{susy01} and \eqref{susy11}. 
For the even-odd part we use the
relation

$$
(J_{\alpha \beta})^\mu{}_\nu \Gamma^\nu =-
\begin{pmatrix}
0&\sigma_{\alpha \beta} \sigma^\mu - \sigma^\mu \bar \sigma_{\alpha \beta} \\
\bar \sigma_{\alpha \beta} \bar \sigma^\mu- \bar 
\sigma^\mu \sigma_{\alpha \beta}&0
\end{pmatrix}.
$$

\noi
A matrix representation of the supersymmetric algebra was also given in
\cite{cornwell}, but with different notations.
The  transformations are now parameterised by

$$
(\Lambda,t, \e)= e^{\frac12 \omega^{\alpha \beta} L_{\alpha \beta}}
e^{t^\alpha P_\alpha} 
e^{\e^\alpha Q_\alpha + \bar Q_{\dot \alpha} \e^{\dot \alpha}},
$$

\noi
with $\e^\alpha, \bar e^{\dot \alpha}$ Majorana spinors.
And since $(\e^\alpha Q_\alpha + \bar Q_{\dot \alpha} \e^{\dot \alpha})^2=0$,
we have for a supersymmetric  transformation

$$
X'= \begin{pmatrix} x'^\mu \\  1 \\ \theta'^\alpha \\ \theta'^{\dot \alpha}
\end{pmatrix}
=e^{\e^\alpha Q_\alpha + \bar Q_{\dot \alpha} \e^{\dot \alpha}}
\begin{pmatrix} x^\mu \\  1 \\ \theta^\alpha \\ \theta^{\dot \alpha}
\end{pmatrix}=
\begin{pmatrix} 
x^\mu +i(\e^\alpha \sigma^\mu{}_{\alpha \dot \alpha}
 \bar \theta^{\dot \alpha} - \theta^\alpha \sigma_\mu{}_{\alpha \dot \alpha}
\bar \e^{\dot \alpha})\\ 1 \\ \theta^\alpha + \e^\alpha \\
\bar \theta^{\dot \alpha} - \bar \e^{\dot \alpha}
\end{pmatrix}
$$

\noi
Several remarks are in order here. 
The minus sign in the transformation of the variable $\bar \theta$ seems to 
be rather surprising. In addition,
since the matrices \eqref{mat} act on
the nine-dimensional ($\ZZ_2-$graded)
 vector space parametrised by $X$, at that level, there
is no need to have anticommuting variables.  As a simple  consequence the
variable $x$ cannot be real. All this will be corrected by the introduction
of Grassmann variables in the next subsection.

\subsection{Superspace}

Since the supersymmetric algebra contains fermionic supercharges its seems
natural to promote the space-time to a superspace by adjoining to the
space-time coordinates fermionic coordinates that we take Majorana
as it was done in \cite{superspace}. We then consider a point in
a superspace as given by $(x^\mu, \theta^\alpha, \bar \theta^{\dot \alpha})$,
together with the conjugate  momenta $(\partial_\mu, 
\frac{\partial}{\partial \theta^\alpha}, 
\frac{\partial}{\partial \bar \theta^{\dot \alpha}})$ which satisfy
(we give the only non-vanishing graded-commutators)

\beqa
\label{oscil}
\left[\partial_\mu, x^\nu\right]= \delta_\mu{}^\nu,\ \ 
\left\{\frac{\partial}{\partial \theta^\alpha}, \theta^\beta\right\}=
\delta_\alpha{}^\beta, \ \ 
\left\{\frac{\partial}{\partial \bar \theta^{\dot \alpha}}, 
\bar \theta^{\bar \beta}\right\}= \delta_{\dot \alpha}{}^{\dot \beta}.
\eeqa

\noi
Differently from the previous subsection we now assume from the
very beginning that the variables $\theta$ are of Grassmann type.
 It should be emphasise that from the properties of spinor indices 
we have the relation 
$\{\frac{\partial}{\partial \theta_\alpha}, \theta_\beta\}=
 -\delta^\alpha{}_\beta$. 
To make contact with the matrices given in \eqref{mat} and the standard 
approach of superspace we also introduce $z, \partial_z$ such that
$[\partial_z,z]=1$. Now,
to construct a differential realisation of the supersymmetric algebra, we
use several properties. Since these properties are not specific to our
superalgebra, we state them to $\g$  a Lie superalgebra which admits a finite
 dimensional 
representation of dimension $m+n$ ($m$ is the dimension of the even part
and $n$ of the odd part). 
Denote

$$
X_i=\begin{pmatrix}X^{00}_i&0\\0&X^{11}_i\end{pmatrix}, \ \
Y_a=\begin{pmatrix}0&Y^{01}_a \\ Y^{10}_a&0 \end{pmatrix}, 
$$

\noi the matrix representation of $\g$ ($X$ are bosonic operators and
$Y$ fermionic operators), satisfying

\beqa
\label{matg}
[X_i, X_j]= f_{ij}{}^k X_k, \ \ 
[X_i,Y_a]= R_{ia}{}^b Y_b, \ \
\{Y_a, Y_b\}= C_{ab}{}^i X_i.
\eeqa

\noi
Introduce $m$ bosonic ($n$ fermionic) oscillators
$(x^A,\partial_A)$ $(\theta^I, \frac{\partial}{\partial \theta^I})$
satisfying the relation \eqref{oscil} and set

$$
X=\begin{pmatrix} x^1 \\ \vdots \\ x^m\end{pmatrix}, \ \
\Theta=\begin{pmatrix} \theta^1 \\ \vdots \\ \theta^n \end{pmatrix}, \ \ 
\partial_X = \begin{pmatrix}\partial_1,\cdots,\partial_m \end{pmatrix}, \ \
\partial_\Theta = \begin{pmatrix}
\partial_{\theta^1}, \cdots, \partial_{\theta^n} \end{pmatrix}.
$$

\noi
Assume further that the vectors $Z=\begin{pmatrix} X \\ \Theta \end{pmatrix}$
and $\partial_Z= \begin{pmatrix} \partial_X & \partial_\Theta\end{pmatrix}$
transform contravariantly (resp. covariantly) under the action of the matrices
$X,Y$. Finally observing that the matrices 
$\tilde X_i = -X_i^t$ and $\tilde Y_a = i Y_a^t$, with
$M^t$ the transpose of the matrix $M$ satisfy the relations \eqref{matg}, one
can show that the operators

$$
{\cal X}_i= X^t \tilde X_i^{00} \partial_X^t +
\Theta^t \tilde X_i^{11} \partial^t_\Theta, \ \
{\cal Y}_a= X^t \tilde Y^{01} \partial^t_\Theta  
+ \Theta^t \tilde Y^{10} \partial_X^t,
$$

\noi
are a differential realisation of $\g$, {\it i.e} satisfy \eqref{matg}.

In our special case, we define (we normalise slightly differently our
operators in order that ${\cal P}_\mu $ leads to $ \partial_\mu$)

\beqa
\label{diff-rep}
{\cal L}_{\mu \nu}= X^t \tilde L_{\mu \nu}^{00} \partial^t_X + 
\Theta^t \tilde L_{\mu \nu}^{11} \partial_\Theta^t&=&
x_\mu \partial_\nu - x_\nu \partial_\mu + 
\theta^\alpha \sigma_{\mu \nu}{}_\alpha{}^\beta \partial_{\theta^\beta} 
- \bar \theta^{\dot \alpha} \bar \sigma_{\mu \nu}{}^{\dot \beta}{}_{\dot \alpha}
\partial_{\bar \theta_{\dot \beta}} \nonumber \\
{\cal P}_\mu = 
-X^t\tilde  P_\mu^{00} \partial^t_X  - \Theta^t P_\mu^{11} \partial^t_\Theta
&=& z \partial_\mu \\
{\cal Q}_\alpha= -i X^t \tilde Q^{01}_\alpha \partial_\Theta^t - i
\Theta^t \tilde Q^{10}_\alpha \partial_X^t
&=& z \partial_{\theta^\alpha} +
i  \sigma^\mu{}_{\alpha \dot \beta}\bar \theta^{\dot \beta} \partial_\mu
\nonumber \\
\bar {\cal Q}_{\dot \alpha}= 
-i X^t \tilde {\bar Q}^{01}_{\dot \alpha} \partial_\Theta^t - i
\Theta^t \tilde {\bar Q}^{10}_{\dot \alpha} \partial_X^t
&=& -z \partial_{\bar \theta^{\dot \alpha}} -
i \theta^{\beta} \sigma^\mu{}_{\dot \alpha  \beta} \partial_\mu 
\nonumber
\eeqa

\noi
which reduces to a differential realisation of
the algebra \eqref{susy00}, \eqref{susy01} and \eqref{susy11} for $z=1$. 
From now on in order to simplify the notations ${\cal Q}$ and $\bar {\cal Q}$
will be denoted $Q$ and $\bar Q$ respectively

\beqa
\label{Q}
Q_\alpha=\partial_{\theta^\alpha} +
i  \sigma^\mu{}_{\alpha \dot \beta}\bar \theta^{\dot \beta} \partial_\mu, \ \
\bar Q_{\dot \alpha}=
-\partial_{\bar \theta^{\dot \alpha}} -
i \theta^{\beta} \sigma^\mu{}_{\dot \alpha  \beta} \partial_\mu
\eeqa

\noi
since no confusion would be possible.
For further use we are looking to some operators $D_\alpha$ and
$\bar D_{\dot \alpha}$ which anticommute with ${ Q}_\beta$ and
$\bar { Q}_{\dot \beta}$:

\beqa
\label{QD}
\left\{Q_\alpha, D_\beta\right\}=0, \ \ 
\left\{\bar Q_{\dot \alpha}, D_\beta\right\}=0, \ \ 
\left\{\bar Q_{\dot \alpha}, \bar D_{\dot \beta}\right\}=0.
\eeqa

\noi
 A simple calculation gives

\beqa
\label{D}
D_\alpha&=& \partial_{\theta^\alpha} -i \sigma^\mu{}_{\alpha \dot \beta} 
\bar \theta^{\dot \beta} \partial_\mu, \nonumber \\
\bar D_{\dot \alpha}&=& \partial_{\bar \theta^\alpha} - i \theta^\beta
\sigma^\mu{}_{\beta \dot \alpha}  \partial_\mu.
\eeqa

\noi
It is easy to see that they also satisfy the supersymmetric algebra, and
in particular we have

$$
\left\{D_\alpha, \bar D_{\dot \alpha}\right\}=-2i \sigma^\mu{}_{\alpha
\dot \alpha} \partial_\mu.
$$

\bigskip
All these operators might be obtained in the more formal way of superspace.
The concept of superspace was first introduced in \cite{superspace}.
Formally, in the same manner as we can see the Minkowski space-time
as a coset of the Poincar\'e group by the Lorentz group, the superspace
can be seen as a coset of the super-Poincar\'e group by the Lorentz
group (see also \cite{west}). This leads naturally to the structure
of supermanifold ({\it i.e.} structures generalising the concept of 
manifold and containing Grassmann variables together with the usual
``bosonic'' coordinates). One can see \cite{manifold} for a more
formal approach (convergence problem, differentiability {\it etc}).
Supermanifolds are intimately related to  
Lie supergroups as  manifolds are related to
 Lie groups (see for instance \cite{cornwell}). A
point in the superspace has coordinates

$$
X=(x^\mu, \theta^\alpha, \bar \theta^{\dot \alpha}).
$$

\noi In the coset approach a point in the superspace is parametrised
by

$$
G(x,\theta,\bar \theta)= e^{x^\mu P_\mu + \theta^\alpha Q_\alpha +
\bar Q_{\dot \alpha} \bar \theta^{\dot \alpha}},
$$

\noi
and the action of a supersymmetric transformation is given by

$$
 G(x,\theta,\bar \theta)G(0,\e, \bar \e),
$$
\noi 
we consider the right action in order to have the correct sign in
the supersymmetric algebra.
Since we are dealing with anticommuting variables we assume further
$\{\theta,\e\}=\{\theta,Q\}=\{\e,Q\}=0$ {\it etc}. 
We recall that $[\e^\alpha Q_\alpha, \bar 
Q_{\dot \alpha}\bar \theta^{\dot \alpha}]= \e^\alpha\{Q_\alpha, \bar Q_{\dot
\alpha}\} \bar \theta^{\dot \alpha}$ {\it etc}. Using the 
Baker-Campbell-Hausdorff  formul\ae \ which reduces to

$$
e^A e^B= e^{A+B+\frac12[A,B]},
$$

\noi since $[[A,B],A]=[[A,B],B]=0$, we finally obtain

$$
 G(x,\theta,\bar \theta)G(0,\e, \bar \e) =
e^{\big[x^\mu +i(\theta \sigma^\mu \bar \e - \e \sigma^\mu \bar \theta)
\big] P_\mu \ + \
\big[\theta^\alpha + \e^{\alpha}\big]Q_\alpha \ + \ 
\bar Q_{\dot \alpha}\big[\bar \theta^{\dot \alpha} + 
\bar \e^{\dot \alpha}\big]}.
$$

\noi
This means that under a supersymmetry transformation we have

\beqa
\label{superspace}
\delta_\e x^\mu &=& i(\theta \sigma^\mu \bar \e - \e \sigma^\mu \bar \theta),
\nonumber \\
\delta_\e \theta^\alpha&=& \e^\alpha, \\
\delta_\e \bar \theta^{\dot \alpha}&=& \bar \e^{\dot \alpha}. \nonumber
\eeqa

\noi
Assume now that the transformations \eqref{superspace} are given by some 
supercharges 

$$
\delta_\e x^\mu = [\e. Q + \bar Q. \bar \e,x^\mu], \ \ 
\delta_\e \theta^\alpha = [\e. Q + \bar Q. \bar \e,\theta^\alpha],
\delta_\e \bar \theta^{\dot \alpha} = [\e. Q + \bar Q. \bar \e,\bar 
\theta^{\dot \alpha}],
$$

\noi gives \eqref{Q} for the supercharges.
We see that the benefit of introducing Grassmann variables is two-fold.
Firstly the unwanted minus sign in the $\bar \theta$ transformation
disappear. Secondly, $\delta x$ is real because of \eqref{conj2}.
\subsection{The superfield formalism: the Wess-Zumino model}
The natural objects which live in the superspace are the superfields.
Superfields
where firstly introduced in \cite{superfield}.
A superfield  $\Phi$ is a field which depends on the
superspace coordinates.  Since the Grassmann variables  
are nilpotent, the superfield has a finite number of
components

\beqa
\label{superphi}
\Phi(x, \theta, \bar \theta)&=&
z(x) + \theta^\alpha \psi_\alpha(x) + \bar \theta_{\dot \alpha} 
\bar \chi^{\dot
\alpha }(x) + \theta^\alpha \theta_\alpha n(x) + \bar \theta_{\dot \alpha}
\bar \theta^{\dot \alpha} m(x)  \nonumber \\
&+&   
\theta^\alpha \sigma_\mu{}_{\alpha \dot \alpha} \bar \theta^{\dot \alpha} 
v^\mu(x) +
\theta^\alpha \theta_\alpha \bar \theta_{\dot \alpha}  \bar \lambda^{\dot
\alpha}(x) + \bar \theta_{\dot\alpha} \bar \theta^{\dot \alpha} \theta^\alpha
\omega_ \alpha(x) + \theta^\alpha \theta_\alpha \bar \theta_{\dot \alpha}
\bar \theta^{\dot \alpha} d(x),  
\eeqa  

\noi
with $z,m,n,d$ complex scalar field's, $v_\mu$ a complex vector field,
 $\psi, \lambda$ left-handed spinors and
 $\bar \chi,\bar \omega$ right-handed spinors.
Since in all supersymmetric calculus one has to be careful with the signs, 
and since for Grassmann variables we have $\theta^\alpha \psi_\alpha=
-\theta_\alpha \psi^\alpha$, when the summation on the spinor indices will be
omitted $\theta. \psi$ means $\theta^\alpha \psi_\alpha$ and
$\bar \theta. \bar \psi$ means 
 $ \bar \theta_{\dot \alpha} \bar \psi^{\dot \alpha}$. In the sequel 
we will  intensively follow this convention. See the reference
of Wess and Bagger \cite{wb} for more details and useful identities
(see also Appendix \ref{conventions}).





A scalar superfield is a superfield which transforms under a supersymmetric
transformation

$$
\Phi'(X')=  \Phi(X).
$$

\noi
At the infinitesimal level

\beqa
\label{transforPhi}
\delta \Phi(X)= \Phi'(X)-\Phi(X) =(\e.Q + \bar Q . \bar \e) \Phi(X).
\eeqa

\noi It turns out that a scalar superfield is a reducible representation
of the supersymmetric algebra. Several types of superfields constructed
from $\Phi$ may be defined.  The chiral superfield is defined by

\beqa
\label{chiral}
\bar D_{\dot \alpha} \Phi=0.
\eeqa

\noi We can have also defined an antichiral superfield $\bar \Phi$
satisfying $D_\alpha \bar \Phi=0$. 
The chiral superfield leads to the first supersymmetric (where
supersymmetry is realised in a linear way) model in field theory
in four space-time dimensions.
It is known as the Zess-Zumino model \cite{wzmod}. In fact this
model was historically constructed in components and not in the
superfield language.

Observing that

$$
\bar D_{\dot \alpha} y^\mu=0, \ \ y^\mu=x^\mu   -i \theta \sigma^\mu 
\bar \theta,
$$

\noi
means that the chiral superfield $\Phi$ depends only on the variables $y$ and
$\theta$. Developing $\Phi(y,\theta)$ using \eqref{spin3} and
\eqref{spin4}  we get

\beqa
\label{chiraldev}
\Phi(y,\theta) &=& z(y)+ \sqrt{2} \theta. \psi(y) - \theta.\theta F(y)
\\
&=&z(x)+ \sqrt{2}\theta.\psi(x)-\theta.\theta \ F(x) 
-i\theta \sigma^\mu \bar \theta \ \partial_\mu z +\frac{i}{\sqrt{2}}
\theta.\theta \partial_\mu \psi(x)\sigma^\mu \bar \theta-
\frac14 \theta.\theta \bar \theta . \bar \theta \ \Box  z. \nonumber
\eeqa

\noi
The minus sign and $\sqrt{2}$ factor are introduced such that the
kinetic part of the Lagrangian is correctly normalised. A chiral
superfield contains two complex scalars $z$ and $F$ and one left-handed Weyl 
spinors $\psi$. The adjoint of $\Phi$ is given by

\beqa
\Phi^\dag(y,\theta)= z^\dag(x) + \sqrt{2} \bar \theta . \bar \psi
-\bar \theta. \bar \theta F^\dag(x) + i \theta \sigma^\mu \bar \theta
\partial_\mu z^\dag(x)+\frac{i}{\sqrt{2}} \bar \theta. \bar \theta 
\theta \sigma^\mu \partial_\mu \bar \psi(x) -\frac14 
\theta.\theta \bar \theta.\theta \ \Box  z^\dag(x), 
\eeqa

\noi
and contains two complex scalars and one right-handed Weyl spinor.
It turns out that $\Phi^\dag$ is an anti-chiral superfield.
To determine the transformation of the chiral superfield under a supersymmetric
transformation, we first observe  that

\beqa
Q_\alpha y^\mu =0, \ \ 
\bar Q_{\dot \alpha} y^\mu= 
-2i \theta^\alpha \sigma^\mu{}_{\alpha \dot \alpha}.
\eeqa

\noi Since $\Phi$ only depends on $y$ and $\theta$, we have

\beqa
Q_\alpha \Phi(y,\theta)&=& Q_\alpha y^\mu \frac{\partial \Phi}{\partial y^\mu}
+ Q_\alpha \theta^\beta \frac{\partial \Phi}{\partial \theta^\beta}=
\frac{\partial \Phi}{\partial \theta^\alpha} \nonumber \\
\bar Q_{\dot \alpha} \Phi(y,\theta)&=&
\bar Q_{\dot \alpha} y^\mu \frac{\partial \Phi}{\partial y^\mu}
+ \bar Q_{\dot \alpha} \theta^\beta \frac{\partial \Phi}
{\partial \theta^\beta}= -2i \theta^\alpha \sigma^\mu{}_{\alpha \dot \alpha}
\frac{\partial \Phi}{\partial x^\mu}.
\eeqa

\noi
The transformation of $\Phi$ is then given by

\beqa
\label{transfo}
\delta_\e \Phi &=& (\e. Q + \bar Q. \bar \e). \Phi\\
&=&\sqrt{2} \e. \psi + \sqrt{2} \theta \left(
-\sqrt{2} \e F -i \sqrt{2} \sigma^\mu \bar \e \partial z\right)+
i \sqrt{2} \theta.\theta \partial_\mu \psi \sigma^\mu \bar \e.
\nonumber
\eeqa

\noi
Expressed in components this gives

\beqa
\label{transfo2}
\delta_\e z &= &\sqrt{2} \e. \psi=
\sqrt{2} \e^\alpha \psi_\alpha, \nonumber \\
\delta_\e \psi_\alpha&=& -\sqrt{2}F \e_\alpha -i 
\sqrt{2} \sigma^\mu{}_{\alpha \bar \alpha} \bar \e^{\dot \alpha} \partial_\mu z,
\\
\delta_\e F&=& -i \sqrt{2} \partial_\mu \psi^\alpha 
\sigma^\mu{}_{\alpha \dot
\alpha} \bar \e^{\dot \alpha}. \nonumber
\eeqa

\noi
Identity \eqref{spin4} has been used to simplify
$-2i \sqrt{2} \theta \sigma^\mu \bar \e \theta \partial_\mu \psi$.
These transformation laws show two points. First, they show that supersymmetry
is a symmetry which maps bosons into fermions and fermions into bosons.
Secondly they show that the highest component of the superfield $\Phi$,
namely $F$, transforms as a total derivative. This is the key point to
construct invariant Lagrangians. Indeed, by definition the product of 
superfields is a superfield. However, the highest component of a
superfield transforms as a total derivative. This means that it is
a good candidate to construct invariant Lagrangians. Two types of
terms may be constructed

\beqa
\label{kinetic}
\Phi^\dag \Phi_{|_{\theta \theta \bar \theta \bar \theta}}=
\partial_\mu z^\dag(x) \partial^\mu z(x) +\frac{i}{2}
\Big(\psi(x) \sigma^\mu  \partial_\mu\bar \psi(x) 
-\partial_\mu \psi(x) \sigma^\mu \bar \psi(x)\Big) + F^\dag F +
\text{derivative~terms},
\eeqa

\noi
where we have used \eqref{spin3} to simplify the $z$ parts
and \eqref{spin4} and \eqref{spin5} for the $\psi$ part.
These terms contribute to the kinetic part of the Lagrangian.
The interacting part is given by the superpotential
$W(\phi)$, which is a polynomial in $\Phi$. This means that 
$W$ is a holomorphic function.
For renormalisation arguments  (see below) $W$  is of degree three.
Using

\beqa
\label{pot}
\Phi_i \Phi_j&=& z_i z_j + \sqrt{2} \theta\Big(z_i \psi_j + z_j \psi_i\Big)-
\theta . \theta \Big(z_i F_j + z_j F_i + \psi_i.\psi_j\Big) \nonumber \\
\Phi_i \Phi_j \Phi_k&=& z_i z_j z_k + \sqrt{2}\theta
\Big(z_i z_j \psi_k + z_i z_k \psi_j +z_j z_k \psi_i\Big)\\
& -& \theta. \theta
\Big( z_i z_j F_k + z_i z_k F_j +z_j z_k F_i +
z_i \psi_j.\psi_k +z_j \psi_i.\psi_k +z_k \psi_i.\psi_j \Big),\nonumber 
\eeqa

\noi
the most general invariant Lagrangian is given by the superpotential

\beqa
\label{W}
W(\Phi)= \alpha^i \Phi_i+\frac12 m^{ij} \Phi_i \Phi_j +
\frac16 \lambda^{ijk} \Phi_i \Phi_j \Phi_k. 
\eeqa

\noi
The Wess-Zumino model is then given by
\beqa
\label{LWZ}
{\cal L}_{\text{W.Z.}}&=&  
(\Phi^\dag{}^i \Phi_i)_{|_{\theta.\theta \bar \theta. \bar \theta}} 
 + W(\Phi)_{\theta.\theta} + W^\dag(\Phi)_{\bar \theta.\bar \theta}
\nonumber \\
&=&\partial_\mu z^i{}^\dag \partial^\mu z_i +\frac{i}{2}
\Big(\psi_i \sigma^\mu  \partial_\mu\bar \psi^i 
-\partial_\mu \psi_i \sigma^\mu \bar \psi^i\Big) + F^i{}^\dag F_i \\
&-&\alpha^i F_i - \frac12 m^{ij} (z_i F_j + z_j F_j + \psi_i.\psi_j) + 
\text{c.c.} \nonumber \\
&-&\frac16\lambda^{ijk}( z_i z_j F_k + z_i z_k F_j +z_j z_k F_i +
z_i \psi_j.\psi_k +z_j \psi_i.\psi_k +z_k \psi_i.\psi_j) +\text{c.c.} 
\nonumber
\eeqa

\noi
The equations of motion for the auxiliary fields $F$ give

\beqa
\label{F}
F^\dag{}^i= \alpha^i + m^{ij} z_j + \frac12 \lambda^{ijk} z_j z_k=
\frac{\partial W(z)}{\partial z_i}.
\eeqa

\noi
If we eliminate $F$ from the Lagrangian and use $W(z)$, the Lagrangian reduces
to

\beqa
{\cal L}_{\text{W.Z.}}&=& 
 \partial_\mu z^i{}^\dag \partial^\mu z_i +\frac{i}{2}
\Big(\psi_i \sigma^\mu  \partial_\mu\bar \psi^i 
-\partial_\mu \psi_i \sigma^\mu \bar \psi^i\Big)\\ 
&-&\frac12 \sum_i \left|\frac{\partial W}{\partial z_i}\right|^2 -
\frac12 \frac{\partial^2 W}{\partial z_i \partial z_j} \psi_i.\psi_i-
\frac12 \frac{\partial^2 \bar W}{\partial z^\dag{}^i \partial z^\dag{}^j}
\bar \psi^i. \bar \psi^i, \nonumber
\eeqa

\noi
$V_F= \frac12 \sum_i \left|\frac{\partial W}{\partial z_i}\right|^2$,
the so-called $F-$terms, contributes to a scalar potential of degree
four (since $W$ is of degree three) and 
$\frac12 \frac{\partial^2 W}{\partial z_i \partial z_j} \psi_i.\psi_i$ 
to Yukawa interactions between fermions and scalars. It has to be emphasised
that all the interaction terms of the Wess-Zumino model are encoded
into the superpotential. Furthermore,
the auxiliary fields gives an
equal number of degrees of freedom (d.o.f) on-shell and off-shell:
\bigskip

\begin{center}

\begin{tabular}{|l|l|l|l|}
\hline
\text{on-shell}&\begin{tabular}{l|l}
                 $z$&2 {\text d.o.f.} \\
                  \hline
                 $\psi$& 2 {\text d.o.f.}\\
                 \hline
             $ F$&0 {\text d.o.f.} \end{tabular}&
\text{off-shell}&\begin{tabular}{l|l}
                 $z$&2 {\text d.o.f.} \\
                 \hline
                 $\psi$& 4 {\text d.o.f.}\\
                 \hline
              $F$&2 {\text d.o.f.} \end{tabular}\\
\hline
\end{tabular}
\end{center}

\noi

To proceed further in a construction of supersymmetric models, one
uses another type of superfields: the real superfield $V$ such that
$V^\dag =V$. This superfield enables us to construct supersymmetric
Yang-Mills theories \cite{superym}, and to couple chiral
superfields to vector superfields in an invariant way. This means that
if the various fields are in some representation of the gauge group, the 
superpotential has to be  invariant.
Finally, it should be observed that
models constructed along these lines are not acceptable physically, since
in their spectrum  they contain bosons and fermions of the same masses.
Since scalar particles with the quantum numbers
of  known fermions   (electron, {\it etc}) have not been observed
supersymmetry has to be broken.

\section{Lie algebras of order $F$}
As it has been mentioned previously,  Lie (super)algebras are binary algebras.
This means that one can always multiply two elements. Indeed, the
multiplication law is given  by  the commutators or
 the anticommutators. In this section we introduce higher order algebras
{\it i.e.} defined by higher order products and try to implement them
for constructing new symmetries in physics. Of course the new structures
considered
would have to respect the principle of Relativistic and Quantum Physics.
Thus we will by-pass the no-go theorems which restrict
drastically the possible extensions of the symmetry of the space-time. 
In this lecture, we construct
some  $F-$ary algebras which can be seen as a possible extension of the
Lie (super)algebras. We then show that these new structures can be applied
in physics in two different ways, leading to higher 
order non-trivial extensions
of the Poincar\'e algebra. It is well known that $(1+2)-$dimensional space-time
is exceptional. In such dimensional space-time,
 one can consider representations which are neither bosons
nor fermions but anyons. An anyon is a particle of arbitrary spin. We
take advantage of this situation to define an $F-$order non-trivial
extension of the Poincar\'e algebra, which maps
a relativistic anyon of spin $s \in \mathbb R$ to a relativistic
anyons of spin $s\pm 1/F$.  In the second application, we play with the
way the Noether theorem is implemented in Quantum Field Theory, and 
construct a non-trivial cubic-extension of the Poincar\'e algebra
for arbitrary dimensional space-time.

\subsection{Lie algebras of order $F$: definition}
The basic idea to construct higher order extensions of the Poincar\'e algebra
is to define higher order extensions of Lie superalgebras.
In supersymmetric theories, the extensions of the Poincar\'e
algebra are obtained from a ``square root'' of the translations,
``$QQ \sim P$''.
It is tempting to consider other alternatives where the new algebra
is obtained from yet higher order roots.
Basically, in such extensions,
 the generators of the Poincar\'e algebra are obtained as   $F-$fold 
symmetric products of more fundamental generators, leading to the
``$F^{\mathrm{th}}-$root'' of translation: ``$Q^F \sim P$'' with $F$
a positive integer.
 It is important to stress that such structures
are not Lie (super)algebras  (even though they  contain a Lie subalgebra),
and as such avoid {\it a priori} the Coleman-Mandula \cite{cm}
as well as  the Haag-Lopuszanski-Sohnius no-go theorems \cite{hls}.
Furthermore, as far as we know, no no-go theorem associated with such types
of extensions has been considered in the literature. 
This can open interesting  possibilities to search for a field theoretic
realization of a non-trivial extension of the Poincar\'e algebra which 
{\it is not the supersymmetric one}. If successful, this might throw
a new light on how to construct physical models.
In small space-time dimensions $D \le 2$, several authors
were able to define along
these lines an extension of supersymmetry called Fractional Supersymmetry
\cite{fsusy}. 
In the same manner than Lie superalgebras are the underlying mathematical
structure of supersymmetry, 
Lie algebras of order $F$ are the mathematical structure associated to
  Fractional Supersymmetry. These algebras were introduced in \cite{flie1,
flie2}.
Subsequently Lie algebras of order three were studied on some 
formal ground. The basis of the
theory of contractions and deformations in the context of Lie
algebras of order three has been studied in \cite{flie3}, some progress
in the classification of Lie algebras (of order three) were undertaken
in  \cite{flie3,cr} and a group together with the parameters of the
transformation for Lie algebras of order three were defined \cite{hopf}.\\

We give now  the abstract mathematical structure which
 generalises the theory of Lie superalgebras
and their  representations. Let $F$ be a positive integer and 
$q=\exp{({2i \pi \over F})}$.
We consider a complex vector space $\g$ together with a linear map 
$\varepsilon$
from $\g$ into itself satisfying $\e^F=1$. 
We set $\g_k$ ($k=0,\cdots, F-1$) the eigenspace 
corresponding to the eigenvalue $q^k$ of $\varepsilon$,
so that $\g=\oplus_{k=0}^{F-1} \g_k$. The map $\varepsilon$ is called the 
grading. If $\g$ is endowed with the following structures we will say that 
$\g$ is  Lie algebra  of order $F$ \cite{flie1}.

\begin{definition}
\label{flie-def}
Let $F\in \mathbb{N}^{\ast }$. A ${\mathbb{Z}}_{F}$-graded ${\mathbb{C}}-$%
vector space ${\mathfrak{g}}={\mathfrak{g}}_{0}\oplus {\mathfrak{g}}%
_{1}\oplus {\mathfrak{g}}_{2}\dots \oplus {\mathfrak{g}}_{F-1}$ is called a
complex Lie algebra of order $F$ if

\begin{enumerate}
\item $\mathfrak{g}_0$ is a complex Lie algebra.

\item For all $i= 1, \dots, F-1 $, $\mathfrak{g}_i$ is a representation of $%
\mathfrak{g}_0$. If $X \in {\mathfrak{g}}_0, \ Y \in {\mathfrak{g}}_i$ then $%
[X,Y]$ denotes the action of $X$ on $Y$ for any $i=1,\cdots,F-1$.

\item For all $i=1,\dots,F-1$, there exists an $F-$linear, $\mathfrak{g}_0-$%
equivariant map 
\begin{equation*}
\{ \cdots \} : \mathcal{S}^F\left(\mathfrak{g}_i\right) \rightarrow 
\mathfrak{g}_0,
\end{equation*}
where $\mathcal{S}^F(\mathfrak{g}_i)$ denotes the $F-$fold symmetric product
of $\mathfrak{g}_i$, satisfying the following (Jacobi) identity 
\begin{eqnarray}  \label{eq:J}
&&\sum\limits_{j=1}^{F+1} \left[ Y_j,\left\{ Y_1,\dots, Y_{j-1},
Y_{j+1},\dots,Y_{F+1}\right\} \right] =0,  
\end{eqnarray}
\noindent for all $Y_j \in \mathfrak{g}_i$, $j=1,..,F+1$.
\end{enumerate}
\end{definition}

We would like to make now several observations. We clearly see that
Definition \ref{flie-def}  reduces to Lie algebras when $F=1$ and
to Lie superalgebras when $F=2$. 
Thus Lie algebras of order $F$ constitute a possible generalisation
of Lie  (super)algebras. Furthermore,
we have given the definition
of Lie superalgebras (Definition \ref{superlie}) and of Lie
algebra of order $F$ (Definition \ref{flie-def}) in a same manner in order
to stress on there similitude. 
It should also be observed that 
 for any $i=1,\ldots,F-1$, the ${\mathbb{Z}}_F-$graded vector spaces 
${\mathfrak{g}}_0\oplus{\mathfrak{g}}_i$ is a Lie algebra of order $F$. 
We call these type
of algebras \textit{elementary Lie algebras of order $F$}.
Finally, as for Lie superalgebras one
can deduce more Jacobi identities from the definition above.
Indeed,  we have

\beq
\left[\left[X_1,X_2\right],X_3\right] +  
\left[\left[X_2,X_3\right],X_1\right] +
\left[\left[X_3,X_1\right],X_2\right] =0,
\tag*{J1} \
\eeq

\beq
 \left[\left[X_1,X_2\right],Y_1\right] +
\left[\left[X_2,Y_1\right],X_1\right] +
\left[\left[Y_1,X_1\right],X_2\right]=0,
\tag*{J2}\
\eeq

\beq
\left[X_1,\left\{Y_1,\dots,Y_F\right\}\right] =
\left\{\left[X_1,Y_1 \right],\dots,Y_F\right\}  +
\dots +
\left\{Y_1,\dots,\left[X,Y_F\right] \right\}.
\tag*{J3}
\eeq

\noi
for any $X_1,X_2, X_3 \in \g_0, \ Y_1,\cdots,Y_F \in \g_1$.
The identity $J_1$  is the consequence of the fact that $\g_0$ is
a Lie algebra, $J_2$ is equivalent to assume that $\g_1$ is a 
representation of $\g_0$ and $J_3$ is related to the $\g_0-$ equivariance
of the mapping ${\cal S}^F(\g_1) \to \g_0$. Only  \eqref{eq:J} is
an extra constraint. 
\\

 A representation of an elementary Lie algebra of
order $F$ is a linear map $\rho : ~ {\mathfrak{g}}={\mathfrak{g}}_0 \oplus {%
\mathfrak{g}}_1 \to \mathrm{End}(V)$, such that (for all $X_i \in {\mathfrak{%
g}}_0, Y_j \in {\mathfrak{g}}_1$)

\begin{eqnarray}  \label{eq:rep}
\begin{array}{ll}
& \rho\left(\left[X_1,X_2\right]\right)= \rho(X_1) \rho(X_2)-
\rho(X_2)\rho(X_1) \cr & \rho\left(\left[X_1,Y_2\right]\right)= \rho(X_1)
\rho(Y_2)- \rho(Y_2)\rho(X_1) \cr & \rho\left(
\left\{Y_1.\cdots,Y_F\right\}\right)= \sum \limits_{\sigma \in S_F}
\rho\left(Y_{\sigma(1)}\right) \cdots \rho\left(Y_{\sigma(F)}\right) \cr%
\end{array}%
\end{eqnarray}

\noindent ($S_F$ being the group of permutations of $F$ elements).
 If $V= V_0 \oplus \cdots \oplus V_{F-1}$ then for all $a
\in\{0,\cdots, F-1\}$, $V_a$ is a ${\mathfrak{g}}_0-$module and we have 

\beqa
\label{module}
\rho({\mathfrak{g}}_1) (V_a) \subseteq V_{a+1}.
\eeqa

\noi
There is no need to impose further conditions, since
the Jacobi identity \eqref{eq:J} is just
a consequence of the associativity of the product in $\text{End}(V)$.

In Definition \ref{flie-def} we have defined complex Lie algebras of order
$F$. Since we want to implement these new structures to extend non-trivially
the Poincar\'e algebra and since this last algebra is a real Lie algebra,
real forms of Lie algebras of order $F$ should be defined.
 A real elementary Lie algebra of order $F$ is
given by a real Lie algebra ${\mathfrak{g}}_0$ and ${\mathfrak{g}}_1$ a real
representation of $\g_0$ which satisfy the axioms of complex elementary Lie
algebras of order $F$. Of course when considering a real form of a complex 
${\mathfrak{g}}$ some structures are lost such that the grading map $%
\varepsilon$. Many examples of Lie algebras of order $F$
(complex and real)  were constructed.
In these lectures, since we study non-trivial higher order extensions of
the Poincar\'e algebra we only give Lie algebras of order $F$ which
are related to the Poincar\'e algebra. There is two types of such extensions
infinite-dimensional and finite-dimensional that will be studied
below.

\subsection{Lie algebras of order $F$ in $(1+2)-$dimensions}
In this subsection we construct explicitly an infinite dimensional
extension of the Poincar\'e algebra in $(1+2)-$dimensions
\cite{fsusy3d}. Studying 
the representations of this extension gives that this higher-order symmetry
is a symmetry between relativistic anyons. We recall firstly some
results upon the Lorentz and the Poincar\'e groups in three dimensions,
and then construct a higher order extension of the Poincar\'e
symmetry.

\subsubsection{Discrete series of $\overline{SO(1,2)}$}
\label{discrete}
Since $SO(1,2)$ is a non-compact Lie group its unitary representations are
infinite dimensional. There are two types of unitary representations.
The discrete and the continuous series. The former are either bounded
from below or bounded from above and the latter are unbounded representations.
These representations were studied by Bargmann \cite{bar} (see also
\cite{ggv-l}).
The generators of the Poincar\'e algebra in $(1+2)-$dimensions 
are  $(P_0, P_1, P_3)$ and $(J_0,J_1,J_2)$ and the algebra is given by

\beqa
\label{poin3}
\left[J_\mu, J_\nu\right]&=& 
-\epsilon_{\mu \nu \sigma} \eta^{\sigma \rho} J_\rho,
\nonumber \\
\left[J_\mu, P_\nu\right]&=& 
-\epsilon_{\mu \nu \sigma} \eta^{\sigma \rho} P_\rho,
 \\
\left[P_\mu, P_\nu\right]&=&0,\nonumber
\eeqa

\noi
with $\eta_{\mu \nu} = \text{diag}(1,-1,-1)$ the three-dimensional
Minkowski metric and $\e_{\mu \nu \rho}$ the Levi-Civita tensor normalised
as follows $\e_{012}=\e^{012}=1$. If we consider the complexified of the 
Poincar\'e algebra and  introduce $L_0=-i J_0, L_\pm=-iJ_1 \mp  J_2$ and
$\Pi_0=-iP_0, \Pi_\pm=-i P_1 \mp P_2$ 
(these unconventional notations come from our normalisation and
from the fact that our structure constants are real for Lie
algebras) the commutation relations reduce to\footnote{In the mathematical
literature one prefers to take $K_0=J_1$ for the Cartan subalgebra
with $K_\pm=J_0 \mp J_2$. We have made of different choice
 $-$ in the complexified
of $\mathfrak{so}(1,2)$ $-$ in such way that the eigenvalues of $L_0$
correspond to the eigenvalues of the $SO(2)$ subgroup of $SO(1,2)$ and
can be identified with the spin of the state.}

\beqa
\label{LP}
\begin{array}{llll}
&\left[L_-, \Pi_-\right]=0&
\left[L_-, \Pi_0\right]=\Pi_-&
\left[L_-, \Pi_+\right]=2\Pi_0\\
\left[L_0, L_\pm\right]=\pm L_\pm&
\left[L_0, \Pi_-\right]=-\Pi_-&
\left[L_0, \Pi_0\right]=0&
\left[L_0, \Pi_+\right]=\Pi_+\\
\left[L_+,L_-\right]=-2 L_0&
\left[L_+, \Pi_-\right]=-2\Pi_0&
\left[L_+, \Pi_0\right]=-\Pi_+&
\left[L_+, \Pi_+\right]=0.\\
\end{array}
\eeqa

Since $\pi_1(SO(1,2))= \mathbb Z$ there exists, in some 
universal covering group noted $\overline{SO(1,2)}$, representations  with
arbitrary spin $s \in \mathbb R$. This means that the eigenvalues of
$L_0$ are not any more integer (bosonic) or half-integer (fermionic). 
There are two types of unitary representations for
 $\overline{SO(1,2)}$ the discrete or the continuous series \cite{bar, ggv-l}.
The only representations that are relevant in the sequel are the
discrete series which are either bounded from below or bounded
from above. We have

\beq
\label{eq:r+}
\begin{array}{ll}
&L_s{}_0 |s_+,n \rangle = (s+n) |s_+,n \rangle \cr
{\cal D}_s^+:&
L_{s+} |s_+,n \rangle = \sqrt{(2s+n)(n+1)} |s_+,n+1 \rangle \cr
&L_{s-} |s_+,n \rangle = \sqrt{(2s+n-1)n} |s_+,n-1 \rangle,
\end{array}
\eeq

\noindent
for representations bounded from below and

\beq
\label{eq:r-}
\begin{array}{ll}
&L_s{}_0 |s_-,n \rangle = -(s+n) |s_-,n \rangle  \cr
{\cal D}_s^-:&L_{s+} |s_-,n \rangle = - \sqrt{(2s+n-1)n}|s_-,n-1 \rangle \cr 
&L_{s,-} |s_-n \rangle = - \sqrt{(2s+n)(n+1)} |s_-,n+1 \rangle. 
\end{array}
\eeq

\noi
for representations bounded from above. 
For the 
first representation we have $L_{s,-} |s_+,0 \rangle =0$ and 
for the second  $L_{s,+} |s_-,0 \rangle =0$. These representations
are of dimensions $2 |s| +1$ if $2s$ is a negative integer but in the
general case we have an infinite number of states.
In both cases the quadratic Casimir
operator $Q=L_0^2-\frac12(L_+ L_-+L_- L_+)$ 
has eigeinvalues $s(s-1)$. Furthermore, when $s <0$ the representations
are not unitary.
There is a well-known way to characterise these representations, using
functions of  complex variables (see the first reference of \cite{ggv-l}).
\\

To define higher order extensions of the Poincar\'e algebra,
taking the supersymmetric extension 
as a guideline, we choose representations for which
 $s=-1/F$ to build a non-trivial extension
of the Poincar\'e algebra. If we observe the  relations (\ref{eq:r+}) and 
(\ref{eq:r-}) with $s=-1/F$, we see an ambiguity in the square
root of $-2/F$. So a priori we have four different representations for 
$s=-1/F$, (two bounded from below/above) with the two choices 
$\sqrt{-1}=\pm i$.
We note ${\cal D}^\pm_{-1/F;\pm}$ (with obvious notations) these 
representations. Next, we can
make the following identifications
\begin{itemize}
\item the dual representation of  ${\cal D}^+_{-1/F;+}$ is obtained through
the substitution $J_\mu \longrightarrow -\left(J_\mu\right)^t$ and is given by
$\left[{\cal D}^+_{-1/F;+}\right]^*\cong{\cal D}^-_{-1/F;+}$;
\item the complex conjugate representation of ${\cal D}^+_{-1/F;+}$ is
defined by $J_\mu \longrightarrow \overline{J_\mu}$
 (we have to be careful when we do such
a transformation because we have {\it by definition 
$L_\pm = -i J_1 \mp J_2$
for any representation})
is given by $\overline{{\cal D}^+_{-1/F;+}}\cong{\cal D}^-_{-1/F;-}$;
\item the dual of the complex conjugate representation of 
${\cal D}^+_{-1/F;+}$ is given by 
$\left[\overline{{\cal D}^+_{-1/F;+}}\right]^*\cong{\cal D}^+_{-1/F;-}$.
\end{itemize}
If we note $\psi_a \in {\cal D}^+_{-1/F;+} ,\psi^a \in {\cal D}^-_{-1/F;+},
\bar{\psi}_{\dot a} \in {\cal D}^-_{-1/F;-}$ and 
$\bar{\psi}^{\dot a} \in {\cal D}^+_{-1/F;-}$  then we  have the following
transformation laws:

\beqa
\psi^\prime_a &=& S_a^{~~b} \psi_b \nonumber \\
\psi^{\prime a} &=& \left(S^{-1}\right)_{b}^{~~a} \psi^b  \\
\bar{\psi}^\prime_{\dot a} &=& \left(\bar S\right)_{\dot a}^{~~ \dot b} 
\bar {\psi}_{\dot b} 
\nonumber \\
\bar{\psi}^{\prime {\dot a}}&=& \left((\bar S)^{-1} \right)_{\dot b}^
{~~ \dot a}  \bar{\psi}^{\dot b}. \nonumber
\eeqa 

\noindent
Furthermore, if we define 
\beq
\psi^a = g^{a \dot a} \bar \psi_{\dot a},
\eeq
\noindent
we can write the following scalar product

\beq
\label{eq:ps}
\varphi^a \psi_a = -\bar \varphi_{\dot 0} \psi_0 + \sum \limits_{a>0} 
\bar \varphi_{\dot a} \psi_a,
\eeq 
\noindent
where the infinite matrix $g^{a \dot a}$ and its inverse $g_{\dot a a}$
are given by ${\mathrm{diag}} (-1,1,\cdots,1,\cdots)$.
The reason why we have a pseudo-hermitian scalar product is because
we are dealing with a non-unitary representation of a
non-compact Lie group. The invariant scalar product gives an explicit
isomorphism between the two representations bounded from below (or above)
($\left(S^{-1}\right)_b^{~~a}= g^{a \dot a} 
\left( \bar S\right)_{\dot a}^{~~\dot b} g_{\dot b b}$).
From now on, we choose  $\sqrt{-2/F}=i\sqrt{2/F}$ for 
representations bounded from below and  $\sqrt{-2/F}=-i\sqrt{2/F}$ for 
those bounded from above.

\subsubsection{Representations of the three-dimensional Poincar\'e algebra}

Particles are  classified according to the values of the Casimir
operators of the Poincar\'e algebra. More precisely, for  a mass $m$
particle
of positive/negative energy, 
the unitary irreducible representations are obtained by studying the little
group leaving the rest-frame momentum 
$\Pi^\alpha=-i P^\alpha=(m,0,0)$ invariant. (In the same manner
we consider $L_\mu=-iJ_\mu$ the angular momentum.) This
stability group in $\overline{SO(1,2)}$
is simply the universal covering group $\RR$ of 
the abelian subgroup of rotations  $SO(2)$ (generated by $L^0$).
As  is well-known, such a group is not quantised. This means that the 
substitution $L^0 \to L^0 + s$ leaves the $SO(2)$ part invariant. But the 
remarkable property of $\overline{SO(1,2)}$ is that the concomitant 
transformation
on the Lorentz boosts $L^i \to L^i + s {  \Pi^i \over  \Pi^0 + m}$  leaves the
algebraic structure (\ref{poin3}) unchanged. Indeed, on account of the 
mass-shell condition $\Pi_0^2 -\Pi_1^2 -\Pi_2^2=m^2$, one proves
that $(L_0+s, L_i+ s\frac{\Pi_i}{\Pi_0+m})$ satisfy the same algebra as
$(L_0,L_i)$. 
Anyway, following the method
of induced representations of groups expressible as  a semi-direct product,
we 
find that unitary irreducible representations for a massive particles are
one dimensional.

The main difference between   $SO(1,2)$, or more precisely the
proper orthochronous Lorentz group, and $SO(3)$ is that  $\Pi_0+m$ never
vanishes with $SO(1,2)$ and $s$ does not need  to be quantised.
In Ref.\cite{jn,p},  a relativistic wave equation for
massive anyons was given.
First,  notice that the two Casimir operators are the
two scalars $P.P$ and $P.J$ and their eigenvalues for a spin$-s$ unitary
irreducible representation are  $-m^2$ and $-ms$, respectively. The  equations
of
motion are then

\beqa
\label{eq:ms}
(P^2 + m^2) \Psi &=& 0 \\
(P.J + sm) \Psi  &=& 0.  \nonumber
\eeqa

However, to obtain  manifestly covariant equations one has to go beyond the
mass-shell conditions (\ref{eq:ms}) given by the induced representation. 
Therefore, we can start with a field which belongs to the appropriate
spin$-s$
representation of the {\it full} Lorentz group instead of the little group.
When $s$ is a negative  integer, or a negative half-integer, this 
representation is not unitary and is  $2|s|+1$
dimensional, and the solution of the relativistic wave equations reduces to
the appropriate induced representation (see \cite{b,jn} for an
explicit calculation in the case $|s|=1,1/2$). When $s$ is an arbitrary
number,
the representation is infinite dimensional and belongs to the discrete
series of $\overline{SO(1,2)}$. A relativistic wave equation for an anyon
in the continuous series was also considered in \cite{p,p2}.
For completeness, we recall  in Appendix
\ref{plu}, how M. Plyushchay obtained a very nice
relativistic wave equation in  \cite{p2}.

\subsubsection{Non-trivial extension of the Poincar\'e algebra in
$(1+2)-$dimensions}
Using the representations (\ref{eq:r-}--\ref{eq:r+}),
and with the sign ambiguity resolved, we can define two
series
of operators, belonging to a non-trivial representation of the Poincar\'e
algebra. We denote now $\sqrt{-1}=i$. 
Note $A^+_{-1/F+n}$ those built from the representation bounded from below
${\cal D}^+_{-1/F;+}$
and $A^-_{-1/F+n}$  the charges of the representation bounded from above
${\cal D}^-_{-1/F;-}$. Recall that $\overline{({\cal D}^+_{-1/F;+})}\cong
{\cal D}^-_{-1/F;-}$.
Using (\ref{eq:r-}, \ref{eq:r+}) we get

\beqa
\label{eq:A}
\left[L_0,A^+_{-1/F+n} \right] &=& (n-1/F)~~ A^+_{-1/F+n}  \nonumber \\
\left[L_+,A^+_{-1/F+n} \right] &=& \sqrt{(-2/F+n)(n+1)}~~ A^+_{-1/F+n+1} 
\nonumber
\\
\left[L_-,A^+_{-1/F+n} \right] &=& \sqrt{(-2/F+n-1)n}~~ A^+_{-1/F+n-1} 
\nonumber \\
&& \\
\left[L_0,A^-_{-1/F+n} \right] &=& - (n-1/F) ~~A^-_{-1/F+n}  \nonumber \\
\left[L_+,A^-_{-1/F+n} \right] &=& - \left(\sqrt{(-2/F+n-1)n}\right)^\star
~~ A^-_{-1/F+n-1} 
\nonumber
\\
\left[L_-,A^-_{-1/F+n} \right] &=& - \left(\sqrt{(-2/F+n)(n+1)}\right)^\star
~~A^-_{-1/F+n+1}.
\nonumber
\eeqa

\noindent
  
We want to combine this algebra (\ref{eq:A}) in a non-trivial way with the
Poincar\'e algebra (\ref{poin3}). With such a choice, $A^+_{-1 \over F}$ (resp.
$A^-_{-1 \over F}$) has  helicity $h=-{1\over F}$ (${1 \over F}$ resp.).
With the above choices for the square roots of the negative numbers, we know
that the representations are conjugate to each other, {\it i.e.} 
$\left(A_{-1/F+n}^+\right)^\dag \equiv A_{-1/F+n}^-$.

By Definition \ref{flie-def}, to associate a Lie algebra of order $F$ with
${\cal D}^\pm_{-1/F, \pm}$, we have to relate 
 ${\cal S}^F({\cal D}^\pm_{-1/F, \pm})$ to ${\cal D}_{-1}$. But here
some care has to be taken, since ${\cal D}^\pm_{-1/F, \pm}$ are
infinite-dimensional representations of $\mathfrak{so}(1,2)$, although
${\cal D}_{-1}$ is finite-dimensional. Indeed, using \eqref{eq:A} we have

$$
\left[L_\mp, (A^\pm_{-\frac{1}{F},\pm})^F\right]=0, \ \
\left[L_0, (A^\pm_{-\frac{1}{F},\pm})^F\right]=
\mp(A^\pm_{-\frac{1}{F}}, \pm)^F.
$$

\noi
This means that $(A^\pm_{-\frac{1}{F},\pm})^F$ might be identified
with  a primitive vector
for the vector representation. 
Consider  hence $\left<(A^\pm_{-\frac{1}{F}})^F\right>
\subset {\cal S}^F\left({\cal D}^{\pm}_{-1/F,\pm}\right)$, the representation
built from the primitive vector $(A^\pm_{-\frac{1}{F}})^F$. For
this representation,  {\it a priori}
 we have

\beqa
\label{verma1}
\left[L_\pm,\left[L_\pm,\left[L_\pm, (A^\pm_{-\frac{1}{F},\pm})^F
\right]\right]\right] \ne 0, 
\eeqa

\noi but using \eqref{LP} we can easily see  that 

\beqa
\label{verma2}
\left[L_\mp,\left[L_\pm,\left[L_\pm,\left[L_\pm, (A^\pm_{-\frac{1}{F},\pm})^F
\right]\right]\right]\right] = 0. 
\eeqa

\noi
This can be represented by means of a diagram. We denote now
$[X,Y]= \text{ad}(X).Y$, $(A^\pm_{-\frac{1}{F},\pm})^F=V_\mp,\
\text{ad}(L_\pm)^n.(A^\pm_{-\frac{1}{F},\pm})^F=V_{ \pm (n-1)}$

\beqa
\label{diag}
\xymatrix{& \ar@<1ex>[dl]^{ L_\mp  }
V_\mp \ar@<1ex>[r]^{ L_\pm  }&
V_0 \ar@<1ex>[l]^{ L_\mp  }
\ar@<1ex>[r]^{ L_\pm }&
V_\pm
\ar@<1ex>[l]^{ L_\mp }
\ar@<1ex>[r]^{ L_\pm }&
V_{ \pm 2} \ar@<1ex>[dl]^{ L_\mp  }
\ar@<1ex>[r]^{ L_\pm  }&
V_{ \pm 3} \ar@<1ex>[l]^{ L_\mp  }
\ar@<1ex>[r]^{ L_\pm }
&
\cdots \ar@<1ex>[l]^{ L_\mp }
\ar@<1ex>[r]^{ L_\pm  }
& V_{\pm n}
\ar@<1ex>[r]^{ L_\pm }
\ar@<1ex>[l]^{ L_\mp }
&\dots
\ar@<1ex>[l]^{ L_\mp  }&
\\
0&&&0&
}
\eeqa

\noi
This means that we can safely impose $V_{n}=0$ for $n= \pm 2, \pm 3, \cdots$.
This can be presented on a more formal ground. See Appendix \ref{indec}
for more details. 
To construct a non-trivial extension of the
Poincar\'e algebra we then proceed as follows. 
We identify $(A^\pm_{-\frac{1}{F},\pm})^F$ with
$\Pi_\mp$ and we impose the relation 
$\text{ad}^3(L_\pm).(A^\pm_{-\frac{1}{F},\pm})^F=0$. Thus we have 
the following isomorphism

$$
{\cal D}_{-1} = \left<\Pi_0,\Pi_+,\Pi_-\right> = \left\{(A^\pm_{-\frac{1}{F},\pm})^F,
\left[L_\pm,(A^\pm_{-\frac{1}{F},\pm})^F\right],
\left[L_\pm,\left[L_\pm,(A^\pm_{-\frac{1}{F},\pm})^F\right]\right]
\right\} 
$$

\noi
 which leads
to the following brackets

\beqa
\label{3dfsusy}
&&\frac{1}{F!}
\left\{A^\pm_{-{1\over F}},\dots,A^\pm_{-{1\over F}} \right\} = \Pi_\mp
\nonumber \\
&&\frac{1}{F!}\left\{A^\pm_{-{1\over F}},\dots,A^\pm_{-{1\over F}},
A^\pm_{1-{1\over F}}
\right\}
 =\pm i \sqrt{{2 \over F}}  \Pi_0 \\
&&   -\frac{(F-1)}{F!} 
\left\{A^\pm_{-{1\over F}},\dots,A^\pm_{-{1\over F}},A^\pm_{1-{1\over F}},
A^\pm_{1-{1\over F}} \right\}
\pm i \frac{\sqrt{ F-2}}{F!}
\left\{A^\pm_{-{1\over F}},\dots,A^\pm_{-{1\over F}},A^\pm_{2-{1\over F}} 
\right\}_F
=   \Pi_\pm \nonumber \\
&&\left[L_\pm,\left[L_\pm,\left[L_\pm, \left(A^\pm_{-{1 \over F}}\right)^F
\right]\right]\right]=0 \nonumber \\
&& ~~~~~~~~~~~~ \vdots \nonumber
\eeqa

How,  can we address the question of the remaining brackets? 
In fact, it is impossible  to find a decomposition

\beq
{\cal S}^F\left({\cal D}^\pm_{-1/F}\right) = {\cal D}_{-1} \oplus 
\D_{-1}^\perp,
\eeq

\noindent
where $\D_{-1}^\perp$ is stable under $SO(1,2)$. Indeed, if there were such a
decomposition, there would exist a $SO(1,2)$ equivariant projection

\beq
\pi:~{\cal S}^F\left({\cal D}^\pm_{-1/F}\right) \longrightarrow {\cal D}_{-1}.
\eeq 

\noindent  
But then 
$X^\pm=\pi\left( {\cal S}^F\left(A^\pm_{-1/F},\cdots,A^\pm_{-1/F},A^\pm_{3-1/F}
\right) \right) \in {\cal D}_{-1}$ would satisfy (see \ref{3dfsusy})

$$\big[L_\mp,\big[L_\mp,\big[L_\mp,X^\pm \big]\big]\big]= 
 i F!\sqrt{2/F}\sqrt{2(1-2/F)} \sqrt{3(2-2/F)} \Pi_\mp\ne 0,$$

\noindent
and this is impossible because in the vector representation ${\cal D}_{-1}$,
$\text{ad}^3(L_-)$ acts as zero. 
This means that the algebra given by \eqref{3dfsusy} is not
 a Lie algebra of order three.
There are in fact two ways to define a Lie algebra of order three. 
Either to embed ${\cal D}_{-1}$ into an infinite dimensional (but
indecomposable) representation of $\mathfrak{so}(1,2)$ or to
 embed the algebra
$\mathfrak{so}(1,2)$ into a infinite dimensional algebra, which here
turns out to be the De Witt algebra (the centerless Virasoro algebra) 
\cite{flie1,vir}. Since this  part is technical and not relevant
for the sequel, it will be developed in Appendix \ref{indec}.

\subsubsection{Representations}
\label{rep-fsusy}
Before studying the representations of the algebra 
given by the quadratic relations \eqref{poin3} and \eqref{eq:A}
and the higher order brackets \eqref{3dfsusy}, let
us draw some general feature. 
First, since $P^2$ commutes with all the generators, all states in
an irreducible representation have the same mass.
Second, define 
an anyonic-number operator (this is in fact the grading map of
the definition of Lie algebras of order $F$)
 $\exp({2i\pi{\cal N}_A})$ which gives
the 
phase $e^{2i\pi s}$ on a spin$-s$ anyon and assume we have 
a finite dimensional representation of our algebra. We have ${\mathrm{tr}} 
\exp({2i\pi{\cal N}_A}) =0$  showing that
in each irreducible representation there are $F$ possible states with
helicity
($s,s\pm{1\over F},\dots, s\pm{ {F-1\over F}}$), where $s$  will be specified
later) and the dimension of the space with a given helicity is
always
the same.  This can be checked proving by that ($\exp({2i\pi{\cal N}_A})
A_s = e^{2i\pi s}  A_s \exp({2i\pi{\cal N}_A})$) and using cyclicity of
the trace

\beqa
&&{\mathrm{tr}} \left( \exp({2i\pi{\cal N}_A})
\left\{A^+_{-{1\over F}},\dots, A^+_{-{1\over F}},A^+_{1-{1\over
F}}\right\}_F\right) \nonumber \\
&=&{(F-1)!} \times {\mathrm{tr}}
\left( \sum \limits_{a=0}^{F-1} e^{2i\pi{\cal N}_A} 
\left(A^+_{-{1\over F}}
\right)^a
\left(A^+_{1-{1\over F}}\right)\left(A^+_{-{1\over F}}\right)^{F-a-1}\right)
\nonumber \\
&=&{(F-1)!}  \times \left(\sum \limits_{a=0}^{F-1} e^{-{2i \pi a \over F}}\right)
{\mathrm{tr}} \left(
\left(A^+_{-{1\over F}}\right)^{F-1}e^{2i\pi{\cal N}_A}
 \left(A^+_{1-{1\over F}}\right)\right)=0.
\nonumber 
\eeqa

To study the representations of the higher order extension of the Poincar\'e
algebra, we follow the method of induced representations of Wigner.
We restrict ourselves to massive representations. The massive representations
are characterised by  $P^\mu P_\mu=-m^2$ 
(recall that $P_\mu$ are antihermitian). Going in the rest-frame where
$P^\mu=(im ,0,0)$ gives  that the only non-vanishing bracket are the ones 
involving the generator $A_{-1/F}^\pm, A_{1-1/F}^{\pm}$. Thus we assume that
$A^\pm_s=0$ with $s \ne -1/F, 1-1/F$ 
{\it i.e.} that the ``active'' charges are $A_{-1/F}^\pm$ and $A_{1-1/F}^\pm$,
and
the little algebra is generated by $1, L_0$ and
$A_{-1/F}^\pm, A_{1-1/F}^{\pm}$.
After appropriate rescaling of the generators, the higher order brackets are
given by

\beqa
\label{little}
(F-1)!&=&\left\{A^\pm_{-1/F},\cdots,A^\pm_{-1/F},A^\pm_{1-1/F}\right\}\\
0&=&\left\{A^\pm_{s_1},\cdots,A^\pm_{s_{F-1}},A^\pm_{s_F}\right\}, 
s_1, \cdots, s_F =-1/F, 1-1/F, \ s_1 + \cdots s_F \ne 0.
\nonumber
\eeqa

\noi
This algebra, called the Clifford algebra of polynomial's is studied in Appendix
\ref{cliff}. A relevant representation is given by

{\tiny
\beqa
\label{eq:q1}
A^+_{-{1\over F}}=
\begin{pmatrix}0&0&0&\ldots&0&0& \cr
                         \sqrt{1(F-1)}&0&0&\ldots&0&0& \cr
                         0& \sqrt{2(F-2)}&0&\ldots&0&0& \cr
                         &\cr
                         \vdots&\vdots&&\ddots&\ddots&\vdots   \cr
                         0&0&\ldots&0&\sqrt{(F-1)1}&0& \end{pmatrix}, 
\ \  
A^+_{1-{1\over F}}=
                \begin{pmatrix} 0&0&0&\ldots&0&1/(F-1)!\cr
                          0&0&0&\ldots&0&0& \cr
                          0&0&0&\ldots&0&0& \cr
                          &\cr\
                          \vdots&\vdots&&&\ddots&\vdots& \cr 
                           0&0&0&\ldots&0&0&
\end{pmatrix}.
\eeqa  
 }

\noi
In \cite{fsusy3d} another equivalent representation was exhibited.
From the basic conjugation we obtain

$$
A^{-}_{-1/F} = (A^{ +}_{-1/F})^\dag, \ \ 
 A^{-}_{1-1/F} = (A^{ +}_{1-1/F})^\dag.
$$

\noi
As a direct consequence, the generators satisfy the additional relations

$$
A^+_{1-1/F}= \frac{1}{(F-1)!^2} (A^-_{-1/F})^{F-1},
$$

\noi
together with the quadratic 
relations

\beqa
\label{eq:sl}
\left[A^-_{-1/F}, A^+_{-1/F}\right]  &= &2  N =2 \mathrm{diag} (\frac{F-1}{2},
\frac{F-3}{2},\cdots,
\frac{1-F}{2}) \\
\left[ N,A^\pm_{-1/F}\right]&=& \mp   A^\pm_{-1/F}. \nonumber
\eeqa

\noi
This means that $iN, \frac{i}{2}(A^+_{-1/F}+ A^-_{-1/F}),
\frac{1}{2}(A^+_{-1/F}- A^-_{-1/F})$ are antihermitian and generate the
finite dimensional unitary
 irreducible representation of $\mathfrak{su}(2)$. Thus the representation
of the little algebra build with $A^\pm_{-1/F}$ will be unitary.
In fact, we obtain an $F-$dimensional representation of
$\mathfrak{su}(2)$. To identify the helicity content of the representation,
we assume that we are starting 
from a vacuum state $\Omega^+_\lambda=\Omega_\lambda$ in the spin$-\lambda$
representation
of $SO(1,2)$ which is annihilated by $A^-_{-1/F}$.  
Thus acting on the vacuum state with the operator $A^+_{-1/F}$ we identify
the helicity content of the representation. Indeed the
states $(A^+_{-1/F})^n \Omega_\lambda, \ n=0,\cdots, F-1$ are of helicity
$h=\lambda-n/F$. We assume further that these states are of positive energy.
Since the representation has to be CPT invariant, we consider also
a conjugated vacuum $\Omega^-_{-\lambda}=\bar \Omega_\lambda$ 
of helicity $h=-\lambda$,
negative energy and annihilated by  $A^+_{-1/F}$. Acting with
 $A^-_{-1/F}$ on the conjugated vacuum, 
we obtain the states of helicity $n/F-\lambda$.
This can be summarised in the following table:

$$\vbox{\offinterlineskip \halign{
\tv# & \cc{#} & \tv# & \cc{#}  & \tv# &
\cc{#} & \tv# & \cc{#} & \tv# & \cc{#}& \tv# \cr
\noalign{\hrule}
&\cc{states}&&\cc{helicity} &&\cc{states}&&\cc{helicity} & \cr
\noalign{\hrule}
&$\Omega_{\lambda}^+$&&$\lambda$&
&$\Omega_{-\lambda}^-$&&$-\lambda$& \cr
\noalign{\hrule}
&$A^+_{-1/F}\Omega_{\lambda}^+$&&$\lambda-1/F$&
&$A^-_{-1/F}\Omega_{-\lambda}^-$&
&$-\lambda+1/F$& \cr
\noalign{\hrule}
&$\vdots$&& && &&$\vdots$& \cr
\noalign{\hrule}
&${\left(A^+_{-1/F}\right)^a }\Omega_{\lambda}^+$&
&$\lambda-a/F$&
&${\left(A^-_{-1/F}\right)^a }\Omega_{-\lambda}^-$&
&$-\lambda+a/F$ &\cr
\noalign{\hrule}
&$\vdots$&& && &&$\vdots$& \cr
\noalign{\hrule}
&${\left(A^+_{-1/F}\right)^{F-1} }\Omega_{\lambda}^+$&
&$\lambda-(F-1)/F$&
&${\left(A^-_{-1/F}\right)^{F-1}}\Omega_{-\lambda}^-$&
&$-\lambda+(F-1)/F$ &\cr
\noalign{\hrule}
}}$$

\noindent
The states of positive energy and helicity ($\lambda,\lambda -{1 \over F},
\dots,\lambda -{F-1 \over F}$) are   CP- conjugated to the states 
of negative  negative energy  and helicity ($-\lambda,-\lambda +{1 \over
F}, \dots,-\lambda +{F-1 \over F}$). 
Having the states on shell, the states of positive energy are boosted and
belong to the representations bounded from below ${\cal D}^+_{\lambda-n/F}$
while those of negative energy correspond to the representations bounded
from above ${\cal D}^-_{\lambda-n/F}$. It should be interesting to have a
explicit Lagrangian where the higher order extensions of the Poincar\'e
algebra are realised, but up to now there are no results in these direction.
As a final remark let us mention that if we had considered only 
one series of operator $A$, say $A^+_{n-1/F}$, the representation would
not have been unitary. 

\subsection{Cubic extensions of the Poincar\'e algebra in any dimensions}
\label{cubic-poincare}
Now, we would like to extend the ideas developed previously in any
space-time dimensions, namely to construct higher order extensions
of the Poincar\'e algebra in $D-$dimensions. We have shown that considering
any  semi-simple Lie algebra $\g$ and any representation ${\cal D}$, 
along the lines of Appendix \ref{indec}, one is able to define a Lie algebra
of order $F$ starting from $\g \oplus \D$ \cite{flie1}. 
In particular this applies to
$\mathfrak{so}(1,D-1) \oplus \D_v$, with $\D_v$ the vector representation of
$\mathfrak{so}(1,D-1)$. However,  this construction leads to infinite
dimensional representations of the Lorentz group which are not exponentiable,
and as a consequence is unacceptable for realistic physical models.
This obstruction is valid as soon as the space-time dimension is greater than  four.
Indeed,
 even in four dimensions, where  the situation is somewhat exceptional,
we cannot construct cubic extensions along the lines of infinite dimensional algebras.
In this case, the little group
for massless particles is $E(2)$, the group of affine transformations  in
the plane. Representing the translational part by zero gives rise to
$SO(2)$, which is not quantised. So we  could have expected  to consider, in
this case, some
massless states with fractional helicity. In \cite{pr}
we defined 
a  relativistic wave equation along
the lines of Appendix \ref{plu}, 
 considering infinite-dimensional
representations corresponding  formally to the massless states
with fractional (real) helicity. 
But the solutions of the relativistic equation,
 however, break down ``spontaneously'' the $(1+3)D$
Poincar\'e invariance to the $(1+2)D$ Poincar\'e invariance
and induce a compactification on a circle that produces a
consistent theory for massive anyons in $D=1+2$.

This seems to be a serious obstacle to define higher order extensions
of the Poincar\'e algebra in the framework of Lie algebras of order $F$
when the space-time dimension is greater than four.
Therefore, if we expect to apply higher order extensions of the Poincar\'e
algebra, we have to follow new lines.
Subsequently, it was realised that finite-dimensional Lie algebras of
order $F$ could be defined. Indeed induction theorems enable us to
define Lie algebras of order $F$ from Lie algebras or Lie 
superalgebras \cite{flie2}. Thus, this opens the way to
obtain higher order extensions of the Poincar\'e algebra. 
However, we should understand in that
context, how
these structures would not be in conflict with the Noether and
spin-statistics theorems. Since we continue to assume that bosonic fields
are quantised by commutation relations although fermionic fields are
quantised by anticommutation relations, 
a fresh look to the Noether theorem should be given in order to implement
Lie algebras of order $F$.

\subsubsection{Noether theorem in higher order algebras}

To understand in which way the Noether theorem can be
implemented in higher order symmetries, we first recall 
some well-known results and give some general features in the
conventional case {\it i.e.}  when the symmetries correspond to a Lie
(super)algebra. 
 According to 
Noether theorem, to all the symmetries there correspond conserved currents.
The symmetries are then  generated by charges which are expressed in 
terms of the fields.
We  actually  start from the classical field theory case,
then go to the quantum case through the usual canonical
quantisation procedure.
Starting from a  general Lagrangian ${\cal L}$
at the classical level,  invariant under 
some Lie (super)algebra $\g$,
one constructs through the standard procedure 
the Noether charges $\hat T_a$ 

\begin{equation}
\label{charge}
\varepsilon^a \hat T_a = 
\int d^3x \frac{\partial {\cal L}}{\partial\partial_0 \Psi} 
\delta_\varepsilon \Psi 
\end{equation}

\noindent
associated with the transformation

\begin{equation}
\delta_\varepsilon \Psi = \varepsilon^aT_a 
\Psi  \label{trans1}
\end{equation}

\noindent
where $T_a$ generate the Lie (super)algebra in some
appropriate (matrix) representation.
Upon use of eq.(\ref{trans1}) in eq.(\ref{charge}) 
one gets

\begin{equation}
\label{charge1}
\hat{T}_a = 
\int d^3x \Pi(x)  T_a \Psi(x) 
\end{equation}

\noindent
where $\Pi(x) = \frac{\partial {\cal L}}{\partial\partial_0 \Psi}$
is the conjugate momentum. Equation (\ref{charge1}) is the general
relation between $\hat T_a$ and $T_a$.
At the quantum level,
$\hat T_a$ and $\Psi$ are operators acting in some Hilbert space
and we have
\begin{equation}
[\varepsilon^a \hat{T}_a, \Psi(x)] = 
\delta_\varepsilon \Psi \label{trans3}.
\end{equation}

\noindent
Thus the
conserved  charges $\hat T_a$ from Noether's currents,
realise the algebra in the sense that

\begin{equation}
\label{dLie}
 [\delta_a, \delta_b] \Psi \equiv 
\delta_a (\delta_b \Psi) -  \delta_b (\delta_a \Psi) =
 f_{a b}{}^{ c} \delta_c \Psi 
\end{equation}

\noindent
where $\Psi$ is a field operator and $\delta$ is {\sl by definition}
given by

\begin{equation}
\label{deltaLie}
\delta_a \Psi \equiv  [ \hat T_a, \Psi].
\end{equation}  

\noindent
Equation (\ref{dLie}) reads then equivalently 

\begin{equation}
\label{Lie}
[\hat T_a, [\hat T_b, \Psi]] - [\hat T_b, [\hat T_a, \Psi]] =
     f_{a b}{}^{ c} [\hat T_c, \Psi]. 
\end{equation}

\noindent
Now due to Jacobi identity ($ [A, [B, C]] + \mbox{cyclic} = 0$), one can recast
the left-hand side of Eq.(\ref{Lie}) in the form 
$ [[\hat T_a, \hat T_b], \Psi]$
to get

\begin{equation}
[[\hat T_a, \hat T_b], \Psi] = f_{a b}{}^{ c} [\hat T_c, \Psi] 
\; \; \; \; \; \; \; \;  \mbox{(for any $\Psi$)} \label{eq4}
\end{equation}

\noindent
meaning that 

\begin{equation}
[\hat T_a, \hat T_b] = f_{a b}{}^{ c} \hat T_c  \label{eq5}
\end{equation}

\noindent
at least on some (sub-)space of field operators $\Psi$. \\

Now consider a Lie algebra of order three where the cubic brackets
are given by

\beqa
\label{3lie}
\left\{Y_i,Y_j,Y_k\right\}= Q_{ijk}{}^a X_a,
\eeqa
all the results above remain true  (Eqs.\eqref{charge} and \eqref{deltaLie})
but  the analogy with the steps 
described above stops at Eq.(\ref{Lie}).
Now the algebra is realised (with $\hat Y_i, \hat X_a$ the conserved charges
associated to \eqref{3lie})

\beqa
\label{F-com}
\left( \delta_i. \delta_j.
\delta_k + \mathrm{perm} \right) { \Psi}& =
\left[\hat  Y_i, \left[\hat Y_j, \left[\hat Y_k,  
{ \Psi} \right]\right]\right]+ \mathrm{perm} =
Q_{ijk}{}^a
\left[\hat X_a, { \Psi} \right],
\eeqa

\noi
but, the
(generalised) Jacobi identities (see Eq. (\ref{eq:J})),   do not allow to  obtain

\begin{equation}
[\left\{\hat Y_i, \hat Y_j,\hat Y_k\right\}, \Psi]
=Q_{ijk}{}^a[\hat X_a, \Psi] 
\label{eq6}
\end{equation}

\noindent
which would have been the analog of Eq.(\ref{eq4}) above. 
In other words, due to the lack of some appropriate Jabobi identities, 
the quantised version of the Noether charges algebra is just (\ref{F-com})
and cannot be cast simply in a ${ \Psi}$ independent form.
 We should stress at this level that the difference with the conventional 
algebras  and algebras of the type \eqref{3lie}
we are pointing out, does not mean the absence of a realisation of 
this algebra  in terms of Noether charges, as we will see latter on.

We thus define the adjoint representation where the generators of
the Lie algebra of order three act on some given operator $\Phi$
which belongs to some endomorphism space $\text{End}(V)$ of some vector 
space $V$. In Quantum Mechanics, $V$ reduces to some Hilbert space.
The adjoint representation of an elementary Lie algebra of order three
is a linear map

\beqa
\label{ad}
\begin{array}{rcl}
\mathrm{ad } :  \g=\g_0 \oplus \g_1 &\to& {\cal L}(\text{End} (V),
\text{End} (V))  \\
g& \mapsto& \text{ad}(g)
\end{array}\\
\eeqa

\noi
such that for all $ \Phi \in \mbox{En}(V)$ we have  $\text{ad}(g).\Phi=
[\text{ad}(g),\Phi]$.
In the adjoint representation,
the algebra is realised through multiple-commutators. 
Furthermore, it is a matter of calculation to check that 
the 
Jacobi identities \eqref{eq:J}  are satisfied.

Another difference with usual quadratic algebras lies on the tensor
product of representations (this has implications on $N-$particles states
at the quantum level). 
Assume that we have two representations $\rho_1$ of
and $\rho_2$ of \eqref{3lie}, then $\rho_1 \ \hat \otimes \ 1 +
 1 \ \hat \otimes \ \rho_2$ is
a representation if $ \hat \otimes$ is the twisted tensor product, 
such that for homogeneous 
elements $G_1,G_2,G_3,G_4$ of $\g$  
we have $(G_1 \hat \otimes G_2)(G_3 \hat \otimes G_4)= q^{|G_2| |G_3|}
G_1 G_3 \hat \otimes G_2 G_4$, where $|G|$ denotes the $\ZZ_3-$grading
of $G$
and $q$ is a cubic primitive root of the unity. 
Indeed, if we calculate 
\beqa
\label{coprod}
\left\{ \rho_1(Y_i) \ \otimes \ 1 + 1 \ \hat \otimes \ \rho_2(Y_i),
\rho_1(Y_j) \ \otimes \ 1 + 1 \ \hat \otimes \ \rho_2(Y_j),
\rho_1(Y_k) \ \otimes \ 1 + 1 \ \hat \otimes \ \rho_2(Y_k)
 \right\}
\eeqa

\noi
(we assume here for simplicity a matrix representation), the mixed terms
like 

$$\left\{ \rho_1(Y_i) \ \otimes \ 1,
\rho_1(Y_j) \ \otimes \ 1 ,
 1 \ \hat \otimes \ \rho_2(Y_k)
 \right\}$$
 vanish because we have $1+q+q^2=0$ and \eqref{coprod} 
reduces to

$$
\left\{ \rho_1(Y_i),
\rho_1(Y_j),
\rho_1(Y_k) \right\} \ \hat \otimes 1 + 1 \ \hat \otimes \
\left\{  \rho_2(Y_i), \rho_2(Y_j), \rho_2(Y_k) \right\} = 
Q_{ijk}{}^a \Big( \rho_1(X_a) \ \hat \otimes \ 1 + 1\ \hat \otimes \
\rho_2(X_a)\Big).
$$

\noi
This result has been established in a more formal way, defining a
coproduct for Lie algebras of order three \cite{hopf}. This twisted tensor
product is in fact a general feature of $n-$ary algebras as Clifford
algebras of polynomial's (see Appendix \ref{cliff}).

To end this subsection, let us mention that
 we have shown that it is possible to endow
a Lie algebra of order three with the structure of a Hopf algebra \cite{hopf}. 
The benefit
of this construction is two-fold. Firstly, the parameters of the
transformation associated to the graded part of an elementary Lie
algebra of order three have been identified, and turn out to be
related to the three-exterior algebra (see Eq.[\ref{3-ext}],
Appendix \ref{cliff}).
Secondly, a group associated to a Lie algebra of order three was
constructed. This latter being simply some matrix group where the
matrix elements belong to the three-exterior algebra, in a straight
analogy with  Lie supergroups associated to Lie superalgebras
\cite{hopf}.

\subsubsection{Lie algebras of order three in any space-time dimensions}

As mentioned previously, it seems  rather difficult to
define higher order extensions of the Poincar\'e algebra
  in the framework of infinite dimensional
Lie algebras of order $F$ when the space-time dimension is bigger than three.
A major progress, in this direction,  was done when it was realised that it
was possible to construct {\it finite-dimensional} Lie algebras of order
$F$ from the usual Lie algebras and Lie superalgebras \cite{flie2},
by an induction theorem.

\begin{theorem} \label{tensor} (M. Rausch de Traubenberg, M. J. Slupinski,
\cite{flie2}).

\noi Let $\mathfrak{g}_{0}$ be a Lie algebra and
$\mathfrak{g}_{1}$ be a $\mathfrak{g}_{0}$-module  such that:

(i) $\g=\mathfrak{g}_{0} \oplus \mathfrak{g}_{1}$ is a Lie algebra
of order $F_1>1$;

(ii) $\mathfrak{g}_{1}$ admits  a $\mathfrak{g}_{0}$-equivariant
symmetric form of order $F_2 \ge 2$.

\noindent Then  $\g=\mathfrak{g}_{0}  \oplus \mathfrak{g}_{1}$
inherits the structure of a Lie algebra of order $F_1 +F_2$.
\end{theorem}

The proof of this theorem is not very difficult, the only delicate point
is to prove the Jacobi identities \eqref{eq:J}. Indeed one can show that
these last identities result
 on some factorisation property \cite{flie2}. This theorem
can be extended to include the $F_1=1$ cases, and as a simple consequence, 
one can define  elementary Lie algebras of order three from any Lie
algebras \footnote{It is also possible to construct Lie algebras of
order $F$ from Lie superalgebras. For instance, in \cite{flie2} a 
quadratic extension of the Poincar\'e algebra has been obtained
from the orthosymplectic Lie superalgebra.}.
Let $\g_0$ be any Lie algebra and let $\g_1$ be its adjoint representation.
Introduce $\{J_a, a=1,\cdots, \text{dim } \g_0\}$
 a basis of $\g_0$,
$\{A_a,a=1,\cdots, \text{ dim } \g_0\}$ the
 corresponding basis of $\g_1$.
The invariant forms are given by the Casimir operators. In particular,
for the Killing form, set
$g_{ab}=Tr(A_aA_b)$ and denote
$f_{ab}{}^c$  the structure constants.
Then one can endow $\g=\g_0\oplus\g_1$ with a
Lie algebra of order $3$ structure given by
\beqa
\label{lie-3lie}
\left[J_a,J_b\right] &=&f_{ab}{}^c J_c, \nonumber \\
\left[J_a,A_b\right] &=&f_{ab}{}^c A_c,  \\
\left\{ A_a,A_b,A_c \right\}&=&g_{ab}J_c+g_{ac}J_b+g_{bc}J_a. \nonumber 
\eeqa

\noi
The algebra \eqref{lie-3lie} enables us to define a cubic extension
of the Poincar\'e algebra in any space-time dimension.
Consider the real Lie algebra $\g_0=\mathfrak{so}(1,D)$
(we could have equally chosen $\g_0=\mathfrak{so}(2,D-1)$) and
define $\g= \mathfrak{so}(1,D) \oplus \text{ad}( \mathfrak{so}(1,D))=
\left<L_{MN}=-L_{NM}, 0 \le M < N \le D+1>\right> \oplus
\left<A_{MN}=-A_{NM}, 0 \le M < N \le D+1\right>$. The algebra
is given by

{\small
\beqa
\label{sitter}
\left[L_{MN}, L_{PQ}\right]&=&
\eta_{NQ} L_{PM}-\eta_{MQ} L_{PN} + \eta_{NP}L_{MQ}-\eta_{MP} L_{NQ},
\nonumber \\
\left[L_{MN}, A_{PQ} \right]&=& \eta_{NQ} A_{PM}-\eta_{MQ} A_{PN} + 
\eta_{NP}A_{MQ}-\eta_{MP} A_{NQ}, \\
\left\{A_{MN}, A_{PQ}, A_{RS} \right \}&=&
 (\eta_{MP} \eta_{NQ} -
\eta_{MQ} \eta_{NP}) L_{RS} 
+ (\eta_{MR} \eta_{NS} - \eta_{MS}
\eta_{NR}) L_{PQ}+ (\eta_{PR} \eta_{QS} - \eta_{PS} \eta_{QR})
L_{MN} \nonumber
\eeqa
}

\noi
with $\eta_{MN}=\text{diag}(1,-1,\cdots,-1)$.
Using vector indices of $\mathfrak{so}(1,D-1)$ coming from the inclusion
$\mathfrak{so}(1,D-1) \subset \mathfrak{so}(1,D)$, 
$\g_0$ is generated by $L_{\mu \nu}, L_{\mu D}$, with 
$\mu, \nu =0,\cdots,D-1$ and the graded part by $A_{\mu \nu}, A_{\mu D}$.
Letting $R \to \infty$
after  the In\"on\"u-Wigner contraction \cite{flie2},

\beqa
\begin{array}{ll}
J_{\mu \nu} = L_{\mu \nu},& P_\mu=\frac{1}{R} L_{\mu D} 
\\
T_{\mu \nu} = \frac{1}{\sqrt[3]{R}} A_{\mu \nu},& 
V_{ \mu }  = \frac{1}{\sqrt[3]{R}} A_{\mu D },
\end{array}
\eeqa

\noindent   
one sees that  
$L_{\mu \nu }$ and $P_\mu$ generate the  Poincar\'e
algebra in $D-$dimensions and that 
$T_{\mu \nu}, V_\mu$ are
 in  respectively  the adjoint and vector representations of
$\mathfrak{so}(1,D-1)$.
This Lie algebra of order three is therefore a non-trivial
extension of the Poincar\'e algebra where translations are cubes
of more fundamental generators. The subspace  generated by 
$L_{\mu \nu}, P_\mu, V_\mu$ is also a Lie algebra of order three
extending the Poincar\'e algebra.
The trilinear symmetric brackets have the simple form 

\beqa
\label{cubic-poin}
\left\{V_\mu, V_\nu, V_\rho \right \}=
\eta_{\mu \nu} P_\rho +  \eta_{\mu \rho} P_\nu + \eta_{\rho \nu} P_\mu,
\eeqa

\noindent
where $\eta_{\mu \nu}=\text{diag}(1,-1,\cdots,-1)$ is  the Minkowski metric.

This algebra has been firstly studied in four-space time dimensions
\cite{cubic1,cubic2,noether} and  then in any dimensions
\cite{p-form}, and it has been realised that it induces naturally
a symmetry on generalised gauge fields or $p-$forms.
An analysis of the implementation of the
Noether theorem in relation with the spin-statistics theorem was
given in \cite{noether} in four dimensions.
Furthermore,
 the cubic algebra \eqref{cubic-poin} is not the only non-trivial
extension of the Poincar\'e algebra one can construct. In fact in 
\cite{flie3} a classification of all cubic extensions of the Poincar\'e
algebra in four space-time dimensions was given. Furthermore in \cite{cr},
along the lines of the classification of Bacry and   L\'evy-Leblond
(where a classification of kinematical algebras was undertaken \cite{bl}), 
we have classified all kinematical Lie algebras of order three in four
dimensions
and  showed that they are related through generalised Inon\"u-Wigner
contractions of the algebra \eqref{sitter} \cite{cr}.

\subsubsection{Representations}
In this section we construct some multiplets of the cubic extension
of the Poincar\'e algebra  \eqref{cubic-poin}. Since, the generators
$V_\mu$ are in the vector representation of the Lorentz algebra, a multiplet
contains states of the same statistics. This is rather different from the
supersymmetric extension of the Poincar\'e algebra. To proceed further, 
from the matrix representations given in \eqref{red} or \eqref{irred}, one
can construct representations of the algebra \eqref{cubic-poin}. The 
 representations are specified by the representation of the vacuum. If the
vacuum is in the trivial representation of the Lorentz group the multiplet 
consists of three spinors 

$$
{\bf \Psi}=\begin{pmatrix} \Psi_1\\ \Psi_2 \\ \Psi_3\end{pmatrix},
$$

\noi
transforming like $\delta_\e {\bf \Psi}=\e^\mu V_\mu {\bf \Psi}.$
An invariant Lagrangian with three spinors was constructed in 
\cite{cubic1,cubic2}.
However, there is more interesting multiplets obtained when the
 vacuum is in the
spinor representation of the Lorentz group \cite{p-form}. In this case, a multiplet 
contains $p-$forms. Generalised gauge fields or $p-$forms which
are fully antisymmetric tensors are  generalisations
of the usual electromagnetic gauge field. The
revival of interest for  the $p$--forms is  mainly due to there
appearance in supergravity or string theory. Furthermore, the
$p$th antisymmetric gauge fields naturally couple to
$(p-1)$--dimensional extended objects. However, the $p$--forms are
known to be relatively rigid in the sense that there is a few
number of consistent interactions for them \cite{h}. Despite these
restrictions $p$--forms may present interesting symmetry
properties, as it is the case, for instance, with the  duality
transformations between  types IIA and  IIB string theories
\cite{A-B}. In this section we investigate a new possible symmetry
among $p$--forms.

Since the properties of spinors and $p-$forms depends on the
space-time dimension, 
from now on, we restrict to the case where $D=4n$. 
The other cases are  analogous.
We thus concentrate on  the case where the vacua are in the spinor
representations of the Lorentz algebra. We take two copies
$\mathbf{\Psi}_\pm, \mathbf{\Lambda}_\pm$ transforming with
$V_\pm$, and two copies of the vacuum in the spinor representation
$\Omega_\pm$, $\omega_\pm$. 
Since $\Psi_\pm$ transforms with \eqref{irred} we have

$$
{\mathbf \Psi_\pm}=\begin{pmatrix}
\Psi_1{}_\pm \\ \Psi_2{}_\mp \\ \Psi_3{}_\pm \end{pmatrix}, 
$$

\noi
with $\Psi_i{}_+$ (resp. $\Psi_i{}_-$)  left-handed (resp. right-handed)
spinors.  Thus for instance we have

$$
\Psi_+ \otimes \Omega_+ = 
\begin{pmatrix} \Psi_1{}_+ \otimes \Omega_+\\ 
\Psi_2{}_- \otimes \Omega_+ \\ \Psi_3{}_+ \otimes \Omega_+\end{pmatrix}. 
$$

\noi
Consequently,
from the decomposition of the product
of spinors  \eqref{spin-p2}
 one gets the four possible multiplets  (See Appendix \ref{pform} 
for notations)

\beqa\label{4-decomposition}
\begin{array}{rcl}
\Xi_{++}&=&\mathbf{\Psi_+} \otimes \Omega_+=\begin{pmatrix}
\Xi_1{}_{++} \cr
 \bar \Xi_2{}_{-+} \cr \Xi_3{}_{++}
\end{pmatrix} = \begin{pmatrix} A_{[0]}\oplus  A_{[2]}\oplus   \cdots\oplus
A_{[2n]_+} \cr
\tA_{[1]}\oplus\tA_{[3]}\oplus\cdots\oplus\tA_{[2n-1]} \cr
\ttA_{[0]} \oplus\ttA_{[2]} \oplus\cdots \oplus\ttA_{[2n]_+} \end{pmatrix},\\
\noalign{\vskip3pt} \Xi_{--}&=&\mathbf{\Psi_-} \otimes
\Omega_- = \begin{pmatrix} \bar \Xi_1{}_{--} \cr
 \Xi_2{}_{+-} \cr \bar
\Xi_3{}_{--}\end{pmatrix}=\begin{pmatrix} A'_{[0]}\oplus A'_{[2]} \oplus\cdots
\oplus A'_{[2n]_-} \cr
\tA'_{[1]}\oplus\tA'_{[3]}\oplus\cdots\oplus\tA'_{[2n-1]} \cr
\ttA'_{[0]} \oplus\ttA'_{[2]} \oplus\cdots
\oplus\ttA'_{[2n]_-}\end{pmatrix},\\
\noalign{\vskip3pt} \Xi_{-+}&=&\mathbf{\Lambda_-} \otimes
\omega_+ = \begin{pmatrix} \xi_1{}_{-+} \cr
 \bar \xi_2{}_{++} \cr \xi_3{}_{-+}
\end{pmatrix} =
\begin{pmatrix}
A_{[1]} \oplus A_{[3]} \oplus\cdots\oplus A_{[2n-1]} \cr
\tA_{[0]}\oplus\tA_{[2]}\oplus\cdots,\tA_{[2n]_+} \cr
\ttA_{[1]}\oplus \ttA_{[3]} \oplus\cdots \oplus\ttA_{[2n-1]}\end{pmatrix},\\
\noalign{\vskip3pt} \Xi_{+-}&=&\mathbf{\Lambda_+} \otimes
\omega_- = \begin{pmatrix} \bar \xi_1{}_{+-} \cr
 \xi_2{}_{--} \cr \bar
\xi_3{}_{ +-} \end{pmatrix} = \begin{pmatrix} A'_{[1]}\oplus A'_{[3]}
\oplus\cdots\oplus A'_{[2n-1]} \cr
\tA'_{[0]}\oplus\tA'_{[2]}\oplus\cdots\oplus\tA'_{[2n]_-} \cr
\ttA'_{[1]}\oplus \ttA'_{[3]} \oplus\cdots \oplus\ttA'_{[2n-1]}\end{pmatrix}.
\end{array}\eeqa 

\noi
From the general property of representations of Lie algebras of order three
\eqref{module}, one can assume that, for instance, 
for the multiplet $\Xi_{++}$, the fields $A$
are in the $(-1)-$graded sector, $\tA$ in the $0-$graded sector
and $\ttA$ in the $1-$graded sector. The same classification
also holds for the other multiplets
Since anti-self dual $2n-$forms are complex (see \eqref{self}), we take all 
the $p-$forms above as being complex. But in $4n-$dimensional space-time
a left-handed spinor is complex conjugate to a right-handed spinor,
this gives us the opportunity to
take the multiplets above as being conjugated $\Xi_{++}^*=\Xi_{--}$
and $\Xi_{+-}^*=\Xi_{-+}$. This gives for the multiplets $\Xi_{++}$
and $\Xi_{--}$

 \beqa
\label{cc} 
\begin{array}{ll}
A_{[2p]}^\star= A^\prime_{[2p]}\,,\quad&
A_{[2n]_+}^\star=A^\prime_{[2n]_-}\,,\\
{\tilde {\tilde A}}_{[2p]}^\star= {\tilde {\tilde
A}}^\prime_{[2p]}\,,\quad&
{\tilde {\tilde A}}_{[2n]_+}^\star={\tilde {\tilde A}}^\prime_{[2n]_-}\,,\\
\tilde A^\star_{[2p+1]}=\tA^\prime_{[2p+1]}.& 
\end{array}
 \eeqa

\noi
We now calculate the transformation law of the different fields.
Since the results are similar for all the multiplets \eqref{4-decomposition},
we only give the results for
 the multiplet $\Xi_{++}$. From the transformation law
$\delta_\e \Xi_{++}= (\e^\mu V^+_\mu \Psi_+) \otimes \Omega_+$,
we obtain

 \beqa
\label{transfo-spin}
\delta_\e \Xi_1{}_{++} &=& \e^\mu \Sigma_\mu \bar \Xi_2{}_{-+}\,,\nonumber \\
\delta_\e \bar \Xi_2{}_{-+} &=& \e^\mu \tilde \Sigma_\mu \Xi_3{}_{++}\,,\\
\delta_\e \Xi_3{}_{++} &=& \e^\mu \partial_\mu
\Xi_1{}_{++}\,.\nonumber \eeqa

\noi 
To calculate \eqref{transfo-spin}, we use the identity
\eqref{prodGamma}. 
We proceed differently for $\Xi_1{}_{++}$ and
$\bar \Xi_2{}_{-+}$, in order to avoid the presence of $\Sigma^{(\ell)}$
or $\tilde \Sigma^{(\ell)}$  with $\ell >2n$.
Starting from  $\delta_\e \bar \Xi_2{}_{-+} = \e^\mu \tilde \Sigma_\mu
\Xi_3{}_{++}$, and using (\ref{p-spin2}) we first have $\delta_\e
\tA_{[2p+1]}=  \e^\mu 2^{-n} \mathrm{Tr} \left(
\tilde \Sigma_\mu \Xi_3{}_{++} {\cal C}_+\Sigma^{(2p+1)}\right)=\e^\mu 2^{-n}
\mathrm{Tr} \left(\Sigma^{(2p+1)}
\tilde \Sigma_\mu \Xi_3{}_{++} {\cal C}_+\right)
$. 
Using (\ref{prodGamma}),
 we calculate $ \Sigma^{(2p+1)} \tilde \Sigma_\mu$. 
Then, from the trace
identities for the $\Sigma$ matrices \eqref{Tr2}, one
obtains the transformations laws for $\tA_{[2p+1]}$.
In order to calculate  $\delta_\e \Xi_1{}_{++} = \e^\mu  \Sigma_\mu
\bar \Xi_2{}_{-+}$, we proceed in the reverse order. Firstly,
using (\ref{spin-p2}) and the identity (\ref{prodGamma}), we
compute the product $\Sigma_\mu \bar \Xi_2{}_{-+}$. Then, using the
trace formulae, we  get the transformation laws of $A_{[2p]}$.
The last case $\delta_\e \Xi_3{}_{++}$ is not difficult to handle.
For instance, this procedure gives for $\delta \Xi_{1++}$ (when $k <n$)

\beqa
\delta A_{[2k]}{}^{\nu_1\cdots \nu_{2k}}&=& \frac{1}{2^n} \text{Tr}\left(
\sum \limits_{p=0}^{n} \frac{1}{(2p+1)!}\left[
\tilde \Sigma^{\nu_1\cdots \nu_{2k}}  \Sigma_{\mu \mu_{2p+1} \cdots \mu_1}
+ (2p+1) \eta_{\mu \mu_{2p+1}}  
\tilde \Sigma^{\nu_1\cdots \nu_{2k}}  \Sigma_{ \mu_{2p} \cdots \mu_1}
\right] \right)\nonumber \\
&\times&\e^\mu \tA_{[2p +1]}{}^{\mu_1\cdots \mu_{2p+1}}.\nonumber
\eeqa

\noi Using the trace formul\ae \ \eqref{Tr2}, 
the first term gives $\tA_{[2k-1]}\wedge \e$
and the second $i_\e\tA_{[2k+1]}$.
Finally we get \cite{p-form}

\beqa 
\label{transfo-form} 
&&\begin{array}{llllll}
\delta_\e A_{[0]}&=& i_\e {\tilde A}_{[1]}
\nonumber \\
& &&\delta_\e {\tilde
A}_{[1]}&=& i_\e {\tilde {\tilde A}}_{[2]} + \ttA_{[0]} \wedge \e
\cr &\vdots && &\vdots \cr 
\delta_\e A_{[2p]}&=& i_\e {\tilde
A}_{[2p+1]} + {\tilde A}_{[2p-1]} \wedge \e & \nonumber \\
&&&\delta_\e {\tilde
A}_{[2p+1]}&=& i_\e {\tilde {\tilde A}}_{[2p+2]} + {\tilde {\tilde
A}}_{[2p]} \wedge \e  \cr &   \vdots && &  \vdots \cr 
 \\
 &&&
\delta_\e {\tilde A}_{[2n-1]}&=& i_\e  {\tilde {\tilde
A}}_{[2n]_+}+ {\tilde {\tilde A}}_{[2n-2]} \wedge \e
 \cr
\delta_\e
A_{[2n]_+}&=&  {\tilde A}_{[2n-1]} \wedge \e -i
 \ {}^\star \hskip -.1truecm
\left
(\tilde A_{[2n-1]} \wedge \e\right) \nonumber \\
\end{array} \nonumber \\
\\
&&\ \ \delta_\e {\tilde {\tilde A}}_{[0]}= \e.\partial A_{[0]}, \ \ \ \ \ \ldots
 \ \ \ \ \
\delta_\e {\tilde {\tilde A}}_{[2n]_+}= \e.\partial A_{[2n]_+}, \nonumber
\eeqa

\noi
the second term in $\delta_\e A_{[2n]}{}_+$ ensure its self-duality.
For the definition and conventions see \eqref{iner2} and \eqref{ext2}.
It is interesting to observe that the transformation \eqref{transfo-form}
have a geometrical interpretation in terms of the  natural operations
on $p-$forms. 
\subsubsection{Invariant Lagrangian}

The transformations (\ref{transfo-form}) suggest that an invariant
Lagrangian should be obtained
coupling the field $A$ with the fields
$\ttA$ and the field $\tA$ with themselves. In other words
to couple $(-1)-$graded sector with $1-$graded sector, and grade $0-$graded
sector with itself.
 Furthermore, if we consider for example the
$\Xi_{++}$ multiplet, in order to have a real Lagrangian, one has
also to take into consideration the conjugate multiplet $\Xi_{--}$
(see (\ref{cc})). For the $\Xi_{++}$ and $\Xi_{--}$ multiplets,
the Lagrangian writes \cite{p-form} 

\beqa\label{lag} {\cal L}&=&{\cal
L}(\Xi_{++}) + {\cal L}(\Xi_{--}) = {\cal L}_{[0]} + \ldots +
{\cal L}_{[2n]}
 + {\cal L}'_{[0]} \ \ + \ldots + {\cal L}'_{[2n]}=\nonumber\\
&=&d A_{[0]}d {\tilde {\tilde A}}_{[0]}
+ \ldots + \nonumber \\
&&- \frac12\frac{1}{(2p+2)!}\,d\tilde A_{[2p+1]}\,d\tilde
A_{[2p+1]} -\frac12  \frac{1}{(2p)!}\,d^\dag \tilde A_{[2p+1]}
\,d^\dag \tilde A_{[2p+1]} \nonumber \\
&&+\frac{1}{(2p+3)!}\,d A_{[2p+2]}\,d{\tilde {\tilde A}}_{[2p +
2]} + \frac{1}{(2p+1)!}\,d^\dag A_{[2p+2]}\,d^\dag {\tilde {
\tilde A}}_{[2p +2]} \nonumber  \\
&&+ \ldots +   \nonumber \\
&&+\frac12 \frac{1}{(2n+1)!}\,d A_{[2n]_+}\,d{\tilde {\tilde
A}}_{[2n]_+} +\frac12 \frac{1}{(2n-1)!}\,d^\dag
A_{[2n]_+}\,d^\dag {\tilde { \tilde A}}_{[2n]_+} \nonumber \\
&&+d A'_{[0]}\,d{\tilde{\tilde A}}'_{[0]}
+ \ldots +\\
&&-\frac12\frac{1}{(2p+2)!}\,d\tilde A'_{[2p+1]}\,d\tilde
A'_{[2p+1]} - \frac12  \frac{1}{(2p)!}\,d^\dag \tilde A'_{[2p+1]}
\,d^\dag \tilde A'_{[2p+1]} \nonumber \\
&&+ \frac{1}{(2p+3)!}\,d A'_{[2p+2]}\,d{\tilde {\tilde A}}'_{[2p
+ 2]} + \frac{1}{(2p+1)!}\,d^\dag A'_{[2p+2]}\,d^\dag {\tilde {
\tilde A}}'_{[2p +2]} \nonumber  \\
&&+ \ldots +   \nonumber \\
&&+\frac12 \frac{1}{(2n+1)!}\,d A'_{[2n]_-}\,d{\tilde {\tilde
A}}'_{[2n]_-} + \frac12 \frac{1}{(2n-1)!}\,d^\dag A'_{[2n]_-}
\,d^\dag {\tilde { \tilde A}}'_{[2n]_-}\,.
 \nonumber
\eeqa

\noi
Here $\omega_{[p]} \omega'_{[p]}$ stands for
$\omega_{[p]}{}_{\mu_1\ldots \mu_p}$\ $\omega'_{[p]}{}^{\mu_1 \ldots
\mu_P}$ , where $\omega_{[p]}$ and $\omega'_{[p]}$ are two
$p$--forms.  For the definition of the exterior derivative $d$
and its  adjoint $d^\dag$ see \eqref{ext} and \eqref{extdag}.
To prove that (\ref{lag}) is invariant under (\ref{transfo-form}),
we firstly note that $\delta_\e {\cal L}(\Xi_{++})$ and
$\delta_\e{\cal L}(\Xi_{--})$ do not  mix. It is thus  sufficient
to check separately their  invariance, which we do here only for
${\cal L}(\Xi_{++})$ as an illustration. Starting from a specific
normalisation for ${\cal L}_{[0]}$, its variation fixes the
normalisation for ${\cal L}_{[1]}$. By a step-by-step process, the
normalisations for ${\cal L}_{[p]}, 0\le p \le 2n$ are also fixed.
At the very end, all the terms of $\delta_\e {\cal L}$ compensate
each other, up to a total derivative. 
The key observations in this compensation process, is to remark that we have

$$
(i_\e. A_{[p+1]}) A_{[p]} = A_{[p+1]}(A_{[p]}\wedge \e),
$$

\noi together with the fact that the Lagrangian can be rewritten in 
a Fermi-like form (see below \eqref{fermi}).
 This shows, the Lagrangian
(\ref{lag}) is invariant.

\smallskip
If one considers the terms involving the (anti--)self--dual
$2n$--form one can have   further simplifications. Indeed, for the
self--dual $2n$--forms we have
$$
\begin{array}{rcl}{\cal L}_{[2n]}&=&
\frac12 \frac{1}{(2n+1)!}\,d
A_{[2n]_+}\,d{\tilde {\tilde A}}_{[2n]_+} + \frac12
\frac{1}{(2n-1)!}\,d^\dag A_{[2n]_+}\,d^\dag {\tilde {\tilde
A}}_{[2n]_+}=\\
\noalign{\vskip4pt} &=& \frac{1}{(2n+1)!}\,d
A_{[2n]_+}\,d{\tilde {\tilde A}}_{[2n]_+} \end{array}
$$
because of the self--duality condition ${}^\star A_{[2n]_+}=i
A_{[2n]_+}$. A more interesting way of regrouping the  terms
involving the self--dual and the anti--self--dual $2n$--forms is to
introduce the real $2n-$forms

\beqa
\label{2n-real} A_1{}_{[2n]}=
\frac{1}{\sqrt{2}}\left(A_{[2n]_+} +A'_{[2n]_-}\right),
\ttA_1{}_{[2n]}= \frac{1}{\sqrt{2}}\left(\ttA_{[2n]_+}
+\ttA'_{[2n]_-}\right), 
\eeqa

\noi such that

\beqa
\label{2nreal}
{\cal L}_{[2n]} + {\cal L}'_{[2n]}&=&
 \frac{1}{(2n+1)!}d A_1{}_{[2n]} d \ttA_1{}_{[2n]} +
 \frac{1}{(2n-1)!}d^\dag A_1{}_{[2n]} d^\dag \ttA_1{}_{[2n]}
\eeqa

\noi
 since $A_{[2n]_+} A'_{[2n]_-} =0$ when $D=1+(4n-1)$, for a
self--dual and an anti--self--dual $2n$--form. The final real
$2n$--forms
 are neither self--dual nor 
anti--self--dual, which is in agreement with the  representation
theory of the Poincar\'e algebra. 
In the same manner, we introduce the real fields
 
\beqa
\label{real} 
\begin{array}{ll}
A_1{}_{[2p]}=\frac{1}{\sqrt{2}}\left(A_{[2p]}+A'_{[2p]}\right),\quad&
A_2{}_{[2p]}=\frac{i}{\sqrt{2}}\left(A_{[2p]}-A'_{[2p]}\right),\\
\noalign{\vskip2pt} \tA_1{}_{[2p+1]}=\frac{1}{\sqrt{2}}
\left(\tA_{[2p+1]} + \tA'_{[2p+1]}\right),\quad&
\tA_2{}_{[2p+1]}=\frac{i} {\sqrt{2}}\left(\tA_{[2p+1]} -
\tA'_{[2p+1]}\right), \\
\noalign{\vskip2pt}
\ttA_1{}_{[2p]}=\frac{1}{\sqrt{2}}\left(\ttA_{[2p]} +
\ttA'_{[2p]}\right),\quad&\ttA_2{}_{[2p]}= \frac{i}{\sqrt{2}}
\left(\ttA_{[2p]} - \ttA'_{[2p]}\right). 
\end{array}
 \eeqa

\noi
With this new fields the Lagrangian is not diagonal, introducing the
fields 

\beqa
 \label{field-diag} 
\begin{array}{lll}\hA_1{}_{[2p]}
=\frac{1}{\sqrt{2}}\left(A_1{}_{[2p]} +
\ttA_1{}_{[2p]}\right),\quad& \hhA_1{}_{[2p]}=
\frac{1}{\sqrt{2}}\left(A_1{}_{[2p]} -
\ttA_1{}_{[2p]}\right),\quad& p=0,\ldots, n\,,\\
\hA_2{}_{[2p]}= \frac{1}{\sqrt{2}}\left(A_2{}_{[2p]} +
\ttA_2{}_{[2p]}\right),\quad& \hhA_2{}_{[2p]}=
\frac{1}{\sqrt{2}}\left(A_2{}_{[2p]} -
\ttA_2{}_{[2p]}\right),\quad& p=0,\ldots, n-1\,,
\end{array}
\eeqa

\noi
we have ($2p \ne 2n$ since for ${\cal L}_{2n}$ we have only one term)

\beqa
\label{p-real}
{\cal L}_{2p+1} + {\cal L}'_{2p+1}=&
-&\frac{1}{2} \frac{1}{(2p+2)!}d \tA_1{}_{[2p+1]} d \tA_1{}_{[2p+1]} 
-\frac{1}{2} \frac{1}{(2p)!}d^\dag \tA_1{}_{[2p+1]} d^\dag \tA_1{}_{[2p+1]}
\nonumber \\
&+&
\frac{1}{2} \frac{1}{(2p+2)!}d \tA_2{}_{[2p+1]} d \tA_2{}_{[2p+1]} 
+\frac{1}{2} \frac{1}{(2p)!}d^\dag \tA_2{}_{[2p+1]} d^\dag \tA_2{}_{[2p+1]}
\nonumber \\
{\cal L}_{2p} + {\cal L}'_{2p}=&
&\frac{1}{2} \frac{1}{(2p+1)!}d \hA_1{}_{[2p]} d \hA_1{}_{[2p]} 
+\frac{1}{2} \frac{1}{(2p-1)!}d^\dag \hA_1{}_{[2p]} d^\dag \hA_1{}_{[2p]}
 \\
&-&
\frac{1}{2} \frac{1}{(2p+1)!}d \hhA_1{}_{[2p]} d \hhA_1{}_{[2p]} 
-\frac{1}{2} \frac{1}{(2p-1)!}d^\dag \hhA_1{}_{[2p]} d^\dag \hhA_1{}_{[2p]}
\nonumber \\
&-
&\frac{1}{2} \frac{1}{(2p+1)!}d \hA_2{}_{[2p]} d \hA_2{}_{[2p]} 
-\frac{1}{2} \frac{1}{(2p-1)!}d^\dag \hA_2{}_{[2p]} d^\dag \hA_2{}_{[2p]}
\nonumber \\
&+&
\frac{1}{2} \frac{1}{(2p+1)!}d \hhA_2{}_{[2p]} d \hhA_2{}_{[2p]} 
+\frac{1}{2} \frac{1}{(2p-1)!}d^\dag \hhA_2{}_{[2p]} d^\dag \hhA_2{}_{[2p]}
\nonumber
\eeqa

Usually, the kinetic term for
a $p$--form  $\omega_{[p]}$ writes, with our conventions for the
metric, $(-1)^p
\frac{1}{2(p+1)!}\,d\omega_{[p]}\,d\omega_{[p]}$.
However, in the Lagrangian written as sum of terms above \eqref{2nreal}
and \eqref{p-real} some of the kinetic terms have the wrong sign.
This implies that these fields have an energy
density not bounded from bellow.
We propose a possible way to construct
a Lagrangian with correct signs for the various kinetic terms,
based on a special choice for the physical fields. 
The main idea is  related to Hodge duality.
However, the duality transformation will act here on the $p-$forms 
with respect to the Lorentz group $SO(1,D-1)$. This should be contrasted with
the case of 
the usual duality transformations (generalising the electric-magnetic duality)
which act on the field strengths with respect to  $SO(1,D-1)$, 
or equivalently on the potentials themselves but with respect
to the little group $SO(D-2)$. 
Firstly, the decomposition \eqref{bi-spin2} is purely conventional, and
we could also have decomposed this product on the set of $p-$forms with 
$p \ge2n$. 
Moreover, looking at the Lagrangians \eqref{2nreal} and \eqref{p-real}
there is a kind of duality between the kinetic term and the gauge fixing term. 
This means that for the field with the wrong signs,  we define the
Hodge dual

\beqa
\label{dual-field}
\begin{array}{rcl} 
\hhA_1{}_{[2p]} &\to&
\hhB_1{}_{[4n-2p]} = {}^\star \hhA_1{}_{[2p]}\,,\\
\noalign{\vskip2pt}
\hA_2{}_{[2p]} &\to& \hB_2{}_{[4n-2p]} =  {}^\star \hA_2{}_{[2p]}\,,\\
\noalign{\vskip2pt}
\tA_2{}_{[2p+1]} &\to& \tB_2{}_{[4n-2p-1]} =
{}^\star \tA_2{}_{[2p+1]}\,,\\
\noalign{\vskip2pt} \hhA_1{}_{[2n]} &\to& \hhB_1{}_{[D-2n]}=
{}^\star \hhA_1{}_{[2n]}, 
\end{array}
\eeqa

\noi
 with $p=0,\ldots, n-1$.
Thus, starting from the Lagrangian (\ref{lag}) and performing the
field redefinitions (\ref{real}), (\ref{field-diag}) and
(\ref{dual-field})  we get a new Lagrangian with the correct signs
(that we do not write but is easy to obtain),
because of relation \eqref{dd}. 
This Lagrangian is given in \cite{p-form}.
In this transformation the kinetic term of $A$
becomes the gauge fixing term (see below Section \ref{gauge-fixe})
of $B$ and {\it vice versa}.
Using \eqref{dual-prod}, one can obtain the  transformations
law  of the new fields. The field content is then
  one one--form, one  three--form, $\ldots$, one
$(4n-1)$--form and two zero--forms, two
two--forms, $\ldots$ and two $4n$--form, 
 all the $p$--forms have a kinetic term and a
gauge fixing term; the only exceptions are the zero--forms, which
have only  kinetic terms, and the $4n$--forms, which have only
gauge fixing terms.
 Let us
emphasise that  the substitutions (\ref{dual-field}) are done with
respect to the gauge fields. This is quite different from the
usual duality transformations (generalising the electric--magnetic
duality) where the duality transformations are done with respect
to the field strengths. 

With the ``duality'' transformations
(\ref{dual-field}) the number of degree of freedom is not the same
for $\omega_{[p]}$ and $\rho_{[D-p]} = {}^\star \omega_{[p]}$,
which is not the case for the usual duality transformations.
Hence, the  two Lagrangians    describe inequivalent  theories.\\

Note finally that the sum of the kinetic terms and the gauge fixing terms
gives rises to  Fermi--like terms. For instance

\beqa
\label{fermi}
\frac12 \frac{1}{(p+1)!} d A_{[p]}  d A_{[p]} + 
\frac12 \frac{1}{(p-1)!} d^\dag A_{[p]}  d^\dag A_{[p]}=
\frac12 \frac{1}{p!} \partial_\mu A_{[p]}  \partial^\mu A_{[p]}, 
\eeqa

\noi and note, in particular that it is more easy to check the invariance
of the Lagrangian \eqref{lag} when it is written as a sum of Fermi-like
terms.

\subsubsection{Gauge invariance}
\label{gauge-fixe}

Let us analyse gauge invariance.
Gauge transformations for $p-$forms, are given by

\beqa
\label{gauge}
A_{[p]} \to A_{[p]} + d \chi_{[p-1]}, \ p >0,
\eeqa

\noi
where $\chi_{[p-1]}$ is a $(p-1)$-form.
We observe that \eqref{lag}  (or the new Lagrangian that
we do not have given) is not invariant.
Indeed, the terms $d^\dag A_{[p]} d^\dag A_{[p]} $ in the Lagrangian
\eqref{lag} fixes partially the gauge, and the Lagrangian is invariant
provided $\chi_{[p-1]}$ satisfies an additional constraint

\beqa
\label{contr}
d^\dag d \chi_{[p-1]}=0.
\eeqa

\noi
This means that this last term is a gauge fixing term, analogous
to the Feynman gauge fixing term for the electromagnetic field.
The peculiar form of our Lagrangian leads also to another gauge invariance.
Since it contains a kinetic term and a gauge fixing term, there is 
some kind of duality between these two terms. It was in fact this 
duality which enabled us to make the field redefinition 
\eqref{dual-field}. At the level of gauge invariance, this duality translates
in the invariance of the Lagrangian under the transformation

\beqa
\label{gauge'}
A_{[p]} \to A_{[p]} + d^\dag  \chi_{[p+1]}, \ p <4n,
\eeqa

\noi
such that $d d^\dag \chi_{[p+1]}=0.$ However, from the Poincar\'e theorem
(since we are in $\mathbb R^{4n}$ there is no topological obstruction),
$d d^\dag \chi_{[p+1]}=0$ implies that there exists a $(p-1)-$form
$\lambda_{[p-1]}$ such that $d^\dag \chi_{[p+1]}= d \lambda_{[p-1]}$.
Thus \eqref{gauge'} is not a new symmetry.
 While in the case of  gauge theories the gauge invariance guarantees that the physical (on-shell) 
quantities are gauge fixing independent, in our case $(d^\dag A_{[p]})^2$
cannot be traded for any other gauge-fixing function since it is imposed by 
the invariance under \eqref{transfo-form},  and is thus expected to affect the physical degrees of 
freedom. This shows that the effective degrees of freedom of $A_{[p]}$ are dictated
only by  the gauge freedom eq.(\ref{gauge}), supplemented by
 $d^\dag d \chi_{[p-1]} =0$. An immediate consequence of the latter
constraint is that the usual Lorentz condition $d^\dag A_{[p]}=0$
cannot be imposed in general to eliminate the unphysical components. 
This means that the way one should eliminate the unphysical components cannot
be handled in a usual manner \cite{ghosts}. On top of that, such a 
condition is not stable under our transformation laws 
(\ref{transfo-form}). For instance,
if we put $\partial_\mu  A_{[1]}^\mu=0$, then 
$\partial_\mu \delta_\e A_{[1]}^\mu=0$ gives 
$\e^\mu \partial_\mu A_{[0]}=0$, which is obviously too strong.

The gauge invariance (\ref{gauge}) and the field equations imply
(for a $p$--form $A_{[p]}$, with $p \le D-2$) $P^\mu
A_{[p]}{}_{\mu \mu_2\cdots \mu_p}=0$ and $P^2=0$ (with $P^\mu$ the
momentum), thus $A_{[p]}$ gives rise to a massless state in the
$p$--order antisymmetric representation of the little group
$SO(4n-2)$. But, in our decomposition, there are  also appearing
$p-$forms with $p=4n-1,\,4n$. Of course these $p$--forms do not
propagate. 
It is interesting to note that
similar phenomena  are well-known  
 in the context of type IIA, IIB string theory
\cite{pol} in $10$ space-time dimensions where $9-$ and $10-$forms appear. 
Actually, subsequent to the early works on  two-forms in
\cite{cs, nambu}, several authors  studied the classical and
quantum properties of the non-propagating $3$- and $4$-forms 
\cite{3f,4f} in four dimensions. 
In particular,  it was pointed out in \cite{4f} that the gauge
fixing term for a  $4$-form takes the form of a kinetic term for
a scalar field, in exact analogy with our results. The gauge condition
for $p-$forms is crucial in order to eliminate the unphysical degrees
of freedom such that
massless $p$--form ($p<D-1$) has $\begin{pmatrix}p\cr D-1\end{pmatrix}$
degrees of freedom off--shell and  $\begin{pmatrix}p\cr D-2\end{pmatrix}$ 
on--shell.
In our case, relation \eqref{contr} is not enough to
eliminate the unphysical degrees of freedom. 
 However, here the situation
is not as simple, since  in our case there is some mixing between
gauge invariance and  \eqref{transfo-form}. This means
that  \eqref{transfo-form} should itself have some role in the elimination
of the superfluous degree of freedom. For a discussion of the compatibility
of the transformation \eqref{transfo-form} and gauge transformation
see \cite{cubic2}.\\

The Lagrangian constructed so far is a free theory. It should be
interesting to construct interacting theories invariant under
\eqref{transfo-form}. It has been proved by a brute force method, that
when $D=4$ and we consider only multiplets of the types 
\eqref{4-decomposition}, no
interacting terms are allowed \cite{cubic2}. In the general case,
it is still an open question to know  whether or not
 interacting terms
compatible with \eqref{transfo-form} exist. \\

Let us mention to conclude this section that invariant Lagrangians
including mass terms were also constructed \cite{p-form}.

\section{Conclusion}
There are many 
mathematical structures generalising Lie superalgebras which can be defined.
In particular one of this structure called
Lie algebras of order $F$ can be used to implement  higher
order extensions of the Poincar\'e algebra.
In these lectures we have studied explicitly two
examples of non-trivial higher order extensions of the Poincar\'e algebra
based on Lie algebras of order $F$,
which are not the supersymmetric ones. 
In the former, we have constructed higher order symmetries
in three space-time dimensions which act on relativistic anyons.
In the latter, cubic extensions of the Poincar\'e algebra  in any
space-time dimensions are obtained. These new symmetries
have a natural geometrical interpretation on generalised
 gauge fields or $p-$forms. Invariant Lagrangians in this last cases have
revealed some problems to be solved: construction of interacting Lagrangian,
relation between the cubic symmetries and gauge invariance.

In order to understand deeper these structures  a program to study 
Lie algebras of order $F$ on a formal way has been investigated.
The basis of the theory of contractions and deformations in the  
 Gerstenhaber sense together with a classification of Lie algebras
associated to $\mathfrak{sl}(2)$ and the four dimensional-Poincar\'e
algebra have been initiated \cite{flie3}. A classification of 
kinematical algebras of order three have been investigated in \cite{cr}.
Since Lie algebras of order three correspond to transformations at the 
infinitesimal level, group associated to ternary algebras 
has been defined in the context of Hopf algebras,
and the parameters of the transformation have been identified \cite{hopf}.
However some  points remains to be studied.
Simple Lie algebras of order three have been defined in \cite{flie2},
however there is no general
classification of simple complex Lie algebras of order $F$ (and
even of order three) analogous to
 the classification of the Lie (super)algebras.
It should be also interesting to have a analogous of a Coleman-Mandula theorem
in that context. Finally, in order to construct interacting Lagrangians
some adapted superspace will  certainly be relevant.

\section*{Acknowledgements}
I would like to acknowledge the organising committee, and in particular
R. Campoamor-Stursberg to offer  me the opportunity to give these lectures,
and for the kind invitation to Madrid.
 
\appendix

\section{Convention and useful identities for spinor calculus
in four dimensions}
\label{conventions}
In this appendix, we collect useful relations and conventions for
the four dimensional spinor calculus.
The metric is taken to be 

\beqa
\eta_{\mu  \nu} = \mathrm{diag}(1,-1,-1,-1)
\eeqa

\noindent
In the $\mathfrak{so}(1,2)=\mathfrak{sl}(2,\mathbb C)$ notations of dotted and
undotted indices for two-dimensional spinors, the spinor
conventions to raise/lower indices are as follows (we have minor differences
as compared to the notations of Wess and Bagger \cite{wb}):
$\psi_\alpha =\varepsilon_{\alpha\beta}\psi^\beta$,
$\psi^\alpha =\varepsilon^{\alpha\beta}\psi_\beta$,
$\bar\psi_{\dot\alpha}=\varepsilon_{\dot\alpha
\dot\beta}\bar\psi^{\dot\beta}$, $\bar\psi^{\dot\alpha}
=\varepsilon^{\dot\alpha\dot\beta}\bar\psi_{\dot\beta}$
with $(\psi_\alpha)^* =\bar\psi_{\dot\alpha}$, 
$\varepsilon_{12} = \varepsilon_{\dot 1\dot 2}=-1$,
$\varepsilon^{12} = \varepsilon^{\dot 1\dot 2}=1$.

\noindent
The $4D$
Dirac matrices,  in the Weyl representation,  are 
\begin{eqnarray}
\label{eq:gamma}
\Gamma_\mu =
 \begin{pmatrix}
 0&\sigma_\mu\\
 \bar\sigma_\mu&0
 \end{pmatrix},
\end{eqnarray}

\noindent
with 

\beqa
\label{eq:spin_rep}
\sigma_\mu=(1,\sigma_i), \bar \sigma_\mu =
(1,-\sigma_i),
\eeqa

\noi
 the $\sigma_i$ ($i=1,2,3$) denoting the Pauli 
matrices. The index structure of the $\sigma_\mu-$matrices is as
follows: $\sigma_\mu \to \sigma_\mu{}_{\alpha \dot \alpha},\; \bar
\sigma_\mu \to \bar \sigma_\mu{}^{\dot \alpha  \alpha}$.
The following relation holds,

\beqa
\label{sigma-sigmab}
\bar\sigma_\mu{}^{\dot\alpha\alpha}=
\sigma_{\mu\,\beta \dot\beta} \varepsilon^{\alpha \beta} 
\varepsilon^{\dot \alpha \dot \beta},
\eeqa

\noi
and \eqref{eq:spin_rep} leads to

\beqa
\label{trace}
\text{Tr}(\sigma^\mu \bar \sigma^\nu)= 2 \eta^{\mu \nu}.
\eeqa

\noindent
Furthermore, the Lorentz generators for the spinor representations are given 
by

\beqa
\label{spin-rep}
\sigma_{\mu \nu }{}_\alpha{}^\beta = \frac{1}{4}\left(
\sigma_\mu{}_{\alpha \dot \alpha} \bar \sigma_\nu{}^{\dot \alpha \beta}-
\sigma_\nu{}_{\alpha \dot \alpha} \bar \sigma_\mu{}^{\dot \alpha \beta}
\right) \nonumber \\
\bar \sigma_{\mu \nu}{}^{\dot \alpha}{}_{\dot \beta} = \frac{1}{4}\left(
\bar \sigma_\mu{}^{\dot \alpha  \alpha}  \sigma_\nu{}_{ \alpha \dot \beta}-
\bar \sigma_\nu{}^{\dot \alpha  \alpha}  \sigma_\mu{}_{ \alpha \dot \beta}
\right)
\eeqa

\noindent
We adopt the usual spinor summation convention

\beqa
\label{spinsum}
\psi. \lambda= \psi^\alpha \lambda_\alpha = - \psi_\alpha \lambda^\alpha, \ \
\bar \psi. \bar \lambda = \bar \psi_{\dot \alpha} \bar \lambda^{\dot \alpha}.
\eeqa

\noi
Since $\psi_{\alpha}^\dag = \bar \psi_{\dot \alpha}$ and 
$\sigma^\dag{}^\mu = \sigma^\mu$, we have

\beqa
\label{conj1}
(\psi. \lambda)^\dag = \bar \lambda. \bar \psi,\\
\label{conj2}
\left(\psi \sigma^\mu \bar \lambda\right)^\dag= \lambda \sigma^\mu \bar \psi.
\eeqa

\noi
For anticommuting Grassmann spinors we have

\beqa
\label{spin1}
\theta^\alpha \theta^\beta &=&- \frac12 \e^{\alpha \beta} \theta. \theta, \\
\label{spin2}
\bar \theta^{\dot \alpha}  \bar \theta^{\dot \beta} &=& \frac12 
\bar \e^{\dot \alpha \dot \beta} \bar \theta. \bar \theta, \\
\label{spin3}
(\theta \sigma^\mu \bar \theta) \ (\theta \sigma^\nu \bar \theta)&=&
\frac12 \theta. \theta \ \bar \theta. \bar \theta \ \eta^{\mu \nu},\\
\label{spin4}
(\theta \sigma^\mu \bar \theta) \ \theta. \lambda&=&
-\frac12 \theta. \theta \ \lambda \sigma^\mu \bar \theta, \\
\label{spin5}
(\theta \sigma^\mu \bar \theta) \ \bar \theta. \bar \lambda&=&
-\frac12 \bar \theta. \bar \theta \ \theta \sigma^\mu \bar \lambda.
\eeqa

\section{Relativistic wave equations for anyons}
\label{plu}
We give now a relativistic wave equation  for anyons.
Following M. Plyushchay it is  based on the  $R-$deformed Heisenberg algebra 
with reflection. This algebra  is defined by generators and relations.
Consider the generators $a^{\pm},{\cal R}$ that satisfy
\cite{rdha}

\beqa
\label{rdha}
[a^-,a^+]=1 + \nu {\cal R}, \ \ 
\left\{R,a^\pm\right\}=0, \ \ 
{\cal R}^2=1,
\eeqa

\noi with $\nu \in \mathbb R$ the deformation parameter.
The operators $a^\pm$ together with the quadratic operators
 generate the
superalgebra $\mathfrak{osp}(1|2)$ whose bosonic part is 
$\mathfrak{sl}(2,\mathbb R) \cong \mathfrak{so}(1,2)$ and 
fermionic part the two-dimensional
(Majorana) spinor representation 

\beqa
\label{osp}
\left\{a^\pm,a^\pm\right\}=4 L_\pm, & \left\{a^+, a^-\right\}= 4 L_0, 
\nonumber \\
\left[L_\pm,a^\mp\right]=\mp a^\mp,& \left[L_0, a^\pm\right]=\pm\frac12 a^\pm,
  \\
\left[L_+,L_-\right]=-2L_0, & \left[L_0,L_\pm\right]=-L_\pm. \nonumber
\eeqa

\noi This is the generalisation of the well-known realisation of the algebra
$\mathfrak{sl}(2,\mathbb R) \cong \mathfrak{so}(1,2)$ by the usual harmonic oscillator. 
When $\nu >-1$, the algebra \eqref{rdha} admits an infinite dimensional 
unitary  representation  (when $\nu = -(2k+1)$ the algebra admits
finite dimensional representations) \cite{p2}

\beqa
\label{rep-rdha}
a^+ \left|n\right> &=& \sqrt{n+1 +\frac{\nu}{2}(1-(-1)^{n+1})} \ \ 
\left|n+1\right>,
\nonumber  \\
a^- \left|n\right> &=& \sqrt{n +\frac{\nu}{2}
(1-(-1)^n)} \ \ \left|n-1\right>, 
\eeqa

\noi 
this representation is
bounded from below since we have $a^- 
\left| 0 \right>=0$, unitary $(a^-{})^\dag=a^+$ and the
vectors $|n>$ are orthonormal $<n|m> = \delta_{mn}$.
We denote ${\cal R}_\nu = \left\{ \left| n \right>, \ n \in
\mathbb N\right\}$ this representation. We also have 
${\cal R} \left| n \right> = (-1)^n 
\left| n \right>$. Furthermore, because of relations \eqref{osp} the
representation of \eqref{rep-rdha} is an irreducible representation of 
$\mathfrak{osp}(1|2)$ but a reducible representation of 
$\mathfrak{sl}(2,\mathbb R)$.
Indeed, it decomposes on the direct sum of the two irreducible representations

\beqa
{\cal R}_\nu = {\cal D}^+_{\frac{1+\nu}{4}} \oplus {\cal D}^+_{\frac{3+\nu}{4}}
\eeqa

\noi
with

\beqa
 {\cal D}^+_{\frac{1+\nu}{4}} &=&  \frac12(1+{\cal R}) {\cal R}_\nu
= \Big\{ \left| 2 n \right>, \ n \in \mathbb N\Big\}, \\
{\cal D}^+_{\frac{3+\nu}{4}} &=&  \frac12(1-{\cal R}) {\cal R}_\nu
= \Big\{ \left| 2 n +1 \right>, \ n \in \mathbb N\Big\}. \nonumber
\eeqa

In order to obtain a relativistic wave equation for relativistic anyons, 
and to make contact with the literature \cite{p2}, in 
this appendix we are working with physical quantities. This means that
differently from the rest of the text we are directly working
with the  tri-momentum and 
 the angular momentum. In particular, this means that they are 
given by
hermitian quantities (although in the rest of the text they are antihermitian).
In order to avoid confusion, we denote here 
$\Pi_\mu = - iP_\mu, L_\mu=-iJ_\mu$, the tri-momentum and angular momentum.
In this basis the Lie algebra $\mathfrak{so}(1,2)$ writes 
$[L_\mu, L_\nu]= i \e_{\mu \nu \rho} \eta^{\rho \sigma} L_{\sigma}$ and
$L_\pm = L_1 \mp i L_2$. 
We also define $\Pi_\pm = \Pi_1 \mp i \Pi_2$.
The Dirac $\Gamma-$matrices in the Majorana
representation are taken to be
$\gamma^0=-i\sigma^2, \gamma^1=i \sigma^2, \gamma^2=-i\sigma^3$
with spinor matrix elements $\gamma^\mu{}_\alpha{}^\beta$, and the spinor
indices can be raised and lowed as in Appendix \ref{conventions}.
We also define the hermitian spinor operators 

$$
L_1=\frac{1}{\sqrt{2}}(a^+ + a^-), \ \ 
L_2=\frac{i}{\sqrt{2}}(a^+ - a^-).
$$ 

\noi
A direct calculation using \eqref{rdha} and \eqref{osp} gives

\beqa
[L_\alpha, L_\beta]&=& -i \epsilon_{\alpha \beta}(1+ \nu {\cal R}), \\
\{L_\alpha, L_\beta\}&=&-4i (L.\gamma)_{\alpha \beta}=
\begin{pmatrix}
L_0+L_1&L_2 \\ L_2&L_0-L_1
\end{pmatrix}.\nonumber
\eeqa

\noi
Finally we introduce the spinor operator

$$
D_\alpha= (\Pi.\gamma)_\alpha{}^\beta L_\beta + \e m L_\alpha,
$$

\noi
with $\e = \pm 1$, and a little algebra gives

\beqa
\label{DD-JD}
L^\alpha L_\alpha&=&-i(1+ \nu {\cal R}), \nonumber \\
D^\alpha D_\alpha&=& -i(\Pi^2 -m^2)(1+\nu {\cal R}),\\
L^\alpha D_\alpha&=&4i(\Pi.L -m \e \frac14(1+\nu {\cal R})).\nonumber
\eeqa

\noi
Now we have all the material and identities to define the spinor set of 
equations. Taking $\left| \psi\right> \in {\cal R}_\mu$ we have 

\beqa
\left|\psi\right>=\left|\psi_+\right>+\left|\psi_-\right>=
\sum\limits_{n=0}^{+ \infty} \psi_{2n}(x) \left| 2n \right> +
\sum\limits_{n=0}^{+ \infty} \psi_{2n+1}(x) \left| 2n+1 \right> 
\eeqa

\noi
and we assume the relativistic wave equations
\beqa
\label{eq-anyons}
D_\alpha \left|\psi\right>=0.
\eeqa

\noi
The covariance of \eqref{eq-anyons} under the $(1+2)-$dimensional Poincar\'e 
group is checked in \cite{p2}. Using the identities \eqref{DD-JD}
a direct calculation gives

\beqa
\label{casimir}
\begin{array}{lll}
(\Pi^2-m^2) \left| \psi_+\right>=0,& \ \ \ &
(\Pi.L-m\e \frac{1+\nu}{4})\left| \psi_+\right>=0,\\
(\Pi^2-m^2) \left| \psi_-\right>=0,&&
(\Pi.L-m\e \frac{1-\nu}{4})\left| \psi_-\right>=0.
\end{array}
\eeqa

\noi
This means that $\left|\psi_\pm\right>$ describe a particle of mass
$m$ and spin $\frac{1 \pm 1}{4}$. Now if we solve equation \eqref{eq-anyons}
in the rest frame we obtain that $\left|\psi_+\right> = \psi_0\left|0 \right>$,
which describes a anyons of mass $m$, energy $ \e m$ and helicity $\frac{1 +\nu}{4}$
although we have 
$\left|\psi_-\right>=0$. Solving the equation in any frame gives

\beqa
\left|\psi_+\right> = \sum \limits_{n=0}^{+ \infty}
\sqrt{\frac{\Gamma(2s+n)}{\Gamma(2s) \Gamma(m+1)}}
\left(\frac{\Pi_1+i\Pi_2}{\Pi_0 + \e m}\right)^n \psi_0 \left|2n\right>.
\eeqa

\noi
This is proven by induction. Assuming 

$$\psi_{2n}=\sqrt{\frac{\Gamma(2s+n)}{\Gamma(2s) \Gamma(m+1)}}
\left(\frac{\Pi_1+i\Pi_2}{\Pi_0 + \e m}\right)^n \psi_0,$$

\noi using 

$$
\frac{\Pi_0-\e m}{\Pi_1-i\Pi_2}= \frac{\Pi_1+i\Pi_2}{\Pi_0+\e m},
$$

\noi
gives the correct value for $\psi_{2n+2}$.
Thus equations \eqref{eq-anyons} describe a relativistic anyon of mass $m$,
helicity $\frac{1+\nu}{4}$ and energy
of sign $\e$. It has one degree of freedom as it should.

\section{Lie algebras of order $F$ associated with anyons}
\label{indec}

In this appendix we give an abstract and an explicit construction
to define a Lie algebra of order $F$ for relativistic anyons.
Recall that the vector representation may be obtained in the Verma module
language formalism \cite{kr}.	
Consider ${\cal U}(\mathfrak{so}(1,2))$ the universal
enveloping algebra. The  Poincar\'e-Birkhoff-Witt theorem
gives

$$
{\cal U}(\mathfrak{so}(1,2))= \Big\{ L_+^m L_-^n L_0^p, \
m,n,n \in \mathbb N \Big\}.
$$

\noi
Consider now  the two-sided ideal
generated by $L_-, L_0+{\bf I}$ (where 
${\bf I}$ denotes the identity of ${\cal U}(\mathfrak{so}(1,2))$) and set 

$$
{\cal V}_{-1} = {\cal U}(\mathfrak{so}(1,2))/I.
$$

\noi
We have, in ${\cal V}_{-1}$ 

$$
L_0.{\bf I}=- {\bf I}, \ \ L_- {\bf I}=0.
$$

\noi Thus ${\bf I}$ is the highest weight representation of ${\cal V}_{-1}$,
and we have 

$$
{\cal V}_{-1} = \Big\{L_+^n, n \in \mathbb N\Big\}.
$$

\noi
But as seen in \eqref{verma2}
($L_- L_+^3=0$)
 and  \eqref{diag}, $M_{-1}=\left<L_+^n, n >2\right>$ is an invariant
subspace of ${\cal V}_{-1}$, and the standard finite dimensional vector
representation is given by

$$
{\cal D}_{-1} = {\cal V}_{-1}/M_{-1}.
$$

\noi

In the same vain, we define the Verma module associated with the
representation  of spin $-1/F$. Recall that among the four
representations of Section \ref{discrete}, 
there are two inequivalent representations, one bounded from below and one 
bounded  from above. Since the construction works equally well on both
representations, in the discussion, we only consider
 the representation bounded
from below. In this language, we have

$$
{\cal D}_{-1/F}^+ = {\cal V}_{-1/F}^+ =
{\cal U}(\mathfrak{so}(1,2))/<L_-, L_0 + \frac{1}{F} {\bf I}>.
$$

\noi
In ${\cal D}_{-1/F}^+$ we have $L_0.{\bf I}=-1/F {\bf I}, \ \
L_- {\bf I}=0$, this means that in ${\cal S}^F({\cal D}^+_{-1/F})$,
we have ${\bf I} \otimes \cdots \otimes  {\bf I}$ ($F-$times) is such that
$L_- .{\bf I} \otimes \cdots \otimes  {\bf I}=0, \ \ 
L_0. {\bf I} \otimes \cdots \otimes  {\bf I}=- 
{\bf I} \otimes \cdots \otimes  {\bf I}$. Thus 
${\bf I} \otimes \cdots \otimes  {\bf I}$ is a primitive vector of
${\cal V}_{-1}$ meaning that there is an $\mathfrak{so}(1,2)-$equivariant
map

$$
i: {\cal V}_{-1} \hookrightarrow {\cal S}^F(\D_{-1/F}^+).
$$
\noi
Taking the coset by $M_{-1}$ we obtain an $\mathfrak{so}(1,2)-$equivariant
inclusion

$$
{\cal D}_{-1} \hookrightarrow {\cal S}^F({\cal D}^+_{-1/F})/i(M_{-1}).
$$

\noi
If we denote by ${\cal S}^F({\cal D}^+_{-1/F})_{\text{red}}= 
{\cal S}^F({\cal D}^+_{-1/F})/i(M_{-1})$ we have

$$
\Big(\mathfrak{so}(1,2) \oplus {\cal S}^F({\cal D}^+_{-1/F})_{\text{red}}\Big)
\oplus  {\cal S}^F({\cal D}^+_{-1/F}),
$$

\noi
is a Lie algebra of order three.. This can
be summarised in the following diagram

$$
\xymatrix{
V_{-1}\ar@<1ex>[r]^{ i \cong  } \ar@{>>}[d] &
\Big< {\mathbf I}^{\otimes^F}_{-\frac{1}{F}}\Big> \ar@{^{(}->}[r]  \ar@{>>}[d]&
{\cal S}^F(\D_{-\frac{1}{F}})  \ar@{>>}[d] \\
\D_{-1}=V_{-1}/M_{-1} \ar@<1ex>[r]^{ i \cong  }&
\Big< {\mathbf I}^{\otimes^F}_{-\frac{1}{F}}\Big>/i(M_{-1}) \ar@{^{(}->}[r]&
{\cal S}^F_{\text{red}}(\D_{-\frac{1}{F}}).
}
$$

\noi
This commutative diagram also shows that there is no mapping from
${\cal S}^F(D_{-1/F})$ into $\D_{-1}$ and thus no Lie algebras of
order $F$ associated with $(\mathfrak{so}(1,2) \oplus \D_{-1}) 
\oplus \D_{-1/F}$. A Lie of algebra of order $F$  can however be defined if
${\cal D}_{-1}$ is extended into an infinite dimensional representation.

To examine some properties of ${\cal S}^F({\cal D}^+_{-1/F})_{\text{red}}$
(indecomposability), and to clarify the abstract construction above, we
obtain the same result in an explicit way. It is well known that
$\mathfrak{so}(1,2)$ can be realised by differential operators

$$
L_-=x \partial_y, L_+=-y \partial_x, L_0=\frac12(y\partial_y - x \partial_x).
$$

\noi
Consider now ${\cal F}$ the vector space of functions on
$\mathbb R_+^{2 *} = \left\{ (x,y) \in \mathbb R^2, x,y>0\right\}$.
The following subspaces of ${\cal F}$

 \beqa
\label{eq:repp}
\D_{-n}&=&\Big<~ x^{2n},x^{2n-1}y,\cdots, x y^{2n-1}, y^{2n} \Big >,\hskip .5cm n \in \NN/2,
\nonumber \\
\D^+_{-\lambda}&=&\Big<~ x^{2 \lambda}  \left( {y \over x }\right)^m, 
m \in \NN \Big >,\hskip .5cm 
\lambda \in \RR \setminus \NN/2, \\
\D^-_{-\lambda}&=&\Big<~ y^{2 \lambda} 
\left( {x \over y }\right)^m, m \in \NN \Big >, \hskip .5cm
 \lambda \in \RR \setminus \NN/2 , \nonumber   
\eeqa

\noi
are representations of $\mathfrak{so}(1,2)$.
The  representation $\D_{-n}$ is the $(2n+1)-$dimensional  irreducible 
representation and the representations $\D^\pm_{-\lambda}$ are infinite
dimensional representations, bounded from below and above respectively.
It is important to emphasise that the representations given in (\ref{eq:repp})
do not have the normalisations  conventionally taken.

To define $S^F\big(\D^{+}_{-1/F}\big)_{{\mathrm {red}}}$, we consider
the multiplication map 
$m_F : {\cal F} \times \cdots \times {\cal F} \rightarrow 
{\cal F}$ given by 

\beq
\label{eq:multi}
 \ m_F(f_1,\cdots, f_F)= f_1 \cdots f_F
\eeq

\noindent
 which is multilinear and 
totally symmetric and hence induces a map $\mu_F$ from
${\cal S}^F({\cal F})$ into ${\cal F}$. Restricting to 
${\cal S}^F\left(D^\pm_{-1/F}\right)$ one sees that

\beqa
\label{eq:FSUSY}
S^F\big(\D^{+}_{-1/F}\big)_{\mathrm {red}} \buildrel{\hbox{def}} \over = 
\mu_F\Big(S^F\big( D^{+}_{-1/F}\big)\Big) 
&=&\Big<x^2 
\left({y \over x} \right)^m, ~ m\in \NN ~~\Big> \supset \D_{-1} \\
S^F\big(\D^{-}_{-1/F}\big)_{\mathrm{red}} \buildrel{\hbox{def}} \over = 
\mu_F\Big(S^F\big( D^{-}_{-1/F}\big)\Big) 
&=&\Big<y^2 
\left({x \over y} \right)^m, ~ m\in \NN ~~\Big> \supset \D_{-1} .\nonumber
\eeqa

\noi
If we construct a diagram analogous to \eqref{diag}
(for $\D^+_{-1/F}$), we have

$$
\xymatrix{&&&&0\\
& \ar@<1ex>[dl]^{ x \partial_y  }
x^2 \ar@<1ex>[r]^{ y \partial_x  }&
xy  \ar@<1ex>[l]^{ x \partial_y  }
\ar@<1ex>[r]^{ y \partial_x}&
y^2
\ar@<1ex>[l]^{ x \partial_y  }
\ar@<1ex>[ur]^{ y \partial_x }&
x^2 \left(\frac{y}{x}\right)^3 \ar@<1ex>[l]^{ x \partial_y  }
\ar@<1ex>[r]^{ y \partial_x   }&
x^2 \left(\frac{y}{x}\right)^4 \ar@<1ex>[l]^{ x \partial_y }
\ar@<1ex>[r]^{ y \partial_x  }
&
\cdots \ar@<1ex>[l]^{ x \partial_y  }
\ar@<1ex>[r]^{  y \partial_x  }
& x^2 \left(\frac{y}{x}\right)^n
\ar@<1ex>[r]^{ y \partial_x }
\ar@<1ex>[l]^{ x \partial_y }
&\dots
\ar@<1ex>[l]^{  x \partial_y  }&
\\
0&&&&
}
$$

\noi
Looking at the representations defined in (\ref{eq:FSUSY}) {\it i.e.}
$S^F\big(\D^{\pm}_{-1/F}\big)_{{\mathrm {red}}}$,
one sees that,  even though $\D_{-1}$ is a subspace stable under
 $\mathfrak{so}(1,2)$ there is no  complement stable under 
$\mathfrak{so}(1,2)$
\cite{fsusy3d}. Indeed,
these representations cannot be built from a primitive vector.
This is due to the fact that $L_{+}^3 \Big(x^2\Big) =0$ 
and consequently we cannot reach $x^{-1}y^3$ from $x^2$
but conversely $L_{-}^3 (x^{-1}y^3)= 6 x^2$. 
This is the reason why there is no $F-$Lie algebra structure on
$\mathfrak{so}(1,2) \oplus \D_{-1}$.
With the normalisations  \eqref{eq:repp} and \eqref{eq:FSUSY},
denoting ${\cal D}_{-1/F}^+=\{A_{-1/F+n}, \ n \in \mathbb N \}$ and
 ${\cal S}^F_{\text{red}}({\cal D}_{-1/F}^+)=\{P_{-1+n}, \ n \in \mathbb N \}$,
the trilinear brackets
are given by 

$$
\left\{A^+_{-1/F+n_1}, \cdots, A^+_{-1/F+n_F}\right\}=P_{-1+n_1 + \cdots n_F}.
$$

\section{Clifford algebras of polynomial}
\label{cliff}

Clifford algebras of a polynomial's is a direct generalisation of usual
Clifford algebras for higher degree polynomials. Consider $p$ a polynomial
of degree $n$ with $k$ variables. Since degree $n$ polynomials are isomorphic
to symmetric tensors of order $n$, we can write

\beqa
\label{cliff-def}
p(x^1,\cdots,x^k) = x^{i_1} \cdots x^{i_n} g_{i_1 \cdots i_n}.
\eeqa

\noi The Clifford algebra of the polynomial $p$ denoted ${\cal C}_p$
is the algebra generated by $k$ primitive elements $g_1,\cdots,g_k$
such that 

\beqa
\label{cliff-gene}
\left\{g_{i_1},\cdots, g_{i_n}\right\}= n !g_{i_1 \cdots i_n}.
\eeqa

\noi
The algebra defined by relations \eqref{cliff-gene} can be real if 
the tensor $g$ is real, or complex. It appears as an example of a
Lie algebras of order $n$ where $\g_0 = \mathbb R$ or $\mathbb C$.
The relations \eqref{cliff-gene} means that
 if we consider $P$ in ${\cal C}_p$ defined by 
$P= x^k g_k$, we have

\beqa
\label{line}
(x^1 g_1 + \cdots x^k g_k)^n= p(x_1,\cdots,x_k).1 \ ,
\eeqa

\noi where $1$ denotes the unity of ${\cal C }_p$ that we omit from now on.
This algebra has been introduced in a formal way by N. Roby
\cite{rr}.  It is a natural generalisation of the usual Clifford algebra,
but this algebra is very different from the usual Clifford algebra.
Since it is defined through $n$-th order relations,
the number of independent monomials increases with polynomial's degree (for
instance, $(g_1 g_2)^k, k \ge 0$ are all  independent). This means that we
do not have enough constraints among the generators to order them in some
fixed way and, as a consequence, ${\cal C}_p$ turns out to be an 
infinite-dimensional algebra. Thus any finite dimensional representation
are non-faithful. Several properties on representations was then 
established such
that the dimension of a representation of Clifford algebras
is a multiple of the degree of the polynomial \cite{ht}, but the 
first systematic way to obtain a matrix representation was given
in \cite{frr}. Subsequently,  an extensive study of the representations 
of Clifford algebras of cubic 
polynomials was undertaken by Revoy \cite{re} and a family of inequivalent 
representations can be obtained.  See also Ref.\cite{r} for many references
on Clifford algebras of polynomial's. 

For a self contained presentation, we give an algorithm to construct a matrix
representation of the Clifford algebra of a given polynomial,
for more details and comments see \cite{frr, r}. In this
process, two basic polynomials will be considered. The sum polynomial
$S(x)= (x^1)^n + \cdots + (x^k)^n$ and the product polynomial
$\pi(x)= x^1 \cdots x^n$.

For the linearisation of the sum polynomial, consider the $2r+1$ following 
matrices (with $r=[k/2]$)

\beqa
\label{eq:pi}
\begin{array}{ll}
\Pi_1=\pi_1 \otimes I^{\otimes^{(r-1)}}~~&
\Pi_2=\pi_2 \otimes I^{\otimes^{(r-1)}} \cr
~~~~~~~~~ \vdots & ~~~~~~~~~ \vdots \cr 
\Pi_{2\ell-1}=\pi_3^{\otimes^{(\ell-1)}}\otimes \pi_1 
\otimes I^{\otimes^{(r-\ell-1)}}~~&
\Pi_{2\ell}=\pi_3^{\otimes^{(\ell-1)}}\otimes\pi_2 
\otimes I^{\otimes^{(r-\ell-1)}} \cr
~~~~~~~~~ \vdots & ~~~~~~~~~ \vdots \cr  
\Pi_{2r-1}=\pi_3^{\otimes^{(r-1)}}\otimes\pi_1~~&
\Pi_{2r}=\pi_3^{\otimes^{(r-1)}}\otimes \pi_2 \cr
\hskip 4.5cm \Pi_{2r+1}= \pi_3^{\otimes^{r}}. 
\end{array} 
\eeqa

\noi
with  the matrices $\pi_1, \pi_2 , \pi_3$ being defined by 
\beq
\label{pi}
\pi_1=\begin{pmatrix}0&1&0&\cdots&0 \cr
                  0&0&1&\cdots&0 \cr
                  \vdots&&\ddots&\ddots&\vdots  \cr
                  0&0&\cdots&0&1 \cr 
                  1&0&0&\cdots&0\end{pmatrix},~ 
\pi_3=\begin{pmatrix}1&0&\cdots&0 \cr
                    0&q&\cdots&0 \cr
                    \vdots&&\ddots& \cr
                     0&0&\cdots&q^{n-1}\end{pmatrix},~
\pi_2=(\sqrt{q}) \pi_3 \pi_1,
\eeq

\noi
$\sqrt{q}$ being there only when $n$ is even and
$q=e^{2 i \frac{\pi}{n}}$. 
Many authors have considered this set of matrices, see
\cite{r} for references.
 It is not difficult to see
that the $\Pi-$matrices satisfy the relation 

\beqa
\label{gca}
\Pi_i \Pi_j=
q \Pi_j \Pi_i, \ \ \Pi_i^n=1, \ \ i <j
\eeqa

\noi
 and as a consequence of the relation root-coefficients
that we have $(x^1 \Pi_1 + \cdots x^k \Pi_k)^n=(x^1)^n + \cdots (x^k)^n$.
The algebra generated by elements satisfying \eqref{gca} generate the
generalised Clifford algebra. This algebra together with its representations
have been classified by Morris \cite{gca}.

The product polynomial is linearised by the matrices $H_{ii+1}$
(with $H_{ij}$ the canonical matrices with a one at the intersection
of the $i-$th line and $j-$th column and zero elsewhere). 
Indeed an easy calculation gives
$(x^i H_{12} + \cdots x^n H_{n1})^n=x^1 \cdots x^n$.

The matrices $\Pi$ and $H$ allow to linearise any polynomial, since an
arbitrary polynomial is a sum of monomials

$$
p(x) = \sum_{\ell=1}^q m_\ell(x).
$$

\noi
Each monomial is a particular case of the polynomial
$(x_1)^{a_1} (x_2)^{a_2} \cdots (x_p)^{a_p}$ which can be linearised by the 
$n \times n$ $H-$matrices

\beqa
&&(x_1)^{a_1} (x_2)^{a_2} \cdots (x_p)^{a_p}= \nonumber \\
&&\left[x_1 \left(\sum\limits_{i=1}^{a_1}   H_{ii+1}\right)  +
     x_2  \left( \sum\limits_{i=1+a_1}^{a_1+a_2} H_{ii+1} \right)  + \cdots+
     x_p \left( \sum\limits_{i=1+a_1+\cdots+a_{p-1}}^n H_{ii+1} \right) 
        \right]^n. \nonumber
\eeqa        

\noi
 Thus any monomial of the polynomial $p$ can be 
linearised as above and we have
$p(x) = \sum_{\ell=1}^q M^n_\ell$, with $M_\ell$,
$q$ \   $n \times n$ matrices.
Introducing  the commuting matrices 
$\tilde M_\ell = I^{\otimes^{\ell-1}}\otimes
M_\ell\otimes I^{\otimes^{q-\ell-1}}$ we have  using the linearisation of
the sum polynomial by the $\Pi-$matrices 

$$
p(x) =  \sum_{\ell=1}^q\tilde M^n_\ell = (\sum_{\ell=1}^q
 (\Pi_\ell\otimes \tilde M_\ell)^n.
$$

\noi
Since $\Pi_\ell \otimes \tilde M_\ell$ are linear in $x^k$ we have
$\sum_{\ell=1}^q
 \Pi_\ell\otimes \tilde M_\ell=x^1 G_1 +\cdots x^k G_k$.
This ends end the process of linearisation. Of course this process 
is far from being unique and many different matrices of different
size can be obtained. See \cite{frr,r} for examples and comments.\\

As an illustration of this process consider the little algebra
\eqref{little} of Section \ref{rep-fsusy}. Equation \eqref{little}
say that the generators $A_{-1/F}$ and $A_{1-1/F}$ generate the Clifford
algebra of the polynomial $x^{F-1} y$. The process above allows to find 
a linearisation of the polynomial by the $n \times n$ matrices

\beqa
\label{mat-clif}
A^+_{-{1\over F}}=
\begin{pmatrix}0&0&0&\ldots&0&0& \cr
                         a_1&0&0&\ldots&0&0& \cr
                         0& a_2&0&\ldots&0&0& \cr
                         &\cr
                         \vdots&\vdots&&\ddots&\ddots&\vdots   \cr
                         0&0&\ldots&0&a_{F-1}&0& \end{pmatrix}, 
\ \  
A^+_{1-{1\over F}}=
                \begin{pmatrix} 0&0&0&\ldots&0&\frac{1}{a_1 \cdots a_{F-1}}\cr
                          0&0&0&\ldots&0&0& \cr
                          0&0&0&\ldots&0&0& \cr
                          &\cr\
                          \vdots&\vdots&&&\ddots&\vdots& \cr 
                           0&0&0&\ldots&0&0&
\end{pmatrix}.
\eeqa

\noi
(For physical reasons the normalisation of these matrices is not
the same of the normalisation which would have been obtained
by the algorithm above 
 {\it i.e} that the non-zero matrix elements  are not
equal to one (see Section \ref{rep-fsusy}).)
The matrix $A^+_{-1/F}$ can be obtained in a different way.
Indeed, \eqref{little} implies that
the first power of $A^+_{-{1\over F}}$ which
is equal to zero  is $F$
(in other words the rank of $A^{+}_{-{1\over F}}$  is $F-1$).
Writing $A^{+}_{-1/F}$ in its Jordan form using the relations
\eqref{little} gives a
solution for $A^+_{1-1/F}$ of the type above. However, it is known for
$F=3$ that there exists other solutions for the matrix $A^{+}_{1-{1/ F}}$
(see \cite{rr,r}). But these matrices would not respect the grading,
see \eqref{module}. Indeed, if  some of the matrix elements which are 
equal to zero in (\ref{mat-clif})  are different from zero, 
 the matrices would not be
consistent with the Poincar\'e
algebra {\it i.e.} 
we obtain equations where both sides do
not have the same helicity.
Thus we take the matrices as in \eqref{mat-clif}.\\

In the same way,
the representation of the cubic extension
of the Poincar\'e algebra in any space-time dimensions
(see Section \ref{cubic-poincare}) are related to
 Clifford algebras of  polynomial's. Indeed, 
writing the R.H.S. of \eqref{cubic-poin} as $g_{\mu \nu \rho}= \frac16
(\eta_{\mu \nu} \delta_\rho{}^\sigma +\eta_{ \nu \rho } \delta_\mu{}^\sigma +
\eta_{\rho \mu} \delta_\nu{}^\sigma )P_\sigma$, we define
$p(x) = g_{\mu \nu \rho }x^\mu x^\nu x^\rho= \frac12 (x.x) (x.P)$.
In \cite{cubic1,cubic2}, along the algorithm above, we have found two
different representations of the Clifford algebra of the polynomial
$\frac12 (x.x) (x.P)$, but only one was compatible with the Poincar\'e algebra
(see  \cite{cubic1}). For the second ones,
writing 

$$
\frac12 (x.x)(x.p)=\frac12(x^\mu 
\Gamma_\mu)^2(x.p)=\frac{1}{2}\begin{pmatrix}
0&\Lambda^{\frac13}x^\mu\Gamma_\mu&0 \\
0&0&\Lambda^{\frac13}x^\mu\Gamma_\mu \\
\Lambda^{-\frac{2}{3}} x^\mu P_\mu&0&0
\end{pmatrix}^3
$$ we obtain

\beqa
\label{red}
V_\mu = \frac{1}{\sqrt[3]{2}}\begin{pmatrix}
0&\Lambda^{\frac13}\Gamma_\mu&0 \\
0&0&\Lambda^{\frac13}\Gamma_\mu \\
\Lambda^{-\frac{2}{3}}P_\mu&0&0
\end{pmatrix},
\eeqa

\noi
where $\Lambda$ is a parameter with dimension of mass (that
we take normalised to one) and $\Gamma_\mu$ are the Dirac 
$\Gamma-$matrices in $D-$space-time dimensions. 
We also renormalise the matrices $V$ in such a way that
the factor $\sqrt[3]{2}$ cancels.
If $D$ is even,
the representation \eqref{red} is reducible. Taking the Dirac
$\Gamma-$matrices in the chiral representation

$$
\Gamma_\mu=\begin{pmatrix}0&\Sigma_\mu \\
\tilde \Sigma_\mu&0
\end{pmatrix},
$$

\noi 
(where $\Sigma_0=\tilde \Sigma_0=1$ and $\tilde \Sigma_i=-\Sigma_i$ are
the generators of the Clifford algebra of $SO(D-1)$)
we obtain

\beqa
\label{irred}
V^+_\mu = \begin{pmatrix}
0&\Sigma_\mu&0 \\
0&0&\tilde \Sigma_\mu \\
P_\mu&0&0
\end{pmatrix}, \ \
V^-_\mu = \begin{pmatrix}
0&\tilde \Sigma_\mu&0 \\
0&0& \Sigma_\mu \\
P_\mu&0&0
\end{pmatrix},
\eeqa
 
\noi
two inequivalent representations.\\

A peculiar Clifford algebra of polynomial is given when $p=0$. In this
case, the algebra \eqref{cliff-gene} reduces to

\beqa
\label{3-ext}
\left\{e_{i_1},\cdots, e_{i_n}\right\}=0.
\eeqa

\noi 
This algebra has been defined in \cite{roby-ext} and called the
$n$-exterior algebra.
An explicit basis of this infinite dimensional algebra was exhibited.
This is clearly a generalisation of the Grassmann algebra.

\section{Some properties on spinors and $p-$forms in 
$(1+(4n-1))-$dimensions}
\label{pform}
In this section, we give a collection of useful identities on $p-$forms
and spinors. 
The Minkowski metric is taken to be $\eta_{\mu \nu}=\text{diag}(1,-1,\cdots,
-1)$, and the Levi-Civita tensor is normalised as follows
$\e_{01 \cdots D-1}=-e^{01\cdots D-1}=1$. For latter convenience, we denote
\begin{eqnarray}
\delta^{(\mu)_\ell}_{(\nu)_\ell} &=& 
\delta^{\mu_1 \cdots \mu_\ell}_{\nu_1 \cdots \nu_\ell} =
\left|
\begin{array}{lll}\delta_{\nu_1}^{\mu_1} &\cdots & 
\delta_{\nu_1}^{\mu_\ell} \cr
       \vdots&&\vdots \cr
              \delta_{\nu_\ell}^{\mu_1} &\cdots & 
\delta_{\nu_\ell}^{\mu_\ell} \cr
\end{array} \right| \nonumber \\ 
\varepsilon_{(\mu)(\nu)} &=&\varepsilon_{\mu_1 \cdots \mu_{2n} 
\nu_1 \cdots \nu_{2n}} \\
\eta_{(\mu)(\nu)}&= &\eta_{\mu_1 \nu_1} \cdots \eta_{\mu_{2n} \nu_{2n}}. \nonumber
\end{eqnarray}

\noi
All the properties below come, for instance, from an explicit
matrix realisation of the Dirac matrices and from the
properties of spinors (see {\it e.g.} \cite{clif}).
\subsection{Dirac matrices}

The Dirac $\Gamma-$matrices
in the chiral representation are given by

$$
\Gamma_\mu=\begin{pmatrix}0&\Sigma_\mu \\
\tilde \Sigma_\mu&0
\end{pmatrix},
$$

\noi 
(where $\Sigma_0=\tilde \Sigma_0=1$ and $\tilde \Sigma_i=-\Sigma_i$ are
the generators of the Clifford algebra of $SO(D-1)$).
We introduce  further the fully antisymmetric matrices:
\begin{eqnarray}
\label{gamma2}
\Gamma^{(\ell)}: \Gamma_{\mu_1 \cdots \mu_\ell} =
\frac{1}{\ell !} \sum \limits_{\sigma } \Gamma_{\mu_{\sigma(1)} }\cdots  
\Gamma_{\mu_{\sigma(\ell )} }   
\end{eqnarray}

\noindent 
which write

\begin{eqnarray}
\Gamma_{ \mu_1 \cdots \mu_{2 \ell}  }& =& \frac{1}{(2\ell) !}\begin{pmatrix}
\Sigma_{\mu_1} \tilde \Sigma_{\mu_2} \cdots \tilde \Sigma_{\mu_{2 \ell}}
+\mathrm{~perm}&0 \cr
0&  \tilde  \Sigma_{\mu_1}  \Sigma_{\mu_2} \cdots \Sigma_{\mu_{2 \ell}}
+\mathrm{~perm}
 \end{pmatrix}
\nonumber \\
&=& \begin{pmatrix} \Sigma_{{\mu_1} \cdots \mu_{2 \ell}}&0 \cr
0&\tilde  \Sigma_{{\mu_1} \cdots \mu_{2 \ell}} \end{pmatrix}
\nonumber \\ 
\\ \nonumber \\
 \Gamma_{ \mu_1 \cdots \mu_{2 \ell + 1 }  }& =&\frac{1}{(2\ell+1) !} \begin{pmatrix}
0& \Sigma_{\mu_1} \tilde \Sigma_{\mu_2} \cdots  \Sigma_{\mu_{2 \ell + 1 }}
+\mathrm{~perm}\cr
  \tilde  \Sigma_{\mu_1}  \Sigma_{\mu_2} \cdots \tilde  \Sigma_{\mu_{2 \ell + 1}}
+\mathrm{~perm}
 \end{pmatrix}\nonumber \\ 
&=&\begin{pmatrix}0 &  \Sigma_{{\mu_1} \cdots \mu_{2 \ell +1 }} \cr
\tilde  \Sigma_{{\mu_1} \cdots \mu_{2 \ell + 1}} &0\end{pmatrix}
\nonumber
\end{eqnarray} 

\noi These matrices are generically denoted by

$$
\Gamma^{(\mu)_\ell}= \Gamma^{\mu_1 \cdots \mu_\ell}, \ \  
\Gamma_{(\mu)_\ell}= \Gamma_{\mu_1 \cdots \mu_\ell} 
$$

\noi
and we have the trace formul\ae

\beqa
\label{Tr2}
\frac{1}{2^{n}} \mathrm{Tr}\left(\Sigma^{(\mu)_{2a}}
\Sigma_{(\nu)_{2b}}\right)&=&
\delta_{ab} \left(\delta^{(\mu)_{2a}}_{(\nu)_{2a}} -i  \delta_{an} 
\varepsilon^{(\mu)(\mu^\prime)}
\eta_{(\mu^\prime)(\nu)} \right) \nonumber \\
\frac{1}{2^{n}} \mathrm{Tr}\left(\tilde \Sigma^{(\mu)_{2a}}\tilde 
\Sigma_{(\nu)_{2b}}\right)
&=&
\delta_{ab} \left(\delta^{(\mu)_{2a}}_{(\nu)_{2a}} +i  \delta_{an}  
\varepsilon^{(\mu)(\mu^\prime)}
\eta_{(\mu^\prime)(\nu)} \right) \nonumber \\
\frac{1}{2^{n}} \mathrm{Tr} \left(\Sigma^{(\mu)_{2a+1}}\tilde 
\Sigma_{(\nu)_{2b+1 }}\right)&=&\delta_{ab} \delta^{(\mu)_{2a+1}}_{(\nu)_{2a+1}}. 
\eeqa

\noi
We have assumed here that $\Sigma^{(\mu)_{2n}}$ projects onto the self-dual
$2n-$forms. Correspondingly we assume that $\tilde\Sigma^{(\mu)_{2n}}$ 
projects onto the anti-self-dual
$2n-$forms. This convention fixes  
the second term in the first equation in \eqref{Tr2}. 
Moreover, we have the following properties

\beqa
\label{prodGamma}
\Gamma_{\mu_1 \cdots \mu_\ell} \Gamma_\mu = 
\Gamma_{\mu_1 \cdots \mu_\ell \mu} + \eta_{\mu_\ell \mu} \Gamma_{\mu_1\cdots  \mu_{\ell -1}}+
\cdots + (-1)^{\ell -1} \eta_{\mu_1 \mu} \Gamma_{\mu_2 \cdots \mu_\ell}.
\eeqa

\subsection{Spinors and $p-$forms}

The Dirac matrices act naturally on spinors.  When the dimension
is even, the spinor space decomposes into
left-handed and right-handed spinors 
${\cal S} = {\cal S}_+ \oplus {\cal S}_-$. Furthermore, 
when $D=4n$, left-handed spinors are in the complex conjugate
representation of right-handed spinors. In a straight analogy with
Appendix \ref{conventions}, we  denote $\Psi_L \in {\cal S}_+$ and 
$\Psi_R \in {\cal S}_-$  by there components $\Psi_L \to \Psi_A$ and
$\Psi_R \to \bar \Psi^{\dot A}$.
This leads to the following index structure for the $\Gamma-$matrices:
 $\Sigma_\mu
\to \Sigma_\mu{}_{A \dot B},
\tilde \Sigma_\mu \to \tilde \Sigma_\mu{}^{\dot A B}$.
The charge conjugation matrix is given by

$$
{\cal C} = \begin{pmatrix} C_+{}^{AB}&0\cr 0&  
C_-{}_{\dot A \dot B} \end{pmatrix}, \ \
{\cal C}^{-1} = \begin{pmatrix} C^{-1}{}_+{}_{AB}&0\cr 0&  
C^{-1}{}_-{}^{\dot A \dot B} \end{pmatrix},
$$

\noi
and allows to raise and lower  the indices $\Psi^A= \Psi_B C_+{}^{AB},
\Psi_A= \Psi^B C^{-1}{}_+{}_{AB}$ {\it etc}.
 
The tensor product of two spinors decomposes on the set of $p-$forms
($[p]$ denotes $p-$forms and $[2n]_\pm$  (anti-)self-dual $2n-$form
see \eqref{self})

\beqa
\label{bi-spin2}
{\cal S}\ \  \otimes {\cal S }\ \ &=& [0] \oplus [1] \oplus \cdots [4n],\nonumber \\
{\cal S}_+ \otimes {\cal S}_+ &=&[0] \oplus [2] \oplus \cdots [2n]_+
\\
{\cal S}_- \otimes {\cal S}_- &=&[0] \oplus [2] \oplus \cdots [2n]_-
\nonumber \\
{\cal S}_+ \otimes {\cal S}_- &=& [1] \oplus [3] \oplus \cdots [2n-1].
\nonumber
\eeqa

  Introducing,
$A_{[p]} \in [p], 0\le p\le 2n-1$ and $A_{[2n]_\pm } \in [2n]_\pm$ 
\eqref{bi-spin2} give
(in the sequel $A_{[p]} \Gamma^{(p)}$ alway means 
 $A_{[p]}{}_{\mu_1 \cdots \mu_p} \Gamma^{\mu_p \cdots \mu_1}$):

\beqa
\label{spin-p2}
 \Xi_{++}=\Psi_+ \otimes  \Psi^\prime_+ &=&
 \sum \limits_{p=0}^{n-1} \frac{1}{(2p)!} A_{[2 p]} \Sigma^{(2p)}
{\cal C}^{-1}_+
+ \frac12\frac{1}{(2n)!} A_{[2n]_+} \Sigma^{(2n)}{\cal C}^{-1}_+ \nonumber \\
 \Xi_{--}= \Psi_- \otimes  \Psi^\prime_-{}&=&
 \sum \limits_{p=0}^{n-1} \frac{1}{(2p)!} A^\prime{}_{[2 p]} 
\tilde \Sigma^{(2p)} {\cal C}^{-1}_-
+ \frac12\frac{1}{(2n)!} A^\prime{}_{[2n]_-} \tilde \Sigma^{(2n)}
{\cal C}^{-1}_-
\nonumber \\ 
\\
\Xi_{+-}=\Psi_+ \otimes  \Psi^\prime_-{}&=&
 \sum \limits_{p=0}^{n-1} \frac{1}{(2p+1)!} A_{[2 p+1]} \Sigma^{(2p+1)}
 {\cal C}^{-1}_-
\nonumber \\
 \Xi_{-+}= \Psi_- \otimes \Psi^\prime_+{}&=&
 \sum \limits_{p=0}^{n-1} \frac{1}{(2p+1)!} A^\prime{}_{[2 p+1]} \tilde 
\Sigma^{(2p+1)} {\cal C}^{-1}_+.
\nonumber 
\eeqa

\noindent
In the first relation in (\ref{spin-p2}) the $2n-$form is self-dual, and 
in the second
the  $2n-$form is anti-self-dual.
Using the trace relations \eqref{Tr2}, one gets
\beqa
\label{p-spin2}
A_{[2p]}&=&  \frac{1}{2^{n}} 
\mathrm{Tr}\left( \Xi_{++}{\cal C}_+\Sigma^{(2p)}\right),
\ \ 
A_{[2n]_+}{} = \frac{1}{2^{n}} \mathrm{Tr}\left(\ \Xi_{++}{\cal C}_+ 
\Sigma^{(2n)}
\right),
\nonumber \\
A^\prime{}_{[2p]}&=&  \frac{1}{2^{n}} 
\mathrm{Tr}\left(
\Xi_{--}{\cal C}_-\tilde \Sigma^{{(2p)}} \right),\ \ 
A{}^\prime_{[2n]_-} = \frac{1}{2^{n}} 
\mathrm{Tr}\left(
\Xi_{--}{\cal C}_-\tilde \Sigma^{{2n}} \right), \nonumber \\
A_{[2p+1]}&=&  \frac{1}{2^{n}} \mathrm{Tr}\left(
\Xi_{+-} {\cal C}_-\tilde \Sigma^{{(2p+1)}} \right),
\\
A^\prime{}_{[2p+1]}&=&  \frac{1}{2^{n}} \mathrm{Tr}\left(
\Xi_{-+} {\cal C}_+\Sigma^{{(2p+1)}}  \right).
\nonumber 
\eeqa

\noindent
where $\mathrm{Tr}\left( \Xi_{+-}{\cal C}_-\tilde \Sigma^{{(2p+1)}}\right)=
\mathrm{Tr}\left( \Xi_{+-}{\cal C}_-\tilde \Sigma_{\mu_1 \cdots \mu_{2p+1}}\right)$ to 
simplify  notations.

Finally, we introduce some operations on $p-$forms.
The Hodge duality is a linear map ${}^\star :  
[p] \longrightarrow [D-p]$.
If $A_{[p]} \in [p], {}^\star A_{[p]} = B_{[D-p]} \in 
[D-p]$ is given by

\beqa
\label{hodge}
B_{[D-p]}{}_{\mu_1 \cdots \mu_{D-p}}= \frac{1}{p!} 
\varepsilon_{\mu_1 \cdots \mu_{D-p} \nu_1 \cdots
\nu_p} A_{[p]}^{\nu_1 \cdots \nu_p},
\eeqa

\noindent
and it is easy to prove that 

\beqa
\label{star2}
{}^\star {}^\star A_{[p]}= (-1)^{(D-1)(p-1)} A_{[p]}.
\eeqa 

\noindent 
When the dimension is even, one can define a (anti-)self-dual $(D/2)-$form:

\beqa
\label{self}
{}^\star A_{[D/2]}= \left\{
\begin{array}{ll}
\pm A_{[D/2]}&\mathrm{~when~}D/2\mathrm{~is~an~odd~number~}({}^{\star^2}=1) \cr
\pm i A_{[D/2]}&\mathrm{~when~}D/2\mathrm{~is~an~even~number~}({}^{\star^2}=-1)
\end{array}\right.
\eeqa

\noindent
This means that (anti-)self-dual $2n-$forms are complex representations of the
Lorentz group when $D=4n$.

Next, introducing $\varepsilon \in [1]$, one defines

\begin{itemize}
\item The inner product

\beqa 
\label{iner1} 
\begin{array}{llll} 
i_\varepsilon: &[p] &\longrightarrow& [p-1] \cr  
&A_{[p]} &\longmapsto& i_\varepsilon A_{[p]},
\end{array}\eeqa

\noindent
in components we have

\beqa
\label{iner2}
 (i_\varepsilon A_{[p]})_{\mu_1 \cdots \mu_{p-1}}= A_{[p]}{}_{\mu_1 \cdots \mu_p}
\varepsilon^{\mu_p}.
\eeqa

\noindent
Notice the difference of convention, useful for our purpose: the summation 
is done on the
last index instead of the first one.

\item The exterior product
\beqa
\label{ext1}
\begin{array}{llll} 
 \wedge:&[p] &\longrightarrow& [p+1] \cr  
&A_{[p]} &\longmapsto&  A_{[p]}\wedge \varepsilon,
\end{array}\eeqa 

 \noindent
in components reads
\beqa
\label{ext2}
(A_{[p]}{}\wedge \varepsilon)_{\mu_1 \cdots \mu_{p+1}}= 
\frac{1}{p!} \delta_{\mu_1 \cdots \mu_{p+1}}^{\nu_1\cdots \nu_{p+1}} A_{[p]}{}_{\nu_1 
\cdots \nu_p}
\varepsilon_{\nu_{p+1}}.
\eeqa
 \item Now, with $A_{[p]} \in [p]$ and ${}^\star A_{[p]}=B_{[D-p]} \in 
[D-p]$, we have that

\beqa
\label{dual-prod}
{}^\star \left(A_{[p]} \wedge \varepsilon\right)&=&(-1)^p i_\varepsilon 
B_{[D-p]}, \\
{}^\star \left(i_\varepsilon A_{[p]}\right)&=&(-1)^{D(p-D)}B_{[D-p]}
\wedge \varepsilon.
\nonumber 
\eeqa
\end{itemize}

Then, one defines the exterior derivative $d$ which maps $[p] \to
[p+1]$ by

\beqa
\label{ext}
(d A_{[p]})_{\mu_1 \cdots \mu_{p+1}}= \frac{1}{p!} \delta_{\mu_1 \cdots 
\mu_{p+1}}^{\nu_1 
\cdots \nu_{p+1}} {\partial_{\nu_1}} A_{[p]}{}_{\nu_2 \cdots \nu_{p+1}}
\eeqa

\noindent
and its adjoint $d^\dag$ which maps  $[p] \to
[p-1]$  is defined by $d^\dag=(-1)^{pD+D} {}^\star d {}^\star$,
and gives

\beqa
\label{extdag}
(d^\dag A_{[p]})_{\mu_2 \cdots \mu_{p}}={ \partial^{\mu_1}} 
 A_{[p]}{}_{\mu_1 \nu_2 \cdots \nu_{p}}.
\eeqa

It can the be shown that we have the following relation

\beqa
\label{dd}
&&\frac{1}{(p+1)!} d A_{[p]}  d A_{[p]}  + \frac{1}{(p-1)!} d^\dag A_{[p]}  
d^\dag A_{[p]} 
\\
&=&(-1)^{D-1}\left( \frac{1}{(D-p-1)!} d^\dag  B_{[D-p]}  
d^\dag   B_{[D-p]}  + 
\frac{1}{(D-p+ 1)!} d B_{[D-p]}  d  B_{[D-p]} \right) 
\nonumber 
\eeqa 
\noi
with $B_{[D-p]}= {}^\star A_{[p]}$.


\baselineskip=1.6pt  
\begin{thebibliography}{99}
%
\bibitem{cm}
 O. W. Greenberg, {\it Coupling of Internal and
Space-Time Symmetries}, Phys. Rev. {\bf 135} (1964) 1447-1450;
L-O'Raifeartaigh, {\it Lorentz Invariance and Internal Symmetry},
Phys. Rev {\bf 139} (1965), B 1052-1062; 
S. Coleman and J. Mandula, {\it All Possible Symmetries of the
$S$-Matrix}, Phys. Rev. {\bf 159} (1967) B 1251-1256.
%
%
\bibitem{gut}
 H.~Georgi and S.~L.~Glashow,
 {\it Unity Of All Elementary Particle Forces},
  Phys.\ Rev.\ Lett.\  {\bf 32} (1974) 438;
H.~Fritzsch and P.~Minkowski,
  {\it Unified Interactions Of Leptons And Hadrons,}
  Annals Phys.\  {\bf 93}  (1975) 193.
F.~Gursey, P.~Ramond and P.~Sikivie,
 {\it A Universal Gauge Theory Model Based On E6,}
  Phys.\ Lett.\  B {\bf 60} (1976) 177;
 R.~Slansky,
  {\it Group Theory For Unified Model Building,}
  Phys.\ Rept.\  {\bf 79}  (1981) 1.
G. G. Ross, {\it Grand Unified Theories}, The Benjanin/Cummings 
Publishing Company, INC, (1985).
%
\bibitem{hls}
R.~Haag, J.~T.~Lopuszanski and M.~F.~Sohnius,
{\it All Possible Generators Of Supersymmetries Of The S Matrix},
{  Nucl. Phys.}  {\bf B88} (1975) 257-274.
%
\bibitem{wei}
 S.~Weinberg,
  {\it The quantum theory of fields.  Vol. 3: Supersymmetry},
{  Cambridge, UK: Univ. Pr. (2000)}.

\bibitem{nam}
 Y.~Nambu,
 {\it Generalized Hamiltonian dynamics,}
  Phys.\ Rev.\  D {\bf 7} (1973) 2405.
%

\bibitem{bg1}
I. Bars   and  M. G\"unaydin, 
 {\it Construction Of Lie Algebras 
And Lie Superalgebras From Ternary Algebras},
  {  J.\ Math.\ Phys.\ }  {\bf 20} (1979)  1977-1993.
%
\bibitem{bg2}
I. Bars I  and M.  G\"unaydin,
{\it  Dynamical Theory Of Subconstituents Based On Ternary Algebras},
{   Phys.\ Rev.\  D } {\bf 22}  (1980)  1403-1413.
%
\bibitem{k1}
L. Vainerman  and R. Kerner, 
{\it On special classes of $n-$algebras},
{ J. Math. Phys}
{\bf 37}  (1995)  2553-2565.
%
\bibitem{k2}
R. Kerner, 
{\it The cubic chessboard. Geometry and physics},{ Class. and Quantum Grav.}
 {\bf 14 (1A)} (1997) A203-A225.
%
\bibitem{k3}
A.  Borowiec,  N. Bazunova   and 
R. Kerner,
{\it Universal Differential Calculus on Ternary Algebras},
{  Lett. Math. Phys.}
{\bf  67}   (2004) 195-206.
%
\bibitem{az}
 J.~A.~de Azc\'arraga and J.~C.~P\'erez Bueno,
 {\it Higher-order simple Lie algebras},
  Commun.\ Math.\ Phys.\  {\bf 184 } (1997), 669.
  [arXiv:hep-th/9605213].
%
\bibitem{r}
M. Rausch de Traubenberg,
 {\it Clifford algebras, supersymmetry and 
$\mathbb Z_n-$symmetries: Applications in
field theory,}  arXiv:hep-th/9802141 (Habilitation Thesis).
%
\bibitem{fi}
V. T. Filippov, 
{\it $n$-Lie algebras} (in Russian), 
   { Sibirsk. Math. Zh} {\bf 26} (1985) 126-140.
%
\bibitem{g}
A. V. Gnedbaye,
{\it Les alg\`ebres $k$-aires et leurs op\'erades} (in French), 
    { C. R. Acad. Sci.  Paris S\'er. I Math.}
{\bf   321}  (1995)  147-152.
%
\bibitem{mv}
P. W. Michor   and  A. M. Vinogradov,
{\it $n$-ary Lie and associative algebras,}
{ Rend. Sem. Mat. Univ. Politec. Torino} {\bf   54} (1996) 373-392.
%
\bibitem{gr}
N. Goze and E. Remm, {\it On n-ary algebras given by Gerstenhaber's products},
arXiv:0803.0553[math.RA]. 
%
\bibitem{bala}
 J.~Bagger and N.~Lambert,
  {\it  Modeling multiple M2's,}
  Phys.\ Rev.\  D {\bf 75}, (2007) 045020
  [arXiv:hep-th/0611108].
%
\bibitem{ff}
 P.~De Medeiros, J.~M.~Figueroa-O'Farrill and E.~M\'endez-Escobar,
  { \it Lorentzian Lie 3-algebras and their Bagger-Lambert moduli space},
  JHEP {\bf 0807} (2008) 111
  [arXiv:0805.4363 [hep-th]];
%
%
\bibitem{flie1} 
M.~Rausch de Traubenberg and M.~J.~Slupinski, \textit{Fractional
supersymmetry and F(th) roots of representations,} J.\ Math.\ Phys.\ \textbf{%
41}, (2000), 4556--4571 [arXiv:hep-th/9904126]. 
%
\bibitem{flie2} M.~Rausch de Traubenberg and M.~J.~Slupinski, \textit{%
Finite-dimensional Lie algebras of order F,} J.\ Math.\ Phys.\ \textbf{43},
(2002), 5145--5160 [arXiv:hep-th/0205113]. 
%
\bibitem{flie3} M.~Goze, M.~Rausch de Traubenberg and A.~Tanasa, \textit{%
Poincar\'e and $\mathfrak{sl}(2)$ algebras of order three},
J. \ Math. \ Phys. {\bf 48} (2007) 093507,
[arXiv:math-ph/0603008]. 
%
%
\bibitem{extpo}
R.~Kerner, 
{\it Z(3) graded algebras and the cubic root of the 
supersymmetry translations},
J. Math. Phys. {\bf 33} (1992) 403;
%
R.~Kerner, 
{\it Z(3) Grading And The Cubic Root Of The Dirac Equation,}
Class. Quantum Grav. {\bf 9} (1992) S137;
%
L.~A.~Wills Toro,
{\it Trefoil symmetries I. Clover extensions beyond Coleman
Mandula theorem},
 { J. Math. Phys.} {\bf 42} (2001) 3915;
%
L.~A.~Wills Toro, L.~A.~ S\'anchez, J.~M.~Osorio and, D.~E.~Jaramillo,
{\it Trefoil symmetry II. Another clover extension},
{ J. Math. Phys.} {\bf 42} (2001) 3935;
%
L.~A.~Wills Toro, 
{\it Trefoil symmetry III. The full clover extension},
{ J. Math. Phys.} {\bf 42} (2001) 3947;
%
L.~A.~Wills Toro, L.~A.~ S\'anchez, and X. Leleu, 
{\it Trefoil Symmetry IV: Basic Enhanced Superspace 
for the Minimal Vector Clover Extension},
{ Int. J. Theor. Phys.}
{\bf 42} (2003) 57;
%
L.~A.~Wills Toro, L.~A.~ S\'anchez, and D. Bleecker,
{\it Trefoil Symmetry V: Class Representations for the
 Minimal Clover Extension},
{ Int. J. Theor. Phys.} {\bf 42} (2003) 73;
%
 J.~Beckers and N.~Debergh,
 {\it Poincare invariance and quantum parasuperfields},
  Int.\ J.\ Mod.\ Phys.\  A {\bf 8}  (1993) 5041;
%
J. Beckers, N. Debergh and A.G. Nikitin, 
{\it On parasupersymmetries and relativistic descriptions for spin 
one particles. 1: The Free context}, Fortsch. Phys. {\bf 43} (1995) 67-80;
%
J. Beckers, N. Debergh, A.G. Nikitin,
{\it On parasupersymmetries and relativistic descriptions for spin
 one particles. 2: The Interacting context with (electro)magnetic fields},
Fortsch. Phys. {\bf 43 } (1995) 81-96; 
%
A.~G.~Nikitin and V.~V.~Tretynyk,
{\it Irreducible Representations Of The Poincar\'e Parasuperalgebra,}
{ J.\ Phys.} {\bf A 28},  (1995) 1655;
%
J. Niederle and A.~G.~Nikitin, 
{\it Irreducible representations of the extended Poincar\'e parasuperalgebra},
{ J.\ Phys.} {\bf A 32},  (1999) 5141.
%
\bibitem{fsusy3d}
M.~Rausch de Traubenberg and M.~Slupinski,
{\it Non-Trivial Extensions of the $3D-$Poincar\'e Algebra
and Fractional Supersymmetry for Anyons},
Mod. Phys. Lett. {\bf A12} (1997) 3051 [hep-th/9609203].
%
%
\bibitem{p-form}
G.~Moultaka, M.~Rausch de Traubenberg and A.~Tanasa,
{\it Non-trivial extension of the Poincar\'e algebra for
antisymmetric gauge fields,}
{ Proceedings of the XIth International Conference Symmetry Methods in 
Physics, Prague 21-24 June 2004},
arXiv:hep-th/0407168.
%
\bibitem{susy}
T.~A.~Gol'fand and E.~P.~Likhtman,
 {\it Extension of the Algebra of Poincar\'e Group Generators and 
Violation of p Invariance},
 { JETP Lett.} {\bf 13} (1971)
32;
%
P.~Ramond, 
{\it Dual Theory for Free Fermions},
{ Phys. Rev.} {\bf D 3} (1971) 2415; 
%
A.~Neveu and J.~Schwarz,
{\it Factorizable dual model of pions},
 { Nucl. Phys. } {\bf B 31} (1971) 86;
%
D.~V.~Volkov and V.~P.~Akulov,
{\it Is the Neutrino a Goldstone Particle?},
 { Phys. Lett.} {\bf B 46} (1973) 109.
%
\bibitem{wzmod}
J.~Wess and B.~Zumino,
{\it Supergauge Transformations in Four-Dimensions},
 { Nucl. Phys.} {\bf B 70} (1974) 39-50.
%
\bibitem{super}
V.~G.~Kac, 
{\it Classification of simple Lie superalgebras}, 
Functional Anal. Appl. {\bf 9} (1975), no. 3, 263--265;
%
V.~G.~Kac, 
 {\it A sketch of Lie superalgebra theory},
{ Comm. Math. Phys. } {\bf 53} (1977) 31;
%
W.~Nahm and M. Scheunert, 
{\it Classification Of All Simple Graded Lie Algebras Whose Lie Algebra Is
 Reductive. 2. Construction Of The Exceptional Algebras},
{ J. Math. Phys.} {\bf 17} (1976)  1626;
%
M. Scheunert, {\it The Theory of Lie Superalgebras: an Introduction},
Lecture Notes in Mathematics 716 (Springer, Berlin);
%
I. Kaplansky, 
{\it Superalgebras},
{ Pacific J. Math} {\bf 86} (1980) 93;
%
P.~G.~O.~Freund and I.~Kaplansky, 
{\it Simple Supersymmetries},
{ J. Math. Phys.} {\bf 17} (1976) 228.
%
%

\bibitem{west}
P. West,  {\it Introduction to supersymmetry and supergravity}
(Singapore, Singapore: World Scientific 1990).
%
\bibitem{wb}
J. Wess, J. Bagger, {\it  Supersymmetry and
Supergravity } (Princeton University Press, 1983).
%
\bibitem{so}
M.~F.~Sohnius, {\it Introducing Supersymmetry}, 
  { Phys.\ Rept.\ }  {\bf 128} (1985), 39-204.
%
\bibitem{f}
P.~G.~O.~Freund,
{\it Introduction To Supersymmetry,}  (Cambridge Univ. Press, 1983).
%
\bibitem{terning}
J. Terning, {\it Modern Supersymmetry}, 
(Oxford, UK: Clarendon 2006).
%
\bibitem{nilles}
H.~P.~Nilles,
{\it Supersymmetry, Supergravity And Particle Physics}
  Phys.\ Rept.\  {\bf 110} (1984) 1-162.
\bibitem{martin}
 S.~P.~Martin, {\it A supersymmetry primer}, arXiv:hep-ph/9709356
(v4, June 2006).
%
\bibitem{aitch}
I.~J.~R.~Aitchison,
  {\it Supersymmetry in particle physics: An elementary introduction},
(Cambridge, UK: Univ. Pr. 2007).

%
\bibitem{ext}
 P.~H.~Dondi and M.~Sohnius,
  {\it Supergauge Transformations With Isospin Symmetry,}
  Nucl.\ Phys.\  B {\bf 81} (1974) 317.
%
\bibitem{ss}
A.~Salam and J.~A.~Strathdee,
  {\it Unitary Representations Of Supergauge Symmetries},
  Nucl.\ Phys.\  B {\bf 80} (1974) 499-505.
%
\bibitem{susyd}
 J.~A.~Strathdee,
  {\it Extended Poincar\'e Supersymmetry},
  Int.\ J.\ Mod.\ Phys.\  A {\bf 2} (1987) 273.
\bibitem{sugra}
 D.~Z.~Freedman, P.~van Nieuwenhuizen and S.~Ferrara,
  {\it Progress Towards A Theory Of Supergravity},
  Phys.\ Rev.\  D {\bf 13},(1976) 3214-3218;
%
S.~Deser and B.~Zumino,
 {\it Consistent Supergravity},
  Phys.\ Lett.\  B {\bf 62} (1976) 335.
%
\bibitem{cornwell}
J.~F.~Cornwell,
{\it Group Theory in Physics, Vol 3: Supersymmetries and Infinite
Dimensional Algebras}, 
(London, UK: Academic 1989  (Techniques of physics, 10)).
%
\bibitem{superspace}
A.~Salam and J.~A.~Strathdee,
  {\it Supergauge Transformations},
  Nucl.\ Phys.\  B {\bf 76} (1974) 477-482.
%
\bibitem{manifold}
B. DeWitt, {\it Supermanifolds}, (Cambridge University Press, Cambridge 1992
$-$second edition$-$).
%
\bibitem{superfield}
 A.~Salam and J.~A.~Strathdee,
  {\it On Superfields And Fermi-Bose Symmetry},
  Phys.\ Rev.\  D {\bf 11} (1975) 1521-1535.
%
\bibitem{superym}
J.~Wess and B.~Zumino,
  {\it Supergauge Invariant Extension Of Quantum Electrodynamics},
  Nucl.\ Phys.\  B {\bf 78} (1974) 1;
%
 A.~Salam and J.~A.~Strathdee,
{\it   Supersymmetry And Nonabelian Gauges},
  Phys.\ Lett.\  B {\bf 51} (1974) 353;
%
S.~Ferrara and B.~Zumino,
  {\it Supergauge Invariant Yang-Mills Theories},
  Nucl.\ Phys.\  B {\bf 79} (1974) 413-421.
%
\bibitem{fsusy}
C.~Ahn, D.~Bernard and A.~ Leclair,
{\it Supersymmetries in perturbed coset  CFTs and integrable soliton theory},
 Nucl. Phys. {\bf B346} (1990) 409;
%
J.~ L.~ Matheus-Valle and Marco A.~ R.~ Monteiro,
{\it Quantum Group Generalization Of The Classical Supersymmetric 
Point Particle},
 Mod. Phys. Lett. {\bf A7}
(1992) 3023;
%
S.~ Durand,
{\it Fractional Superspace Formulation Of Generalized Mechanics},
 Mod. Phys. Lett {\bf A8}  (1993) 2323 [hep-th/9305130];
%
N.~ Debergh, J. Phys.
{\it  On a q deformation of the supersymmetric Witten model},
 {\bf A26} (1993) 7219;
%
N.~ Mohammedi, 
{\it Fractional Supersymmetry},
 Mod. Phys. Lett. { \bf A10} (1995) 1287 [hep-th/9412133];
%
L.~ P.~ Colatto and J.~ L.~ Matheus-Valle, 
{\it On q deformed supersymmetric classical mechanical models},
J. Math. Phys. {\bf 37} (1996)
6121 [hep-th/9504101];
%
J.~A.~ de Azc\`arraga and A.~J.~ Macfarlane, 
{\it Group Theoretical Foundations Of Fractional Supersymmetry},
J. Math .Phys. {\bf 37}
(1996) 1115 [hep-th/9506177];
N.~ Fleury and M. Rausch de Traubenberg,
{\it Local Fractional Supersymmetry For Alternative Statistics},
 Mod. Phys. Lett. {\bf A11} (1996)
899 [hep-th/9510108];
%
%
R.~S.~Dunne, A.~J.~Macfarlane, J.~A.~de Azc\'arraga and J.~C.~P\'erez Bueno,
{\it Geometrical foundations of fractional supersymmetry,}
{  Int. J.  Mod. Phys. } {\bf A 12} (1997) 3275;
[hep-th/9610087];
%
 M.~Daoud and M.~Kibler,
 {\it A Fractional Supersymmetric Oscillator and Its Coherent States,}
  arXiv:math-ph/9912024;
%
K.~Aghababaei Samani and A.~Mostafazadeh,
{\it Quantum Mechanical Symmetries and Topological Invariants},
{ Nucl.\ Phys.\ }  {\bf B595} (2001) 467
[hep-th/0007008];
%
A.~Perez, M. Rausch de Traubenberg and P. Simon,
{\it 2D Fractional Supersymmetry For Rational Conformal Field Theory: 
Application For Third Integer Spin States},
 Nucl. Phys. {\bf B482} (1996) 325 [hep-th/9603149];
%
M. Rausch de Traubenberg and P. Simon, 
{\it 2D Fractional Supersymmetry And Conformal Field Theory 
For Alternative Statistics},
Nucl. Phys. {\bf B517} (1998) 485
[hep-th/9606188];
%
J.~ L.~Matheus-Valle and Marco A.~ R.~Monteiro, 
{\it Quantum group generalization of the heterotic QFT},
 Phys. Lett. {\bf B300}
(1993) 66;
%
E.~H.~Saidi, M.~B.~Sedra and J.~Zerouaoui, 
{\it On D = 2 (1/3, 1/3) Supersymmetric Theories 1},
Class. and
Quantum Grav.  {\bf 12} (1995) 1567;
%
E.~H.~Saidi, M.~B.~Sedra and J.~Zerouaoui, 
{\it On D = 2 (1/3, 1/3) Supersymmetric Theories 2},
Class.Quant.Grav.{
 \bf 12} (1995) 2705;
%
H.~Ahmedov and O.~F.~Dayi,
{\it Non-Abelian fractional supersymmetry in two dimensions},
{ Mod.\ Phys.\ Lett.\ }  {\bf A15} (2000) 1801
[math.qa/9905164];
%
H.~Ahmedov and O.~F.~Dayi,
{\it Two dimensional fractional supersymmetry from the quantum 
Poincar\'e  group at roots of unity}, 
{ J.\ Phys.\ } {\bf A32} (1999) 6247
[math.qa/9903093];
%
F.~Kheirandish and M.~Khorrami,
{\it Logarithmic two-dimensional spin-1/3 fractional supersymmetric  
conformal field theories and the two point functions},
{ Eur.\ Phys.\ J.\ } {\bf C18} (2001) 795
[hep-th/0007013];
%
F.~Kheirandish and M.~Khorrami,
{\it Two-dimensional fractional supersymmetric conformal field theories and  
the two-point functions},
{   Int. J .\ Mod.\ Phys.}\ {\bf A 16} (2001) 2165
[hep-th/0004154];
%
S.~ Durand, 
{\it Fractional superspace formulation of generalized superVirasoro algebras},
Mod. Phys. Lett. {\bf A7}  (1992) 2905 [hep-th/9205086];
%
A.~T.~Filippov, A.~P.~Isaev and A.~B.~Kurdikov,
{\it ParaGrassmann extensions of the Virasoro algebra},
{ Int.\ J.\ Mod.\ Phys.\ }  {\bf A8} (1993) 4973 [hep-th/9212157].
%
%
\bibitem{cr}
R.~Campoamor-Stursberg and M.~Rausch de Traubenberg,
  {\it Kinematical superalgebras and Lie algebras of order 3,}
  J.\ Math.\ Phys.\  {\bf 49} (2008) 063506
  [arXiv:0801.2630 [hep-th]].
%
\bibitem{hopf}
M.~Goze and M. Rausch de Traubenberg,
 {\it Hopf algebras for ternary algebras and groups},
  arXiv:0809.4212 [math-ph].
%
\bibitem{bar}
V.~Bargmann,
{\it Irreductible unitary representations of the Lorentz group},
 Ann. of Math {\bf 48} (1947) 568-640.
\bibitem{ggv-l}
I. Gel'fand, M. Graev and N. Vilenkin,
{\it Generalized functions} (Academic, New-York, 1966) Vol 5;
S. Lang $SL_2 \mathbb R$, (Springer, Berlin, 1985).
%
%
\bibitem{b}
B.~ Binegar,
{\it Relativistic Field Theories In Three-Dimensions}
  J.\ Math.\ Phys.\  {\bf 23} (1982) 1511.
%
\bibitem{jn}
R.~Jackiw and V.~P.~ Nair, 
 {\it Relativistic wave equations for anyons},
  Phys.\ Rev.\  D {\bf 43} (1991) 1933.
%
\bibitem{p}
M.~S.~Plyushchay,
 {\it Relativistic particle with torsion, Majorana equation and fractional
spin},
  Phys.\ Lett.\  B {\bf 262} (1991) 71;
M.~S.~Plyushchay, 
{\it The Model of relativistic particle with torsion},
  Nucl.\ Phys.\  B {\bf 362} (1991) 54;
M.~S.~Plyushchay, 
{\it Fractional spin: Majorana-Dirac field},
  Phys.\ Lett.\  B {\bf 273} (1991) 250;
%
 J.~L.~Cortes and M.~S.~Plyushchay,
  {\it Linear differential equations for a fractional spin field,}
  J.\ Math.\ Phys.\  {\bf 35} (1994) 6049
  [arXiv:hep-th/9405193].
%
\bibitem{p2}
 M.~S.~Plyushchay,
  {\it R-deformed Heisenberg algebra, anyons and d = 2+1 supersymmetry},
  Mod.\ Phys.\ Lett.\  A {\bf 12} (1997) 1153
  [arXiv:hep-th/9705034].
%
\bibitem{vir}
M.~Rausch de Traubenberg,
 {\it Fractional supersymmetry and infinite dimensional Lie algebras}, 
  Nucl.\ Phys.\ Proc.\ Suppl.\  {\bf 102} (2001) 256
  [arXiv:hep-th/0109106].
%
%
\bibitem{pr}
 S.~M.~Klishevich, M.~S.~Plyushchay and M.~Rausch de Traubenberg,
 {\it Fractional helicity, Lorentz symmetry breaking and anyons},
  Nucl.\ Phys.\  B {\bf 616} (2001) 419
  [arXiv:hep-th/0101190].
%
\bibitem{cubic1}
N.~Mohammedi, G.~Moultaka and M.~Rausch de Traubenberg,
{\it Field theoretic realizations for cubic supersymmetry,}
Int.\ J.\ Mod.\ Phys.\ A {\bf 19},  (2004), 5585--5608 [arXiv:hep-th/0305172].
%
\bibitem{cubic2}
G.~Moultaka, M.~Rausch de Traubenberg and A.~Tanasa,
{\it Cubic supersymmetry and abelian gauge invariance,}
Int.\ J.\ Mod.\ Phys.\ A {\bf 20}, (2005), 5779--5806 [arXiv:hep-th/0411198].
%
\bibitem{noether}
 M.~Rausch de Traubenberg,
 {\it Four dimensional cubic supersymmetry},
 Pr. Inst. Mat.
Nats. Akad. Nauk Ukr. Mat. Zastos., 50, Part 1, 2, 3, Natsional. Akad. Nauk 
Ukraïni, \~Inst. Mat., Kiev, 2004,
pp. 578-585,
  [arXiv:hep-th/0312066].
%
%
\bibitem{bl}
H.~Bacry and J.~L\'evy-Leblond,  {\it Possible kinematics}, J. Math.
Phys. {\bf 9} (1968) 1605-1614;
 J. F. ~Cari\~{n}ena, M. A. ~del
Olmo and M. ~Santander, {\it Kinematic groups and dimensional analysis},
J. Phys. A: Math. Gen. {\bf 14} (1981) 1-14.
%
%
\bibitem{h}
M.~Henneaux and B.~Knaepen,
{\it  All consistent interactions for exterior form gauge fields},
 Phys. Rev. \textbf{D 56} (1997) 6076,
 [arXiv:hep-th/9706119];
 M.~Henneaux and B.~Knaepen,
{\it A Theorem on first order interaction vertices for 
free p form gauge fields},
 Int.
J. Mod. Phys. \textbf{A15} (2000) 3535-3548
 [arXiv:hep-th/9912052].
%
\bibitem{A-B}
P.~Meessen and T.~Ortin,
{\it An Sl(2,Z) multiplet of nine-dimensional type II supergravity theories},
  Nucl.\ Phys.\  B {\bf 541}, 195 (1999)
  [arXiv:hep-th/9806120].
%
\bibitem{ghosts}
R.~K.~Kaul,
{\it Quantization Of Free Field Theory Of Massless Antisymmetric Tensor Gauge
Field Of Second Rank},
{ Phys.\ Rev.\ }  {\bf D 18} (1978) 1127.
%
\bibitem{pol}
J.~Polchinski,
{\it Dirichlet Branes and Ramond-Ramond charges},
 Phys. Rev. Lett. \textbf{75} (1995) 4724
[\texttt{arXiv:hep-th/9510017}]; 
%
J.~Polchinski and Y.~Cai,
{\it Consistency of Open Superstring Theories},
 Nucl.
Phys. \textbf{B296} (1988) 91.
%
P.~K.~Townsend,
{\it Covariant Quantization Of Antisymmetric Tensor Gauge Fields},
{\ Phys.\ Lett.\  }  {\bf B88} (1979) 97;
%
M.~A.~Namazie and D.~Storey,
{\it On Secondary And Higher Generation Ghosts},
{\ J.\ Phys.\ }  {\bf A13} (1980) L161;
%
T.~Kimura,
{\it Quantum Theory Of Antisymmetric Higher Rank Tensor Gauge Field In Higher
Dimensional Space-Time},
{ Prog.\ Theor.\ Phys.\  } {\bf 65} (1981) 338;
%
W.~Siegel,
{\it Hidden Ghosts},
{\ Phys.\ Lett.\ }  {\bf B 93} (1980) 170;
%
H.~Hata, T.~Kugo and N.~Ohta,
{\it Skew Symmetric Tensor Gauge Field Theory Dynamically Realized 
In QCD U(1) Channel},
{ Nucl.\ Phys.\  }  {\bf B 178} (1981) 527;
%
J.~Thierry-Mieg,
{\it BRS Structure Of The Antisymmetric Tensor Gauge Theories},
{ Nucl.\ Phys.\  }  {\bf B 335} (1990) 334.
%
\bibitem{cs}
E.~Cremmer and J.~Scherk,
{\it Spontaneous Dynamical Breaking Of Gauge Symmetry In Dual Models,}
{ Nucl.\ Phys.\ }  {\bf B 72} (1974) 117-124.
%
\bibitem{nambu}
S.~Deser, 
{\it Lagrangian forms of the dynamical theory of currents},
{ Phys. Rev.} {\bf 187} (1969) 1931-1969;
M.~Kalb and P.~Ramond,
{\it Classical Direct Interstring Action,}
{ Phys. \ Rev.\ }  {\bf D9} (1974) 2273;
Y.~Nambu,
{\it Magnetic And Electric Confinement Of Quarks,}
{ Phys.\ Rept.\ }  {\bf 23} (1976) 250.
%
%
\bibitem{3f}
E.~Sezgin and P.~van Nieuwenhuizen,
{\it Renormalizability Properties Of Antisymmetric Tensor Fields Coupled To
Gravity,}
{ Phys.\ Rev.\ }   {\bf D 22}, 301 (1980);
P.~K.~Townsend,
{\it Classical Properties Of Antisymmetric Tensor Gauge Fields,}
CERN-TH-3067;
{\it Lecture given at 18th Winter School of Theoretical Physics, Karpacz,
 Poland, Feb 18 - Mar 3, 1981}.
%
%
\bibitem{4f}
W.~Siegel,
{\it Quantum Equivalence Of Different Field Representations,}
Phys.\ Lett.\ B {\bf 103} (1981) 107.
%
%
\bibitem{rdha}
Y.~Ohnuki, S.~Kamefuchi,
{\it Quantum Field Theory and Parastatistics}
(Springer Verlag, 1982);
A.~J. Macfarlane, {\it Generalized Oscillator Systems
and Their Parabosonic Interpretation}, in Proc. {\it
International Workshop on Symmetry Methods in Physics},
Eds. A.~N.~Sissakian, G.~S.~Pogosyan and
S.~I.~Vinitsky (JINR, 1994) p. 319;
M.~S.~Plyushchay,
{\it Deformed Heisenberg algebra with reflection},
 Nucl.\ Phys.\  B {\bf 491}, 619 (1997)
  [arXiv:hep-th/9701091].
%
\bibitem{kr}
V.~G.~Kac and A.~K.~Raina, {\it Bombay Lectures on Highest Weight 
Representations of
Infinite Dimensional Lie Algebras}, World Scientific (Singapore, New Jersey, Hong Kong, 1987).
%
\bibitem{rr}
N.~ Roby, 
{\it Algèbres de Clifford des formes polynomes}  (in French), 
C. R. Acad. Sc. Paris {\bf 268} (1969) A484--A486;
Ph. Revoy,  
{\it Nouvelles algèbres de Clifford}, (in French),
C. R. Acad. Sc. Paris {\bf 284} (1977) A985--A988.
%
\bibitem{ht}
D.~H.~Haile et S.~Tesser, 
{\it On Azumaya algebras arising from Clifford algebras},
J. Alg. {\bf 116} (1988) 372.
%
%
\bibitem{frr}
N. Fleury and M. Rausch de Traubenberg,
{\it Linearization of Polynomials}, J. Math. Phys. {\bf 33} (1992)
3356;
N. Fleury and M. Rausch de Traubenberg, 
{\it Finite dimensional representations of Clifford algebra
of polynomials}, Adv. Apl. Clif. Alg. {\bf 4} (1994)
123.
%
\bibitem{re}
Ph. Revoy, 
{\it Alg\`ebre de Clifford d'un polyn\^ome} (in French),
Adv. Appl. Cliff. Alg. {\bf 3} (1993) 39-55.
%
\bibitem{gca}
A.~O.~Morris, {\it On a generalized Clifford algebra},
Quart. J. Math., Oxford {\bf 18} (1967) 7--12;
A.~O.~Morris, {\it On a generalized Clifford algebra. II},
Quart. J. Math., Oxford {\bf 19} (1968) 289--299. 

%
\bibitem{roby-ext} N. Roby, \textit{L'alg\`ebre $h-$ext\'erieure d'un module
libre} (in French), Bull. Sc. Math. \textbf{94} (1970) 49-57.
%
\bibitem{clif} M. Rausch de Traubenberg,  {\it Clifford algebras
in physics}, to appear in Proceedings of the 7th International
Conference on Clifford algebras and their applications. (ICCA),
May 2005, Toulouse,  arXiv:hep-th/0506011.

%


\end{thebibliography}
\end{document}